\documentclass[aps,prx,twocolumn,superscriptaddress,floatfix,10pt]{revtex4-1}
\usepackage[utf8]{inputenc}
\usepackage{natbib}
\usepackage{graphicx}
\usepackage{latexsym}
\usepackage{amssymb}
\usepackage{amsmath}
\usepackage{amsfonts}
\usepackage{amsthm}
\usepackage{bm}
\usepackage{floatrow}
\usepackage[caption=false]{subfig}
\usepackage{bbm}
\usepackage{enumitem}
\usepackage{hyperref}
\usepackage{lipsum}
\usepackage{braket}
\usepackage{tikz}
\usepackage{pgfplots}
\usepackage{comment}
\usepackage{ifthen}
\usepackage{ragged2e}
\usepackage{multirow}
\usepackage[export]{adjustbox}
\usetikzlibrary{arrows,decorations.pathreplacing,decorations.markings,arrows.meta,patterns,3d}
\usepackage[normalem]{ulem} 
\usepackage{cancel}

\theoremstyle{definition}

\definecolor{arpit}{RGB}{127,0,0}

\newcommand{\F}[1]{{}^#1\!F}
\newcommand{\Fi}[1]{{}_#1\mspace{-1mu}F}

\newcommand{\swap}{\mathrm{SWAP}}

\hypersetup{
  colorlinks = true,
  urlcolor = blue,
}
\pgfplotsset{compat=1.18}

\begin{document}

\title{Sequential Quantum Circuits as Maps between Gapped Phases}
\date{\today}


\author{Xie Chen}
\affiliation{Walter Burke Institute for Theoretical Physics, Caltech, Pasadena, CA, USA \looseness=-1}
\affiliation{Department of Physics, Caltech, Pasadena, CA, USA}
\affiliation{Institute for Quantum Information and Matter, Caltech, Pasadena, CA, USA \looseness=-1}

\author{Arpit Dua}
\affiliation{Department of Physics, Caltech, Pasadena, CA, USA}
\affiliation{Institute for Quantum Information and Matter, Caltech, Pasadena, CA, USA \looseness=-1}

\author{Michael Hermele}
\affiliation{Department of Physics and Center for Theory of Quantum Matter, University of Colorado, Boulder, CO 80309, USA}

\author{David T. Stephen}
\affiliation{Department of Physics, Caltech, Pasadena, CA, USA}
\affiliation{Institute for Quantum Information and Matter, Caltech, Pasadena, CA, USA \looseness=-1}
\affiliation{Department of Physics and Center for Theory of Quantum Matter, University of Colorado, Boulder, CO 80309, USA}

\author{Nathanan Tantivasadakarn}
\affiliation{Walter Burke Institute for Theoretical Physics, Caltech, Pasadena, CA, USA \looseness=-1}
\affiliation{Department of Physics, Caltech, Pasadena, CA, USA}

\author{Robijn Vanhove}

\affiliation{Department of Physics, Caltech, Pasadena, CA, USA}
\author{Jing-Yu Zhao}
\affiliation{Institute for Advanced Study, Tsinghua University, Beijing 100084, China \looseness=-1}

\begin{abstract} 
Finite-depth quantum circuits preserve the long-range entanglement structure in quantum states and map between states within a gapped phase. To map between states of different gapped phases, we can use \textit{Sequential Quantum Circuits} which apply unitary transformations to local patches, strips, or other sub-regions of a system in a sequential way. The sequential structure of the circuit on the one hand preserves entanglement area law and hence the gapped-ness of the quantum states. On the other hand, the circuit has generically a linear depth, hence it is capable of changing the long-range correlation and entanglement of quantum states and the phase they belong to. In this paper, we discuss systematically the definition, basic properties, and prototypical examples of sequential quantum circuits that map product states to GHZ states, symmetry-protected topological states, intrinsic topological states, and fracton states. We discuss the physical interpretation of the power of the circuits through connection to condensation, Kramers-Wannier duality, and the notion of foliation for fracton phases.
\end{abstract}

\maketitle


Understanding the nature of entanglement in ground state wave functions has led to important developments in the theoretical understanding and classification of quantum phases of matter. In particular, for zero-temperature gapped phases, it was understood that gapped ground states connected by a finite-depth local unitary quantum circuit -- quantum circuits composed of a finite number of layers of non-overlapping local unitaries -- have the same `long-range entanglement' structure and are hence in the same phase \cite{Chen2010}. This understanding has been helpful in the classification or systematic construction of gapped phases in various dimensions.

To map between ground states of different gapped phases, what kind of quantum circuits do we need? It has been proven in Ref.~\onlinecite{Bravyi2006} that, to map from a product state (ground state of a trivial phase) to either the GHZ state (ground state of a symmetry breaking phase) or a topological state, a quantum circuit of at least linear depth is needed. The intuition is simply that we need an `effort' or more precisely time that grows linearly with the total system size in order to establish the long-range correlation in the GHZ state or the long-range entanglement in the topological states. On the other hand, however, applying a generic linear depth circuit to a product state could easily lead to too much correlation and entanglement that cannot be accommodated in gapped ground states. Therefore, the question becomes: which subset of linear (or higher) depth circuits can map between gapped phases and not beyond? 

\begin{figure}[ht]
    \centering
    \includegraphics[scale=0.3]{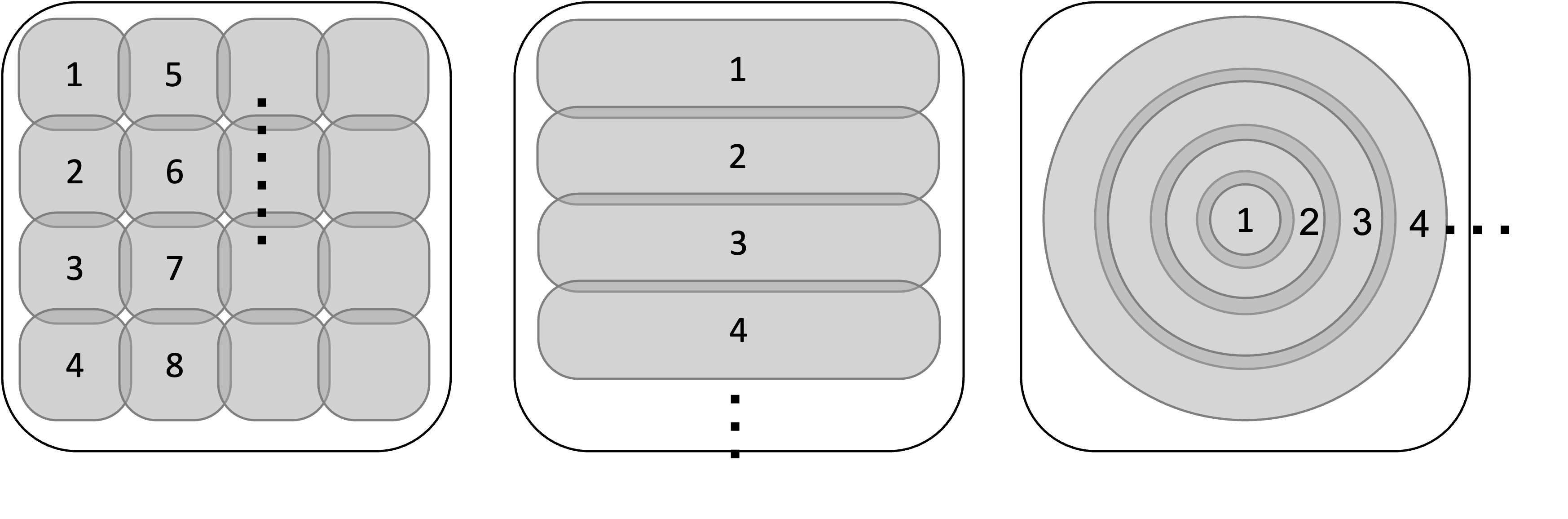}
    \caption{Examples of Sequential Quantum Circuits in 2D where (a) a sequence of local unitary gates are applied to local patches, or a sequence of finite depth quantum circuits are applied to (b) strips or (c) annuli.}
    \label{fig:sequential}
\end{figure}

The \textit{Sequential Quantum Circuit} \cite{Schon2005,Schon2007,Banuls2008,Wei2022} (SQC) is such a subset. As the name suggests, sequential quantum circuits are local unitary quantum circuits that act in a sequential way. As shown in Fig.~\ref{fig:sequential}, a sequence of local unitary quantum gates or finite depth quantum circuits can be applied to local patches, strips, annuli, or other lower dimensional sub-regions in the system one at a time. To cover the whole system, the circuit depth scales as the number of sub-regions, which can be constant, linear, or higher in the linear size of the system. As the action of each layer is restricted to one sub-region, it ensures that each local region in the system gets acted upon by only a finite number of local unitaries in the whole circuit. We use this as the defining feature of a sequential quantum circuit. That is,

\textit{A quantum circuit (composed of local unitary gates) is called a Sequential Quantum Circuit if each local degree of freedom is only acted upon by a finite number of gates in the circuit.}

Finite depth circuits are hence examples of sequential quantum circuits, although only a small subset.
Sequential quantum circuits have been used to propose efficient generation methods for matrix product states and a large class of tensor product states and extensively discussed \cite{Schon2005,Schon2007,Banuls2008,Wei2022}.
They also play an important role in various quantum protocols. See for example Ref.~\onlinecite{Lin2021, Lamata2008,Saberi2011}. 

A direct consequence of the sequential structure is that, if the initial quantum state satisfies the entanglement area law, the final state also satisfies the entanglement area law. Therefore, gapped systems remain gapped under the action of SQC. (When we say a state is gapped, it means the state is the gapped ground state of a local Hamiltonian.) On the other hand, the linear depth gives SQC the power to generate or change the long-range correlation and long-range entanglement in the quantum state, hence mapping between different quantum phases. 

In this paper, we discuss the sequential quantum circuit that generates various gapped quantum phases, including symmetry breaking phases (section~\ref{sec:SB}), symmetry protected topological phases (section~\ref{sec:SPT}), $2+1$D topological phases (section~\ref{sec:TO2D}), $3+1$D topological phases and Walker Wang models (section~\ref{sec:TO3D}),  and fracton phases (section~\ref{sec:fracton}). In section~\ref{sec:qca}, we show that all locality preserving unitary operators, \textit{i.e.} quantum cellular automata, can be realized as sequential quantum circuits. We discuss circuits acting in different dimensions and in the presence of different symmetries. We note that some sequential circuits in the context of generating gapped phases of matter have been discussed in previous literature \cite{Schon2005,Liu2022, Huang2015, Ho2019, Chen2022arxiv}. We interpret the power of the SQC by connecting the circuit action to Kramers-Wannier duality between symmetric and symmetry-breaking phases, condensation on gapped boundaries of topological / Walker Wang models, the foliation structure of fracton models, etc. Note that to demonstrate that a sequential quantum circuit can connect a generic state in one gapped phase to a generic state in another gapped phase, we only need to show how to map between particular chosen states (usually fixed point states) in the two phases. The mapping from a generic state to the fixed point state in the same phase can be accomplished by an additional finite depth circuit.


\section{Map to symmetry breaking phases}
\label{sec:SB}

In this section, we consider the mapping between symmetric and symmetry-breaking phases. We will first focus on the prototypical example of the $1+1$D transverse field Ising model, before generalizing it to all dimensions and all finite symmetry groups. We show that the circuit preserves the global $Z_2$ symmetry of the $1+1$D transverse field Ising model, but does not preserve locality.

\begin{figure}[ht]
    \centering
\includegraphics[scale=0.25]{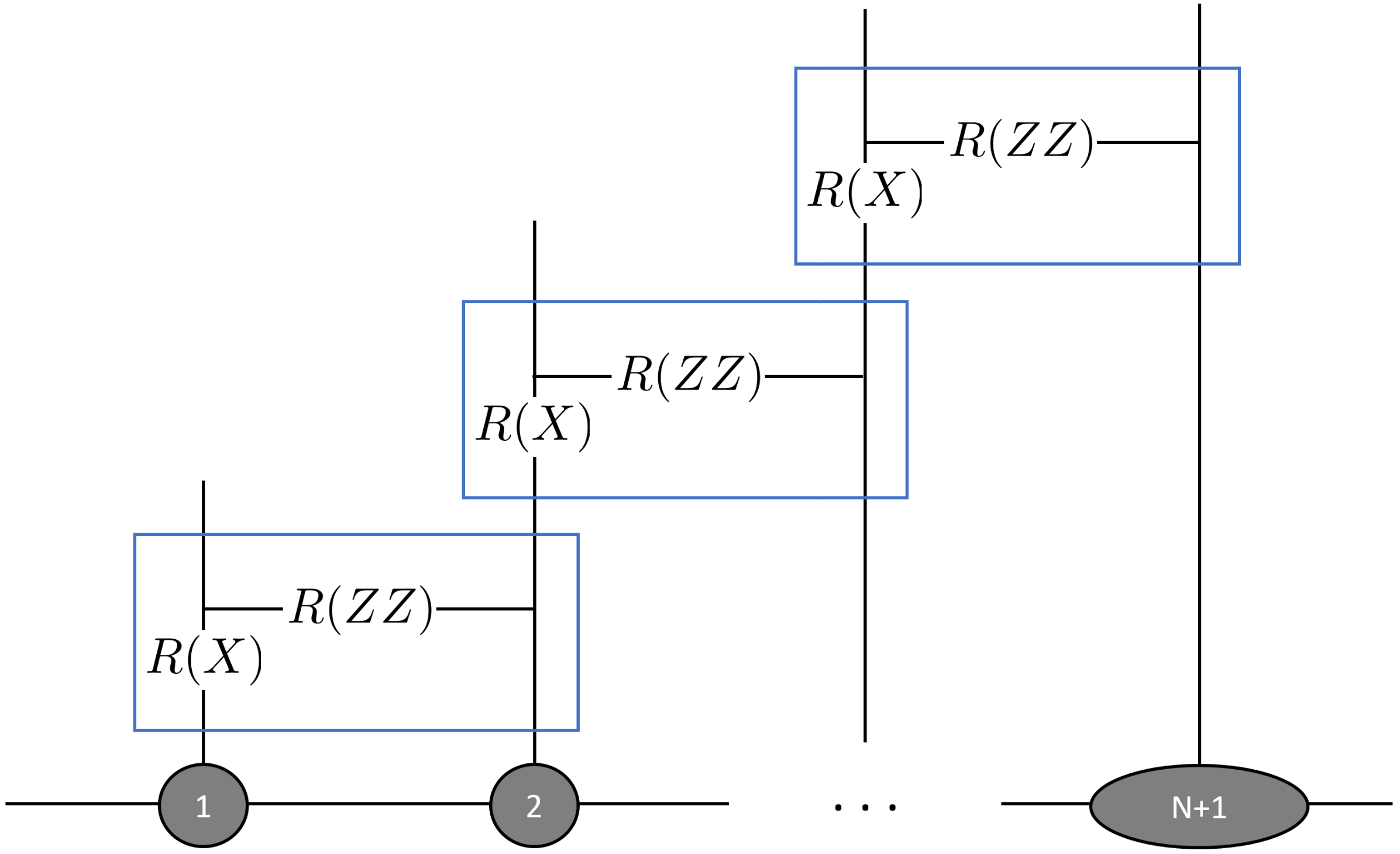}
    \caption{Sequential circuit that maps from the symmetric state to the symmetry breaking GHZ state of the $1+1$D transverse field Ising model. Each blue box represents the unitary $U_{i,i+1}$ in Eq.~\eqref{eq:U1D}. Site $N+1$ is the same as site $1$.}
    \label{fig:SBcircuit}
\end{figure}

The transverse field Ising model in $1+1$D,
\begin{equation}\label{eq:IsingHam}
H = -J\sum_i Z_iZ_{i+1} -B\sum_i X_i,
\end{equation}
has a symmetric phase ($B>J>0$) and a symmetry breaking phase ($J>B>0$) with respect to the global $Z_2$ symmetry $\prod_i X_i$. The fixed point wave functions of the symmetry preserving phase and the symmetry breaking phase are,
\begin{equation}
\begin{aligned}
&|\psi_{\text{SP}}\rangle = |++...+\rangle, \\
&|\psi_{\text{SB}}\rangle = \frac{1}{\sqrt{2}}|00...0\rangle + \frac{1}{\sqrt{2}}|11...1\rangle,
\end{aligned}
\end{equation}
where $|+\rangle = \frac{1}{\sqrt{2}}\left(|0\rangle + |1\rangle\right)$ and we have chosen the symmetrized GHZ state to represent the symmetry breaking ground space. To motivate the sequential circuit, it is insightful to map the above Hamiltonian to fermions using the Jordan-Wigner transformation, 
\begin{align}
    Z_i Z_{i+1} &\rightarrow i \tilde \gamma_i \gamma_{i+1},\\
    X_i &\rightarrow i \gamma_i \tilde \gamma_i,
\end{align}
where $\gamma_i = c_i+c^\dagger_i$ and $\tilde \gamma_i = -i(c_i-c^\dagger_i)$.
Under such a mapping, the symmetric and symmetry-breaking phases map to the trivial and nontrivial Kitaev chain phases respectively. It is now clear that the sequential circuit that maps between these two phases is composed of Majorana swaps~\cite{Huang2015},

\begin{align}\label{eq:sy2sbRo}
    \mathcal U_F =  e^{\frac{\pi}{4}\tilde \gamma_N \gamma_{1} } \prod_{i=N-1}^1  e^{\frac{\pi}{4} \gamma_{i+1} \tilde \gamma_{i+1} }  e^{\frac{\pi}{4}\tilde \gamma_i \gamma_{i+1} }.
\end{align}
Here, the ordering is chosen such that from right to left the product goes from $1$ to $N-1$.
Mapping back to the spin operators, and defining,
\begin{align}
     R(\mathcal O) \equiv  e^{-\frac{i\pi}{4} \mathcal O},
\end{align}
the desired circuit is therefore,
\begin{align}\label{eq:sy2sb}
    \mathcal U =    R(Z_1 Z_{N} ) \prod_{i=N}^1  U_{i,i+1},
\end{align}
where,
\begin{align}
    U_{i,i+1} = R(X_{i+1})  R( Z_i Z_{i+1}),
    \label{eq:U1D}
\end{align}
as shown in Fig.~\ref{fig:SBcircuit}.

We note that a convenient property for $R(\mathcal O)$ is that for Pauli operators $P$ and $Q$,
\begin{align}
    R(Q) P R(Q)^\dagger = \begin{cases}
    P ;& [P,Q]=0\\
    iPQ; & \{P,Q \}=0
    \end{cases}.
\end{align}
Using the above property, one can immediately see that conjugation by $U_{i,i+1}$ sends $X_{i+1}$ to $Z_i Z_{i+1}$.

The complete mapping of operators using the unitary \eqref{eq:sy2sb} is
\begin{equation}\label{eq:sy2sb_ope}
\begin{array}{lll}
X_1 & \to & X_1X_2...X_{N-1}X_N \cdot Z_1Z_N \\
X_i & \to & Z_{i-1}Z_i, \ i=2,...,N\\
X_1X_2...X_N & \to & X_1X_2...X_N \\
Z_1 & \to & Z_1\\
Z_i & \to & X_iX_{i+1}...X_N \cdot Z_1, i = 2...N \\
Z_iZ_{i+1} & \to & X_i, i =2...N\\
Z_1Z_2 & \to &  X_2...X_N
\end{array}
\end{equation}
Note that in the $\prod_i X_i=1$ subspace, the mapping sends $X_i \rightarrow Z_{i-1}Z_i$ and $Z_i Z_{i+1} \rightarrow X_i$ as desired. A similar circuit has been discussed in Ref.~\onlinecite{Ho2019}.

We want to point out some important features of this circuit. 
\begin{itemize}
\item $|\psi_{\text{SP}}\rangle$ maps to $|\psi_{\text{SB}}\rangle$ and vice versa. 

\item In the bulk of system, the transverse field term maps to the Ising term and vice versa. Near the two end points, the correspondence is lost and some terms map to nonlocal terms. 

\item Applying the circuit twice generates translation by one site in the bulk of the system. 

\item The circuit is of linear depth, which saturates the lower bound required to generate long-range correlation in the GHZ state from a product state. 

\item The circuit is symmetric. That is, the circuit is composed of local gates (gate sets in the blue boxes) that commute with the global $Z_2$ symmetry $\prod X_i$. 

\item The circuit is not locality-preserving, as for example the operator $X_1$ maps to $Z_1Z_N\cdot X_1X_2...X_N$. This is also necessary because it is known that locality-preserving unitaries in one spatial dimension are either finite depth circuits or translation, neither is able to generate long-range correlation and map from symmetric to symmetry-breaking phases.\footnote{If we perform the Jordan Wigner transformation and map the transverse field Ising model to the Majorana chain, the mapping between the two phases can be realized by translation by a single Majorana mode.} 

\end{itemize}

\begin{figure}[ht]
    \centering
    \includegraphics[scale=1.2]{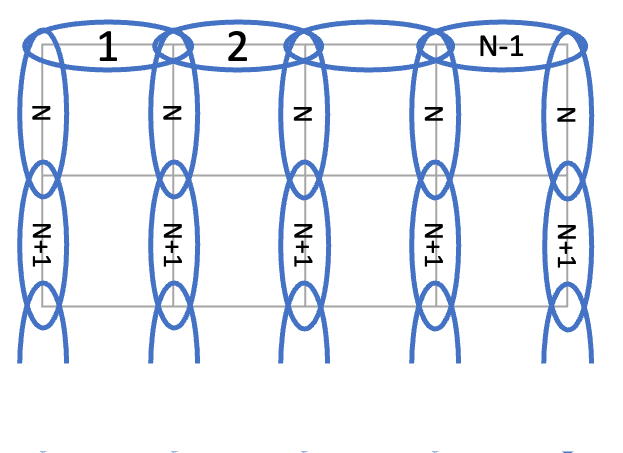}
    \caption{Sequential circuit that maps from symmetric state to the symmetry breaking GHZ state of the $2+1$D transverse field Ising model on a $N\times M$ lattice. The blue ovals correspond to the gate set in the blue boxes in Fig.~\ref{fig:SBcircuit}. The numbers in the ovals indicate the layer numbers in the sequential circuit. The horizontal gates act on the top layer only.}
    \label{fig:SBcircuit2d}
\end{figure}

In appendix~\ref{ap:SBnonabelian}, we generalize the circuit to all finite groups. Higher dimensional versions of the circuit can be built starting from the $1+1$D version. For example, in $2+1$D, to map from the symmetric state to the symmetry breaking state with $Z_2$ symmetry, we can use the circuit shown in Fig.~\ref{fig:SBcircuit2d}. 

\section{Map to symmetry-protected topological phases}
\label{sec:SPT}

\begin{figure*}
    \centering
    \includegraphics[width=\linewidth]{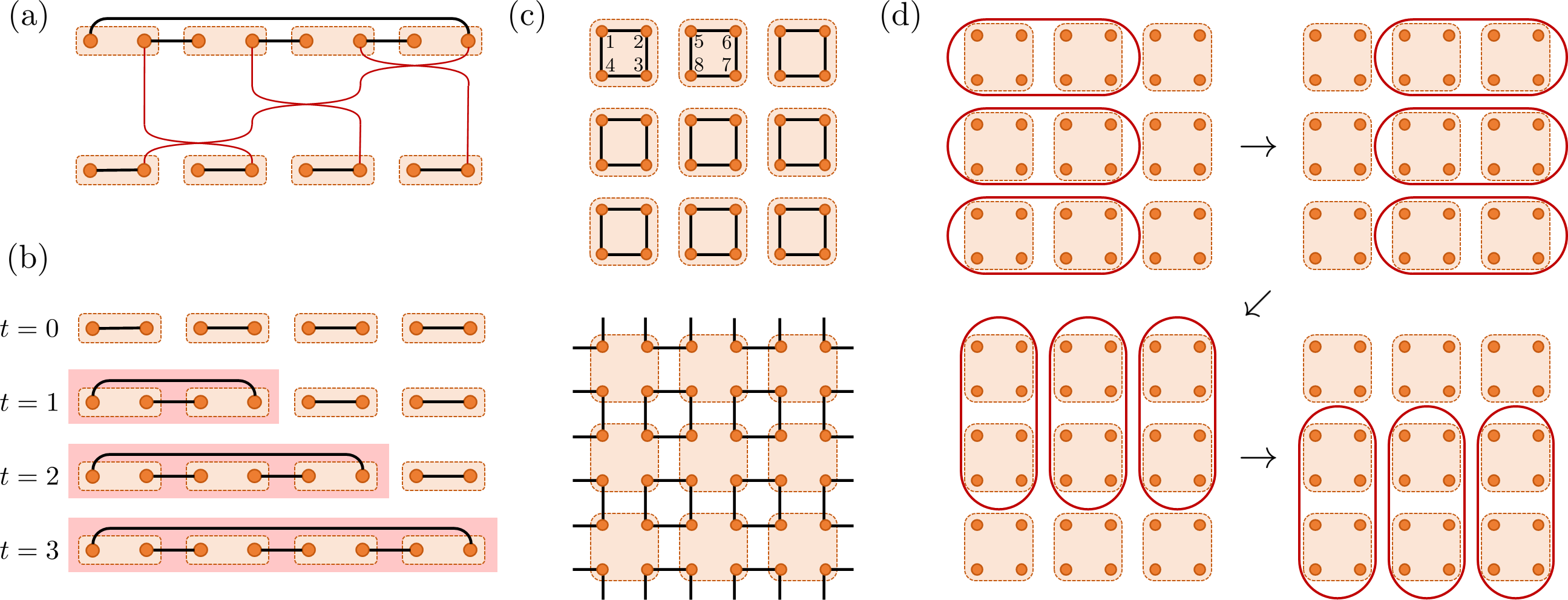}
    \caption{(a) The trivial (bottom) and non-trivial (top) fixed-point states for 1D SPTs on a ring of length 8. Circles indicate qubits, and two circles connected by a black line are in the maximally entangled state $|\Omega\rangle$. The red lines indicate the linear-depth circuit of $\swap$s which map between the two states. (b) The 1D states created after applying each of the three gates in (a) to the initial trivial state. The red region indicates the region spanned by the non-trivial SPT with periodic boundaries, which grows with each applied gate. (c) The trivial (upper) and non-trivial (lower) fixed-point 2D SPT states. The shaded regions depict single sites, and solid lines indicate which qubits are entangled. (d) The linear-depth circuit which maps between the trivial and non-trivial 2D SPT states. Each ellipse indicates one application of $\swap^{CCZ}$.}
    \label{fig:spt}
\end{figure*}

In this section, we construct symmetric, linear-depth SQCs which generate fixed points of symmetry-protected topological (SPT) phases from symmetric product states. We focus on the bosonic SPT phases described by group cohomology \cite{Chen2013}. 

We begin with 1D SPT phases, which are described by an element $[\omega]$ of the second cohomology class $H^2(G,U(1))$ of the protecting symmetry $G$ \cite{Chen2011,Schuch2011,Pollmann2010}. We can construct fixed-point states realizing such phases by picking a representative cocycle $\omega(g,h)$ and constructing a $d$-dimensional projective representation $V(g)$ such that $V(g)V(h) = \omega(g,h)V(gh)$ \cite{Chen2011}. Define the $d^2$-dimensional linear representation $U(g)=V(g)\otimes V(g)^*$ where $*$ is complex conjugation. Trivial and non-trivial representative fixed-point states with this symmetry can be defined on a chain of $N$ sites, where the Hilbert space of each site has dimension $d^2$ and consists of two $d$-dimensional particles. In the trivial state, the two particles within each site are in the maximally entangled state $|\Omega\rangle =\sum_{i=0}^{d-1}|ii\rangle$, so the state is a product state. In the non-trivial state, the right particle of one site is maximally entangled with the left particle of the next site. This is pictured in Fig.~\ref{fig:spt}(a).
These states are both symmetric under the symmetry $U(g)^{\otimes N}$. It was shown in Ref.~\onlinecite{Huang2015} that these two states cannot be related by a finite-depth symmetric circuit, but that they can be related by a linear-depth symmetric circuit of $\swap$ gates, as
pictured in Fig.~\ref{fig:spt}(a). Each $\swap$ gate commutes with the symmetry $U(g)^{\otimes N}$, as only particles which transform under the representation $V(g)^*$ are swapped, so this is a linear-depth symmetric SQC mapping a trivial SPT state to a non-trivial SPT state. 

It is interesting to consider the effect of truncating this SQC. Namely, if we decide to stop applying gates at some point in the circuit, what state is left over? This is illustrated in Fig.~\ref{fig:spt}(b), which shows that, after applying some of the gates in Fig.~\ref{fig:spt}(a), the resulting state is a non-trivial 1D SPT on periodic boundary conditions (PBC). This is necessary in order to ensure that the symmetry is preserved after each step. Each gate in the SQC serves to extend the region of the lattice occupied by the non-trivial SPT, and the circuit ends when this region is the entire lattice. This will contrast with the circuits we construct for topological phases in section \ref{sec:TO2D}, in which truncating the SQC can result in a droplet of topological order with open boundaries and a particular gapped boundary to vacuum.

Now we consider 2D SPT phases, focusing on one example and giving the general case in Appendix \ref{app:2dspt}. We consider the CZX model, which is a fixed-point representative of a 2D SPT phase with $\mathbb{Z}_2$ symmetry \cite{Chen2011a} and is defined on a square lattice with four qubits per site, as pictured in Fig.~\ref{fig:spt}(c). Similar to the 1D fixed-point states, the qubits are divided into a tensor product of four-qubit entangled states $\frac{1}{\sqrt{2}}(|0000\rangle + |1111\rangle)$ which are contained within one site in the trivial SPT fixed-point and shared between four sites in the non-trivial SPT fixed-point, see Fig.~\ref{fig:spt}(c). The symmetry acts on each site as $U_{CZX} = CZ_{12}CZ_{23}CZ_{34}CZ_{41} X_1X_2X_3X_4$ where the four qubits within the site are numbered clockwise. It is straightforward to check that $U_{CZX}^2=1$, and that the trivial and non-trivial states are both symmetric under $U_{CZX}^{\otimes N}$ where the tensor product is over all $N$ sites. Therefore, $U_{CZX}$ defines a $\mathbb{Z}_2$ global symmetry of the models.

As these two states are in different phases with respect to the $\mathbb{Z}_2$ symmetry \cite{Chen2011a}, there is no finite-depth symmetric circuit that relates them. To construct a linear-depth SQC, we might first try to mimic the 1D case and use $\swap$ gates to shift the four-qubit entangled states from within the sites to between the sites. This will map between the two states. However, in this case, the $\swap$ gates do not commute with the $Z_2$ symmetry since they drag around the $CZ$ gates involved in $U_{CZX}$. To remedy this, we can dress the $\swap$ gates with additional operations that restore the symmetry without affecting the action of the gate on the fixed-point states. Define the dressed $\swap$ gate acting on two adjacent sites as
\begin{equation}
\begin{aligned}
    \swap^{CCZ} &= \swap_{26}\swap_{37} \\
    &\times CCZ_{265}CCZ_{261}CCZ_{378}CCZ_{374}
\end{aligned}
\end{equation}
where the qubits are numbered as in Fig.~\ref{fig:spt}(c) and $CCZ$ is the controlled-controlled-$Z$ gate which acts on three qubits as $CCZ|ijk\rangle=(-1)^{ijk}|ijk\rangle$.
These $CCZ$ are placed such that the additional phase factors created by commuting the $\swap{}$'s in $\swap^{CCZ}$ past the $CZ$'s in $U_{CZX}$ are canceled by phase factors created by commuting the $CCZ$'s in $\swap^{CCZ}$ past the $X$'s in $U_{CZX}$.
One can indeed check that $\swap^{CCZ}$ acting on any neighbouring pair of sites commutes with $U_{CZX}$, as is shown for a more general case in Appendix \ref{app:2dspt}. 

Now we can use these symmetric $\swap^{CCZ}$ gates to map between the two states. Beginning with the trivial state, we first apply the gates sequentially between each pair of adjacent columns, noting that all gates within a given column can be performed in parallel such that the depth of this step is linear. Then we apply $90^\circ$-rotated versions of the gates sequentially to each row, which also has linear depth. This results in the non-trivial SPT states. It is important that, at each step, the extra $CCZ$ gates included in $\swap^{CCZ}$ cancel out pairwise when acting on the state, such that they do not affect the final state. This procedure is pictured in Fig.~\ref{fig:spt}(d). 

The general construction of dressing $\swap$ operators with additional phases to make symmetric gates is described in Appendix \ref{app:2dspt} for arbitrary 2D SPT phases. From that construction, and our general understanding of fixed-point SPT states in all dimensions \cite{Chen2013}, it is clear that we should be able to construct similar circuits for SPT phases within the group cohomology classification in all dimensions. Indeed, the construction of symmetric SQCs for Quantum Cellular Automata (QCA), described in Sec.~\ref{sec:qca}, can be applied to obtain symmetric SQCs for all SPT states that can be created via a (non-symmetric) FDQC, which includes all group cohomology SPTs \cite{Chen2013} and some beyond cohomology SPTs \cite{Fidkowski2020}.


\section{Map to 2+1D topological phases}
\label{sec:TO2D}

In this section, we discuss the mapping from product states to $2+1$D  topological states with gappable boundary -- the string-net states \cite{levin2005string}. We discuss first the circuit for generating Toric Code (TC) ground states and then generalize to all string-nets. A key feature of our general construction is that we can write down SQCs which generate string net models with arbitrary gapped boundaries to vacuum, as well as models with periodic boundary conditions. 


\begin{figure}
    \centering
    \includegraphics[width=\linewidth]{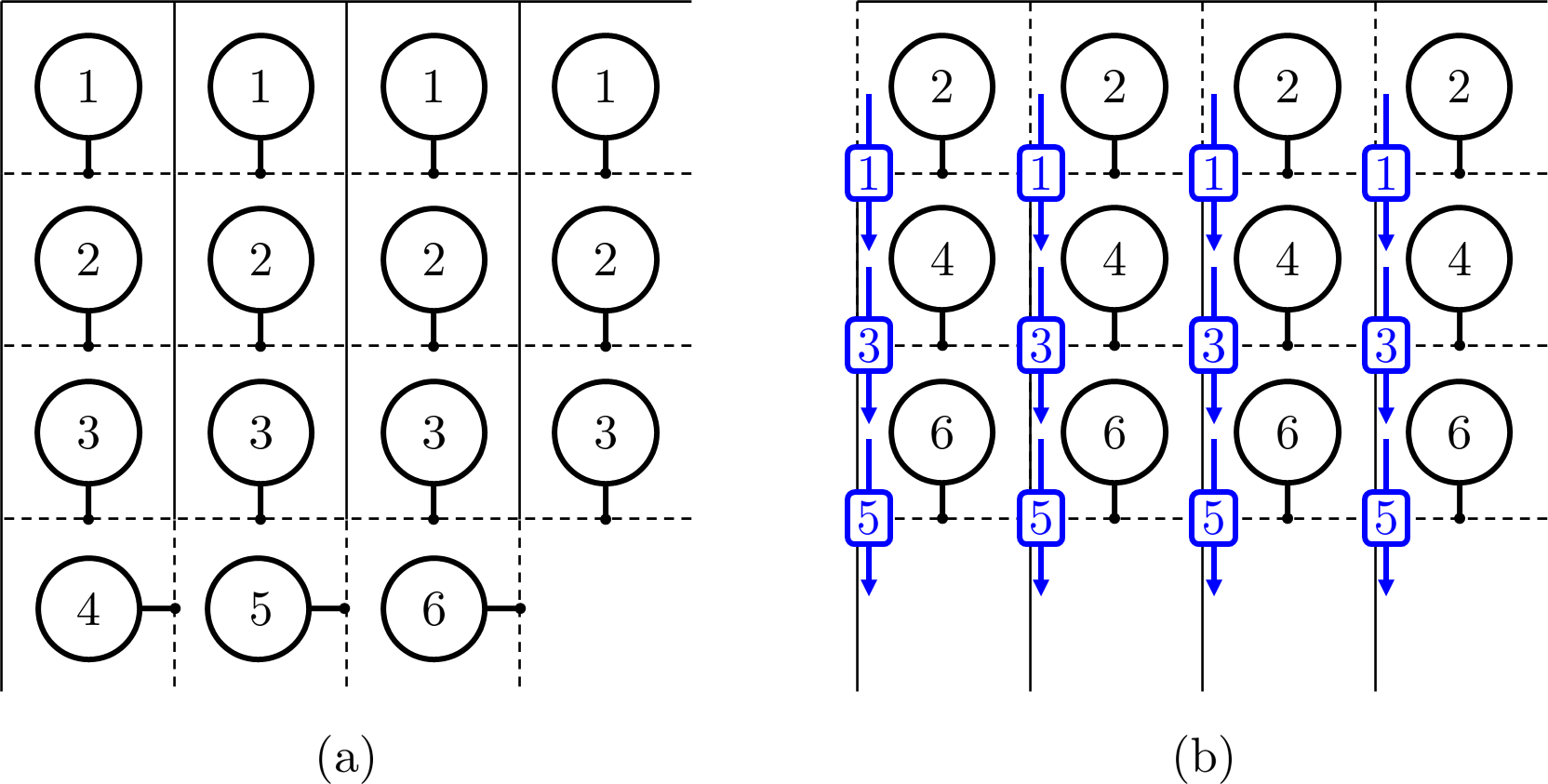}
    \caption{Quantum circuits for creating Toric Code ground states with $L_x=L_y=4$. The lattice has periodic boundary conditions in both directions. Each circle represents the gate defined in Eq.~\ref{mapping_between_gate_loop_notation}, while the blue arrows represent CNOT gates coupling two adjacent vertical edges. Qubits on solid (dashed) edges are initialized in the state $|0\rangle$ ($|+\rangle$), then the gates are applied in the order shown. (a) Circuit which leaves a smooth boundary when truncated. The gates in the bottom row ``zip'' the smooth boundaries together to create the periodic boundary conditions. (b) Circuit which leaves periodic boundaries when truncated.}
    \label{fig:tc_circuits}
\end{figure}

\begin{figure*}
    \centering
    \includegraphics[scale=0.85,page=24]{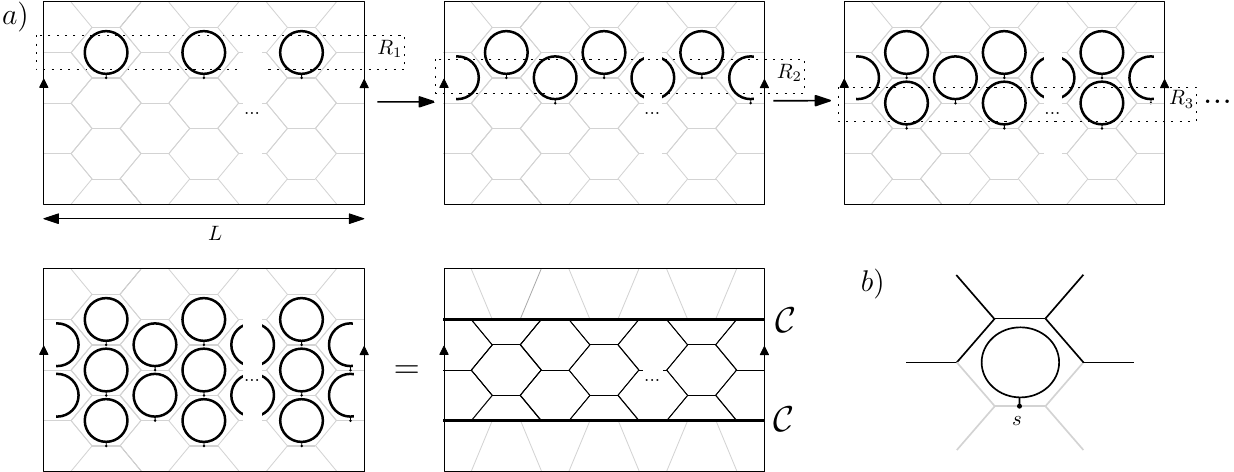}
    \caption{a) We start with a honeycomb lattice with periodic boundary conditions in the horizontal direction and all edges initialized on the trivial object $\mathbf{1}$ (grey). We generate a bulk string-net ground state with two boundaries labeled by $\mathcal{C}$ (`smooth' boundary) by applying the controlled plaquette projector C-$B_p$ (\ref{controlled-Bp}), sequentially, row by row. b) In every step of the circuit, at least three of the edges are trivial, initially, and the bottom edge serves as the control edge.}
    \label{string-netCircuit}
\end{figure*}

\subsection{2+1D Toric Code}
\label{sec:2dTC_seq_circuit}

We start with the ground state of the 2+1D Toric Code \cite{Kitaev2003}. The model has qubit DOFs on edges of a square lattice, and the Hamiltonian is,
\begin{equation} \label{eq:h2dtc}
H_{\text{2dTC}} = -\sum_{\text{vertex}}\prod_{e \ni \text{vertex}} Z_e - \sum_{\text{plaquette}} \prod_{e\in \text{plaquette}} X_e.
\end{equation}
We discuss two distinct SQCs in detail. It is convenient to define the following notation for a unitary gate,
\begin{equation}
    \includegraphics[scale=0.4]{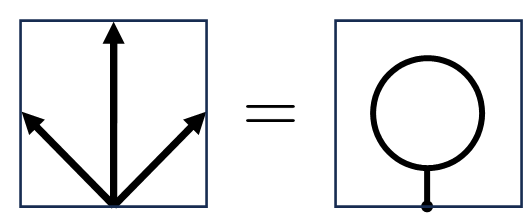}
\label{mapping_between_gate_loop_notation}
\end{equation}   
where the arrows represent CNOT gates acting on the qubits on the edges of the lattice, with the arrow pointing from the control qubit to the target qubit. This gate will be generalized for string-net models in the next section. 

The first circuit closely follows the construction given in Ref.~\onlinecite{Liu2022}, and is pictured in Fig.~\ref{fig:tc_circuits}(a). The circuit consists of two distinct parts. First, we apply parallel gates to rows of the lattice sequentially. Each gate generates one plaquette term of $H_{\text{2dTC}}$. After each row of gates is applied, we are left with a TC state on a cylinder with periodic boundaries in the $x$ direction and smooth \cite{kitaev2012models} boundaries to vacuum (\textit{i.e.} the qubits that are still in a product state) on the top and bottom edges. In other words, each row of gates pushes the gapped boundary one row further into the vacuum, expanding the region spanned by the topological order. The second part of the circuit involves the gates acting on the final row. We act on each plaquette in the final row (except the last) sequentially from left to right. This has the effect of `zipping' the smooth boundaries on the top and bottom of the cylinder, resulting in a TC ground state on the torus. The specific ground state which is obtained is the $+1$ eigenstate of the $Z$-type Wilson loops winding in both directions. 
The overall depth of the SQC scales like $\sim L_x+L_y$ where $L_x$, $L_y$ are the lengths in the $x$ and $y$ directions. 
We can use a similar circuit on the dual lattice to prepare the eigenstate $X$-type Wilson loops in both directions, such that truncating the circuit results in a rough \cite{kitaev2012models} boundary to vacuum.

The second circuit is pictured in Fig.~\ref{fig:tc_circuits}(b). The key difference compared to the first circuit is that, after each step in the circuit, we are left with a TC state with periodic boundaries in both directions. That is, there is never a gapped boundary to vacuum. Rather, the first two steps of the circuit prepare a TC state on a thin torus, and each subsequent row of gates serves to extend the region over which the periodic Toric Code state is defined, similar to the circuits for SPT states defined in the previous section.
Interestingly, because the final `zipping' step is not required (since the periodic boundaries in the vertical direction are present from the start), the depth of this circuit scales only as $\sim L_y$. Accordingly, the ground state of the TC which this circuit prepares is the $+1$ eigenstate of both the $X$ and $Z$-type Wilson loops that wrap around the torus in the $y$ direction.
We remark that the circuit in Fig.~\ref{fig:tc_circuits}(b) can also be run such that all blue CNOTs are first applied in sequence, and then the plaquette operators in Eq.~\ref{mapping_between_gate_loop_notation} are applied in sequence. The first step prepares a stack of 1D GHZ states, and the second step merges them into the TC ground state, with each step having depth $\sim L_y$.

\subsection{Levin-Wen string-net models} \label{sec:string-nets}

In this section, we show how the construction in the previous section for the 2D TC can be generalized to arbitrary Levin-Wen string-net states \cite{levin2005string}, writing down a sequential circuit that generates the bulk string-net ground state on a cylinder and on the torus. In the first case, the state is created row-by-row from a product state, creating and moving arbitrary gapped boundaries in the process like in the previous section. We generalize the unitary circuit that generates the `smooth' boundary at the ends of the cylinder constructed in Ref.~\onlinecite{Liu2022} to arbitrary gapped boundaries. We show the general framework here and present the technical details in appendix \ref{ap:SNstates}. The general construction follows from Kitaev and Kong \cite{kitaev2012models} and bears many resemblances to the (finite-depth) quantum circuit written down in Ref.~\onlinecite{lootens2022mapping}, mapping between different Morita equivalent string-nets. \\

The Hilbert space of string-net models consists of configurations on a hexagonal lattice, with the edges labeled by simple objects $a,b,...$ of a unitary fusion category $\mathcal{C}$. The simple objects obey fusion constraints: $a \times b = \sum_c N_{ab}^c \ c$, which are assumed to be multiplicity free here ($N_{ab}^c \in {0,1}$). There exists a unit object $\mathbf{1}$, such that $\forall \ a, \mathbf{1}\times a = a$. The string-net Hamiltonian is a sum of commuting projectors, generalizing Eq.~\ref{eq:h2dtc}:

\begin{equation} \label{eq:string-netH}
    H_{\text{SN}} = -\sum_{v} A_v - \sum_{p} B_p.
\end{equation}
In the ground state, the vertex terms ($A_v$) enforce the fusion rules at every vertex and the plaquette terms $(B_p)$ give dynamics to the string-net by projecting every plaquette on the trivial anyon sector 
\begin{equation} \label{eq:Bp}
B_p = \sum_s \frac{d_s}{D} B_p^s, \ \ (B_p)^2 = B_p,
\end{equation}
with $d_s$ the quantum dimensions of the simple objects $s$ and $D = \sum_s (d_s)^2$ the total quantum dimension. The action of $B_p$ can be schematically drawn as 
\begin{align} \label{eq:plqauetteOp}
B_p \Ket{\includegraphics[valign=c,page=16]{figuresSN}} = \Ket{\includegraphics[valign=c,page=17]{figuresSN}}.
\end{align}
After the weighted sum of $s$-loops (\ref{eq:Bp}) is created in the middle of the plaquette, the loops are fused in the lattice and every vertex is recoupled until the hexagonal basis states are recovered (Eq. \ref{eq:Sloop}). An explicit ground state on the torus can be obtained by initializing all the edges on the trivial object, a state which trivially satisfies the vertex constraints, and then applying the plaquette operator on every plaquette. We remark that these plaquette operators are not unitary, so this does not directly lead to a circuit to prepare the ground state. \\

Similarly, to define our circuit, we start with an initial state with all edges fixed on the trivial object.
We leave the lattice unoriented here for simplicity and choose an explicit orientation in appendix \ref{ap:SNstates}. Just like for the Toric Code, the circuit is constructed such that a control edge can be assigned for every individual plaquette operator. This edge is not yet entangled in the lattice and is still in the trivial state $\ket{\mathbf{1}}$. To merge the edge into the latttice, it is first mapped to $\ket{\mathbf{1}} \rightarrow \sum_{s}\frac{d_s}{D}\ket{s}$, after which it is used as the control to draw the loops around the added plaquette. The controlled-$B_p$ (C-$B_p$) action is similar as the action of $B_p^s$, but with the control edge treated as in the trivial state \cite{Liu2022}:
\begin{align} \label{controlled-Bp}
\begin{split}
\Ket{\includegraphics[valign=c,page=51,scale=0.5]{figuresSN}} = \sum_{s}\frac{d_s}{D} \ (\text{C-}B_p)\Ket{\includegraphics[valign=c,page=52,scale=0.5]{figuresSN}}=\\
\sum_{s}\frac{d_s}{D}B_p^s \Ket{\includegraphics[valign=c,page=53,scale=0.5]{figuresSN}} = \sum_{s}\frac{d_s}{D}\Ket{\includegraphics[valign=c,page=54,scale=0.5]{figuresSN}}.
\end{split}
\end{align}
The C-$B_p$ operator now acts as a unitary with the same action as $B_p^s$, provided at least one edge is in the trivial state initially (grey).  This is shown in Ref.~\onlinecite{Liu2022, wang2022renormalization}. The proof relies on the unitarity of the $F$-symbols of the fusion category $\mathcal{C}$ (\ref{eq:Fsymbol}).
The full circuit is shown in Fig. \ref{string-netCircuit}. A cylinder is generated with boundaries labeled by $\mathcal{C}$. This boundary is the generalization of the `smooth' boundary in the Toric Code for general string-nets with Hamiltonian Eq.~\ref{eq:string-netH}, where both the bulk- and boundary edges are labeled by objects in $\mathcal{C}$. The effect of applying operators in a new row is to push the boundary down and enlarge the bulk string-net ground state. \\

\begin{figure}
    \centering
    \includegraphics[width=\linewidth,page=26]{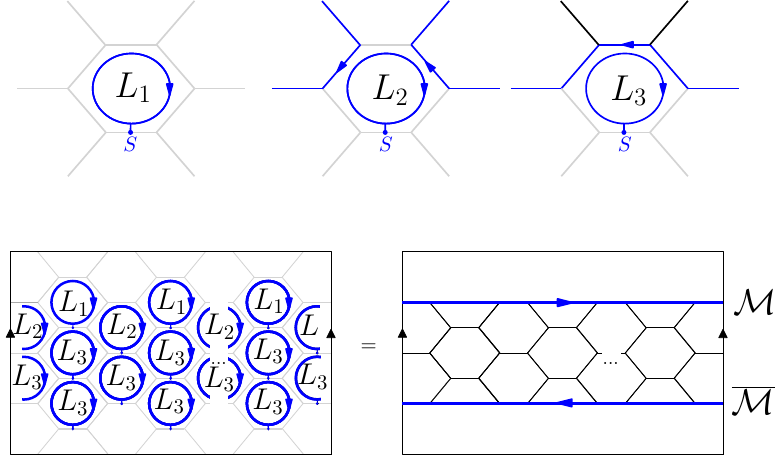}
    \caption{The general sequential circuit to generate the string-net ground state with boundaries labeled by $\mathcal{M}$ by acting row-by-row on the plaquettes with a controlled-$B_p^{\mathcal{M}}$. We start with an initial state with all edges fixed on the trivial object in the category $\mathcal{C}_{\mathcal{M}}^{*}$ (grey). We need three different plaquette operators: $L_1$ and $L_2$ for the first two rows and $L_3$ for all rows below that. The precise action of the three operators are shown in Eqs. \ref{eq:Sloop}-\ref{eq:L3}. In the bulk, the loops are fused in the lattice to the string-net labeled by $\mathcal{C}$ (by virtue of Eq. \ref{eq:resIdBlue}), but the boundary edges are labeled by objects in $\mathcal{M}$ (top) and $\overline{\mathcal{M}}$ (bottom).}
    \label{generalString-netCircuit}
\end{figure}

Following Kitaev and Kong, the string-net boundaries with edges labeled by objects in $\mathcal{C}$ are not the only possible ground state solution of Eq.~\ref{eq:string-netH}. More generally, the edges can carry labels in a $\mathcal{C}$-module category $\mathcal{M}$ (indicated by blue lines in the diagrams). Given the input category $\mathcal{C}$ and the choice of boundary $\mathcal{M}$, it is always possible to define the `Morita dual' category $\mathcal{C}_{\mathcal{M}}^{*}$ and consider $\mathcal{M}$ as a $(\mathcal{C}_{\mathcal{M}}^{*}, \mathcal{C})$-bimodule category \cite{etingof2016tensor}. This is the structure we require to define our general quantum circuit, the details of which are explained in appendix \ref{ap:SNstates}. \\

\begin{figure*}
    \centering
    \includegraphics[scale=0.85,page=31]{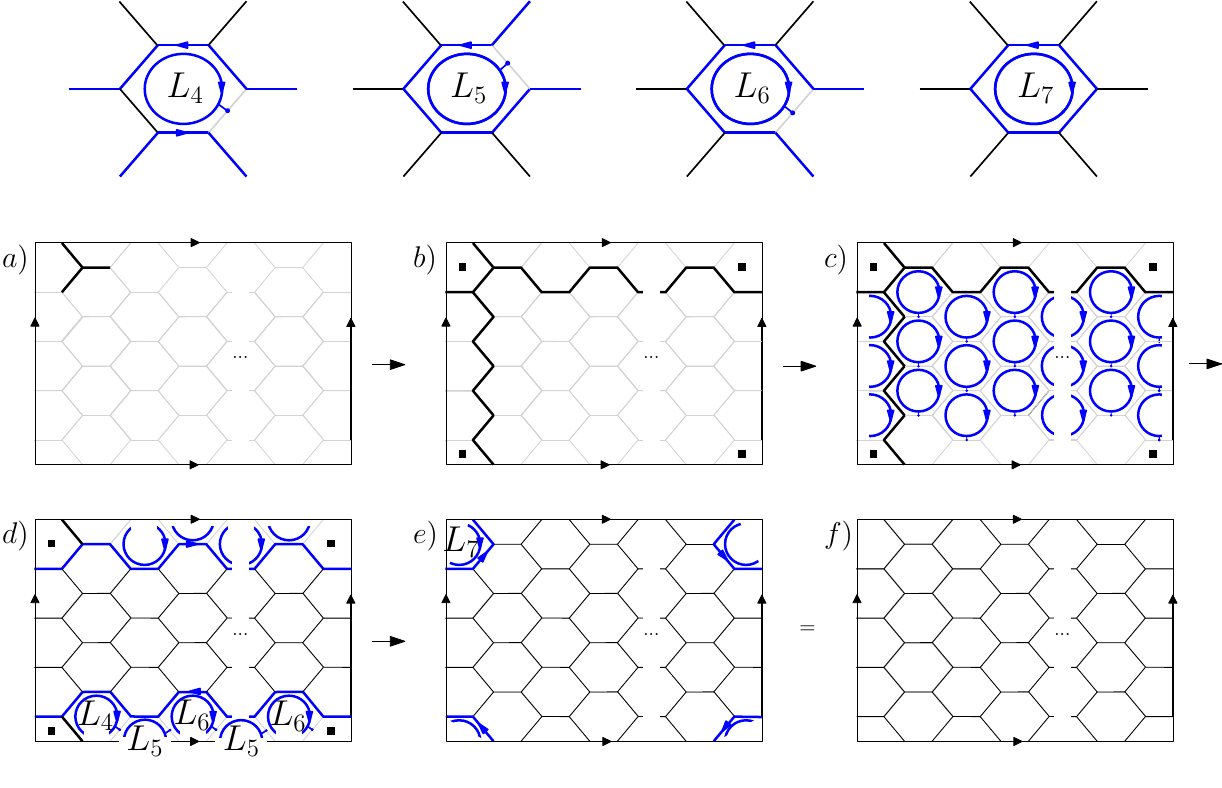}
    \caption{The procedure for generating the string-net ground state on the torus. a) The ground state of the string-net with input category $\mathcal{C}_{\mathcal{M}}^{*}$ is created on the minimal torus by fixing three edges according to Eq.~\ref{eq:minimalTorus}. b) The minimal torus is expanded by copying the initial edges along the two non-contractible cycles of the torus (this is a sequential circuit). The black boxes identify the same plaquette at the four corners. c) The sequential circuit is used to generate the bulk ground state (using the bottom edge as the control) using Eqs. \ref{eq:Sloop}-\ref{eq:L3}. d) For the last row, the circuit needs to be altered, using a side edge as the control (Eqs. \ref{eq:L4}-\ref{eq:L6}). e) and f) The plaquette operator Eq.~\ref{eq:L7} can now be applied on the last plaquette to obtain the torus ground state. This operator is now unitary since we are guaranteed to be in the torus ground state already.}
    \label{torusCircuit}
\end{figure*}

The opposite category $\overline{\mathcal{M}}$ is a $(\mathcal{C}, \mathcal{C}_{\mathcal{M}}^{*})$-bimodule category, with objects that are in one-to-one correspondence with those of $\mathcal{M}$, but with their orientation reversed.
One can now construct a weighted sum of simple objects in $\mathcal{M}$ ($B_p^{\mathcal{M}}$):
\begin{align} \label{generalBM}
B_p^{\mathcal{M}} \Ket{\includegraphics[valign=c,page=16]{figuresSN}} &= \Ket{\includegraphics[valign=c,page=18]{figuresSN}}, \\
B_p^{\mathcal{M}} &= \sum_S \frac{d_S}{D} B_p^{S,\mathcal{M}},
\end{align}
with $d_S$ the quantum dimensions of the objects in $\mathcal{M}$. When applied to all plaquettes, $B_p^{S,\mathcal{M}}$ maps the ground state of a string-net with input category $\mathcal{C}_{\mathcal{M}}^{*}$ to that of a string-net with input category $\mathcal{C}$ (the operator $B_p^{\overline{\mathcal{M}}}$ does the reverse). \\

The initial trivial state, on which our general sequential circuit will act, is interpreted as a state with all edges fixed on the trivial object in the Morita dual category $\mathcal{C}_{\mathcal{M}}^{*}$. The objects in $\mathcal{M}$ have an action on the objects in $\mathcal{C}_{\mathcal{M}}^{*}$ ($\mathcal{C}_{\mathcal{M}}^{*} \times \mathcal{M} \rightarrow \mathcal{M}$), such that on the boundary, the loops are fused with the trivial edges ($\in \mathcal{C}_{\mathcal{M}}^{*}$) to obtain boundaries with $\mathcal{M}$ labels. In the bulk, loops are fused pairwise on every edge $\overline{\mathcal{M}} \times \mathcal{M} \rightarrow \mathcal{C}$ (by virtue of \ref{eq:resIdBlue}), such that the desired bulk string-net $\mathcal{C}$ is recovered. The general sequential circuit is shown in Fig.~\ref{generalString-netCircuit}. It is the generalization of the one shown in Fig.~\ref{string-netCircuit} for arbitrary gapped boundaries labeled by $\mathcal{M}$. The circuit is slightly more complicated as we need different plaquette operators on different sublattices for the first two rows. The general picture, however, is very similar. The same isometry argument can be invoked here for the general plaquette operator, the proof of which we omit, but is a generalization of the one given in Ref.~\onlinecite{Liu2022, wang2022renormalization}, and relies on the unitarity of all the $F$-symbols involved (see Appendix \ref{ap:SNstates}).\\

Note that $\mathcal{C}$ is always a valid choice for the module category over itself ($\mathcal{M} = \mathcal{C}$), in which case the Morita dual is simply $\mathcal{C}$ itself. This case corresponds to the original circuit and the `smooth' boundary is recovered. The Toric Code is recovered from this general picture by choosing $\mathcal{C} = \text{Vec}_{\mathbb{Z}_2}$. The `smooth' and `rough' boundaries are produced by the sequential circuit by choosing $\mathcal{M} = \mathcal{C} = \text{Vec}_{\mathbb{Z}_2}$ and $\mathcal{M} = \text{Vec}$ (the category of vector spaces, with only one trivial object) respectively \cite{lootens2021matrix}.\\

We finish this section by showing how the circuit needs to be adapted for generating the string-net ground state on the torus. We can try to close the cylinder in Fig. \ref{generalString-netCircuit} by applying a new row with periodic boundary conditions, using the side edges as the control edges this time (because the bottom edges of the new row are already occupied after the circuit for the first row). However, we cannot completely close the cylinder, as there are no unoccupied edges left to use as the control on the very last plaquette. The solution is explained in Ref.~\onlinecite{wang2022renormalization} and we use a generalization of it here to construct the torus ground state in different representations labeled by $\mathcal{M}$. The string-net ground state is created on a minimal torus first (a torus with two vertices, one plaquette and three edges), by initializing the edges on the trivial object in $\mathcal{C}$ and applying the plaquette operator in the reverse direction of Eq.~\ref{generalBM} on the one plaquette (Eq.~\ref{eq:minimalTorus}). The result is a minimal torus ground state of a string-net with input category $\mathcal{C}_{\mathcal{M}}^{*}$. After this initial step, the original sequential circuit is now used to add plaquettes to obtain the ground state for a string-net with input category $\mathcal{C}$ in the bulk. For the last row, the circuit needs to be altered, using side edges as control qudits similar to the Toric Code circuit. The plaquette operator can now be applied on the last plaquette since we are guaranteed to be in the torus ground state already. The procedure is shown in Fig. \ref{torusCircuit}. The general circuit can be used to generate a Projected Entangled Pair State (PEPS) representation of the string-net ground state, in which case the representation from Ref.~\onlinecite{lootens2021matrix} is recovered, generalizing the original representation from Ref.~\onlinecite{buerschaper2009explicit,gu2009tensor}. Note that in the abelian group case, the minimal torus is trivial, and for $\mathcal{M}=\text{Vec}_{\mathbb{Z}_2}$ we recover the Toric Code sequential circuit (on a hexagonal lattice). \\

Just like for the Toric Code in Fig.~\ref{fig:tc_circuits}(b), we can change the string-net circuit such that after each step, we are left with a string-net with periodic boundary conditions in both directions (without any open boundaries). The periodic boundary condition in the vertical direction can be enforced in the circuit, without applying a last row that zips the boundaries together, by first initializing a one-row torus groundstate and enlarging the torus in every step. 

\section{Map to 3+1D topological phases and Walker-Wang models}
\label{sec:TO3D}

We now describe SQCs for preparing topological phases in 3+1D. For the $3+1$D Toric Code, there are two ways to view the bulk wavefunction: either as a condensate (equal-weighted superposition) of membrane-like excitations or as a condensate of loop-like excitations\cite{Hamma2005}. Correspondingly, there are two types of gapped boundary: a rough boundary that condenses the charges, and a smooth boundary that condenses the loop-like fluxes. In this section, we discuss two sequential quantum circuits for generating the $3+1$D Toric Code from product states. The first one generates a membrane condensate in the bulk with a rough boundary while the second one generates a loop condensate in the bulk with a smooth boundary. The first type of circuit can be generalized to other $3+1$D Dijkgraaf-Witten gauge theories while the second type of circuit can be generalized to other Walker-Wang models, as we discuss later in this section. 

\subsection{Point Charge Condensed Boundary}

\begin{figure}
    \centering
    \includegraphics[scale=0.4]{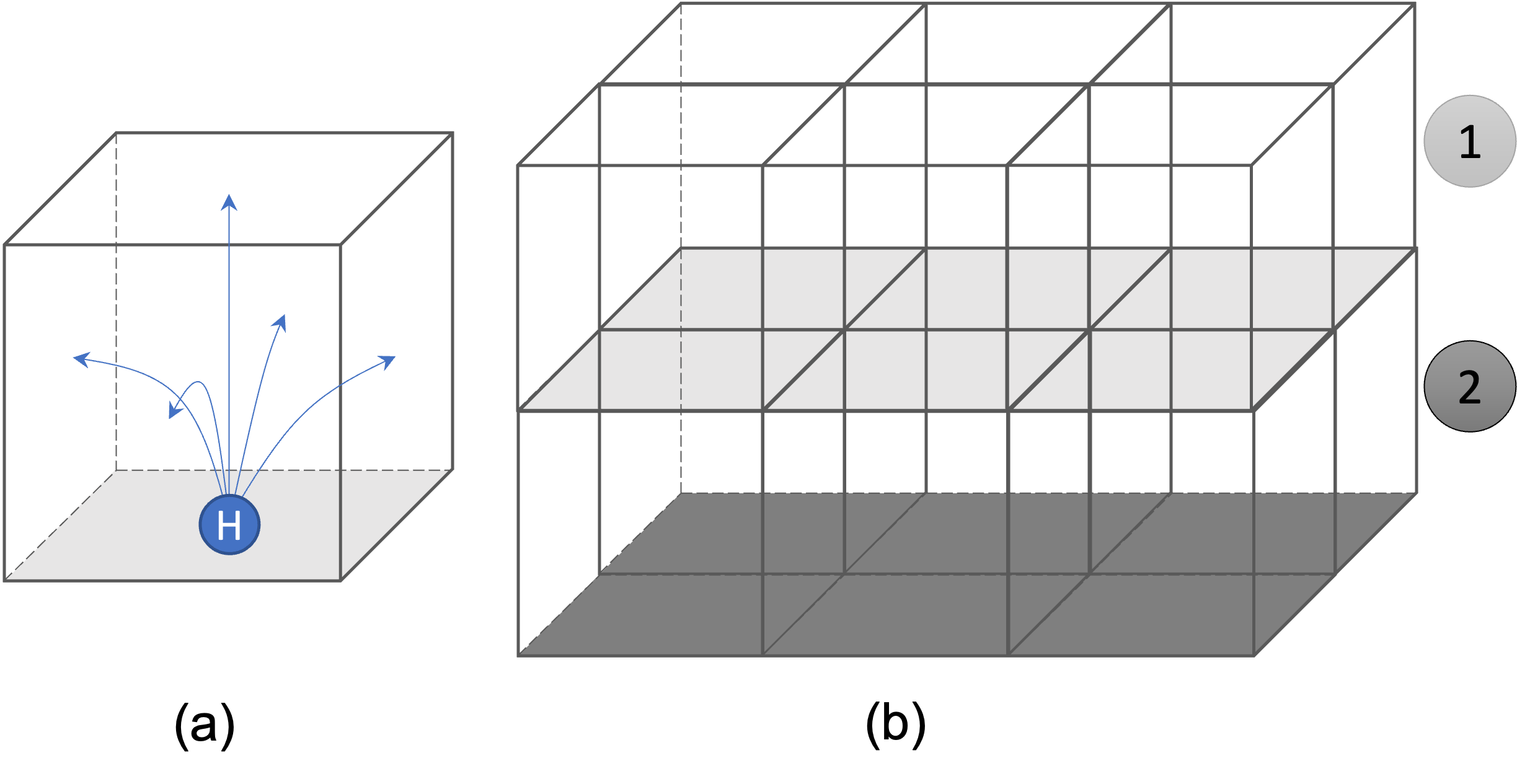}
    \caption{Sequential circuit that generates the $3+1$D Toric Code with charge condensed boundary. (a) Within each cube, the bottom plaquette is first transformed by a Hadamard gate and then used as the control qubit for a set of controlled-Not operations targeting all the other plaquettes in the same cube. (b) Such operations are applied to all cubes in layer 1 first, then to layer 2 beneath it, and so on.}
    \label{fig:3DTCmem}
\end{figure}

Consider the $3+1$D Toric Code defined on a cubic lattice. In the membrane condensate picture, the $Z_2$ DOFs are on each plaquette. The Hamiltonian contains two types of terms, one associated with each cube, and one associated with each edge.
\begin{equation}
H_{\text{3dTC}} = -\sum_{\text{cube}}\prod_{p\in \text{cube}} X_p - \sum_{\text{edge}} \prod_{p \ni \text{edge}} Z_p
\end{equation}
Starting from a product state Hamiltonian $H=-\sum_p Z_p$, the Toric Code can be generated with a sequential circuit as shown in Fig.~\ref{fig:3DTCmem}. Within each cube (Fig.~\ref{fig:3DTCmem}(a)), the bottom plaquette is first transformed by a Hadamard gate that exchanges $X_p$ and $Z_p$. It is then used as the control qubit for a set of controlled-Not operations targeting all the other plaquettes in the same cube. After these operations, the $Z_p$ term on the bottom plaquette gets mapped to the cube term $\prod_{p\in\text{cube}}X_P$ in the Toric Code Hamiltonian. The sequential circuit acts by applying this transformation first to all the cubes (at the same time) in layer 1, then to all the cubes in layer 2 underneath layer 1, and so on. This circuit is a direct generalization of the circuit generating 2D Toric Code discussed in section~\ref{sec:TO2D}. 

Applying the circuit up to layer $n$ generates a gapped boundary between the $3+1$D Toric Code and the vacuum state. It is easy to see that this is the `rough' boundary of the Toric Code (`rough' in the dual lattice) where the gauge charge excitation condenses. 

The sequential circuit described in Fig.~\ref{fig:3DTCmem} can be generalized to all Dijkgraaf-Witten (DW) gauge theories in $3+1$D. DW gauge theories are membrane condensates in $3+1$D where different membrane configurations can come with different phase factors (for Toric Code the phase factors are all $1$). To generate the membrane condensate, we can generalize the sequential circuit in Fig.~\ref{fig:3DTCmem} from $Z_2$ DOF to group $G$ labeled DOF and then supplement each elementary closed membrane generation operation in (a) with the appropriate phase factor. Similar to the string-net circuit discussed in section~\ref{sec:TO2D}, the elementary closed membrane generation operations in the same layer can be shown to commute with each other. Therefore, the DW states can be generated layer by layer with a linear depth circuit. If the circuit is terminated at a certain layer, the gapped boundary is a condensate of the bosonic gauge charges, just like in the Toric Code case.  

\subsection{Flux Loop Condensed Boundary}
\label{sec:loop_cond3D}

A different sequential circuit can give rise to the flux loop condensed boundary for the $3+1$D Toric Code. The circuit generates a loop condensate in the bulk. To describe this circuit, we take the dual lattice of that in Fig.~\ref{fig:3DTCmem} so that DOF are associated with each edge. The Toric Code Hamiltonian is then expressed as
\begin{equation}
H_{\text{3dTC}} = -\sum_{\text{vertex}}\prod_{e \ni \text{vertex}} X_e - \sum_{\text{plaquette}} \prod_{e\in \text{plaquette}} Z_e
\end{equation}

\begin{figure}
    \centering
    \includegraphics[scale=0.5]{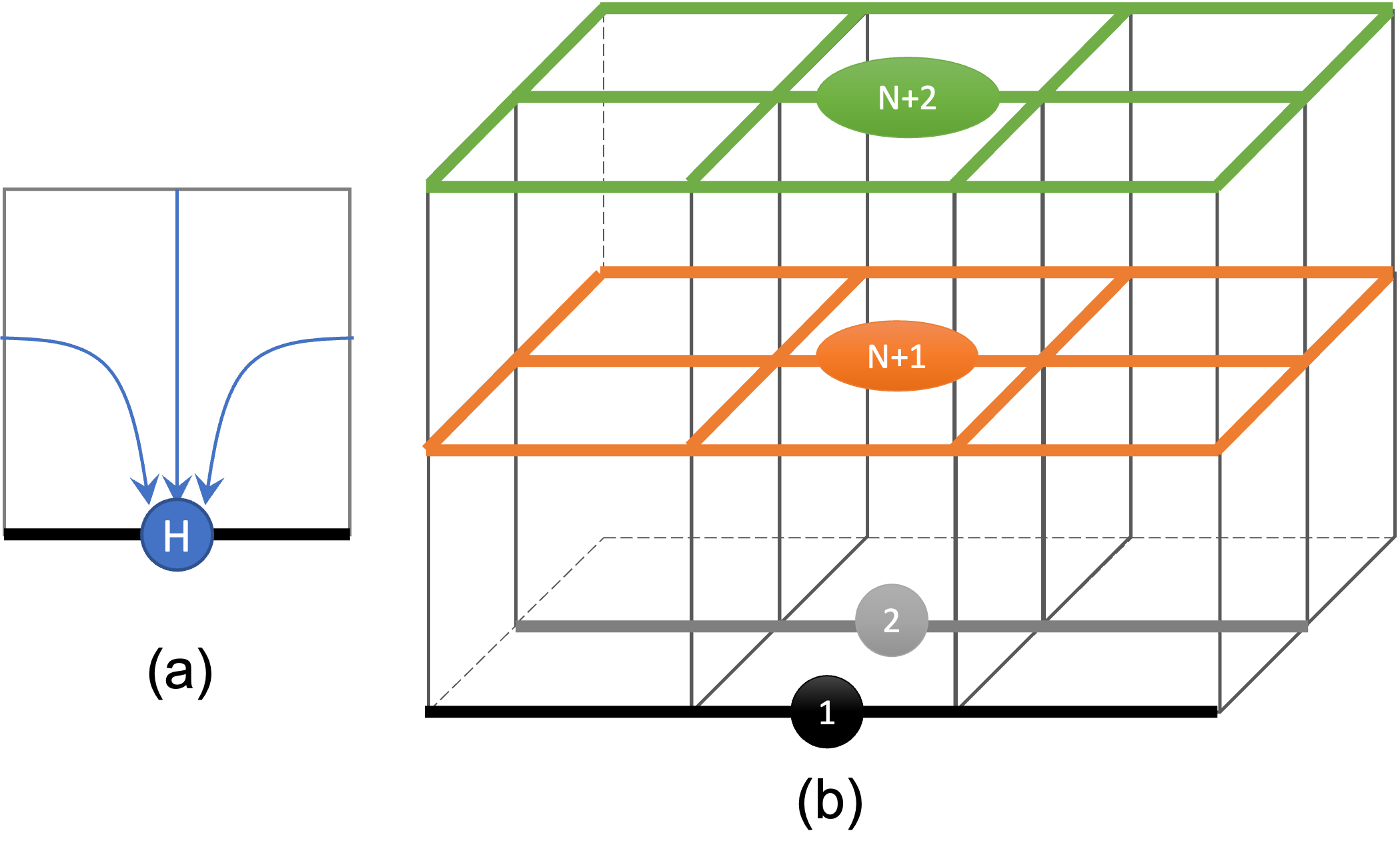}
    \caption{Sequential circuit that generates the $3+1$D Toric Code with flux loop condensed boundary. (a) Within each plaquette, a Hadamard gate is first applied to a chosen edge (the thick black one) which is then used as target for a set of controlled-Not gates with all the other edges in the plaquette as the control. (b) Such operations are first applied row by row to all the plaquettes in the bottom surface }
    \label{fig:3DTCloop}
\end{figure}

We start with a trivial product state with Hamiltonian $H=-\sum X_e$. The circuit to obtain the Toric Code is shown in Fig.~\ref{fig:3DTCloop}. 
Within each plaquette (Fig.~\ref{fig:3DTCloop}(a)), a Hadamard gate is first applied to a chosen edge (the thick black one) which is then used as target for a set of controlled-Not gates with all the other edges in the plaquette as the control. After these operations, the $X_e$ term on the target edge gets mapped to the plaquette term $\prod_{e\in \text{plaquette}} Z_e$ in the Toric Code Hamiltonian. The sequential circuit acts by applying this transformation first to all the plaquettes in the first row on the bottom surface with the black edges as the target, then to all the plaquettes in the second row with the grey edges as the target, and so on for all $N$ rows in the bottom surface. Then to move up, apply the transformation in (a) to all the vertical plaquettes immediately above the bottom surface with the orange edges as the target in step $N+1$. This generates the plaquette term for all the involved vertical plaquettes. After this step, the plaquette terms on the orange plane are automatically satisfied because the product of the six plaquette terms around a cube is the identity. Therefore, repeating step $N+1$ allows us to expand the topological region and push the boundary upward.

It can be checked that, if the sequential circuit is terminated at step $N+m$, the gapped boundary at the $m$th plane is the `smooth' boundary type where the flux loop condenses. 

\subsection{Walker-Wang with Loop Condensed Boundary}

The circuit in Fig.~\ref{fig:3DTCloop} can be generalized to construct all Walker-Wang (WW) models in $3+1$D \cite{Walker2012}. The ground state of any WW model is a loop condensate, although the loops are more complicated than that in the Toric Code. To construct the ground state wavefunction from a product state, we can choose a target edge in each plaquette and use it to draw loops in each plaquette in a way similar to that described in section~\ref{sec:TO2D} for the string-nets. The only difference is that, in the WW wavefunctions, when we draw loops, there can be extra phase factors due to edges lying over or under the plaquettes. Applying these operations in each plaquette in the sequence shown in Fig.~\ref{fig:3DTCloop}(b) generates the WW wavefunction. Note that, the plaquette terms in the WW model satisfy the same local constraint as that in the Toric Code model: the product of the six plaquette terms around a cube is identity. For Toric Code, this constraint is satisfied in the whole Hilbert space while for a generic WW model, this constraint is only satisfied in the closed loop subspace. To see why this constraint holds, notice that the product of plaquette terms in the WW model in a closed surface measures the strings that go through the surface. For a topologically trivial closed surface like the surface of a cube, in the closed loop subspace, all strings going into the surface must come out. Therefore, the net flux going through a cube is zero in the closed loop subspace -- the product of all the plaquette terms must be identity. Therefore, generating the plaquette terms in the bottom and side faces of a cube automatically gives rise to the plaquette term on the top face. The circuit depicted in Fig.~\ref{fig:3DTCloop} hence proceeds in the same way as for Toric Code. The gapped boundary created with a circuit terminating at step $N+m$ has the corresponding $2+1$D topological order of the WW model.


\section{Map to fracton phases}
\label{sec:fracton}
\begin{figure*}
\subfloat[\label{fig:Xcubestabilizers}]{\includegraphics[width=0.25\linewidth]{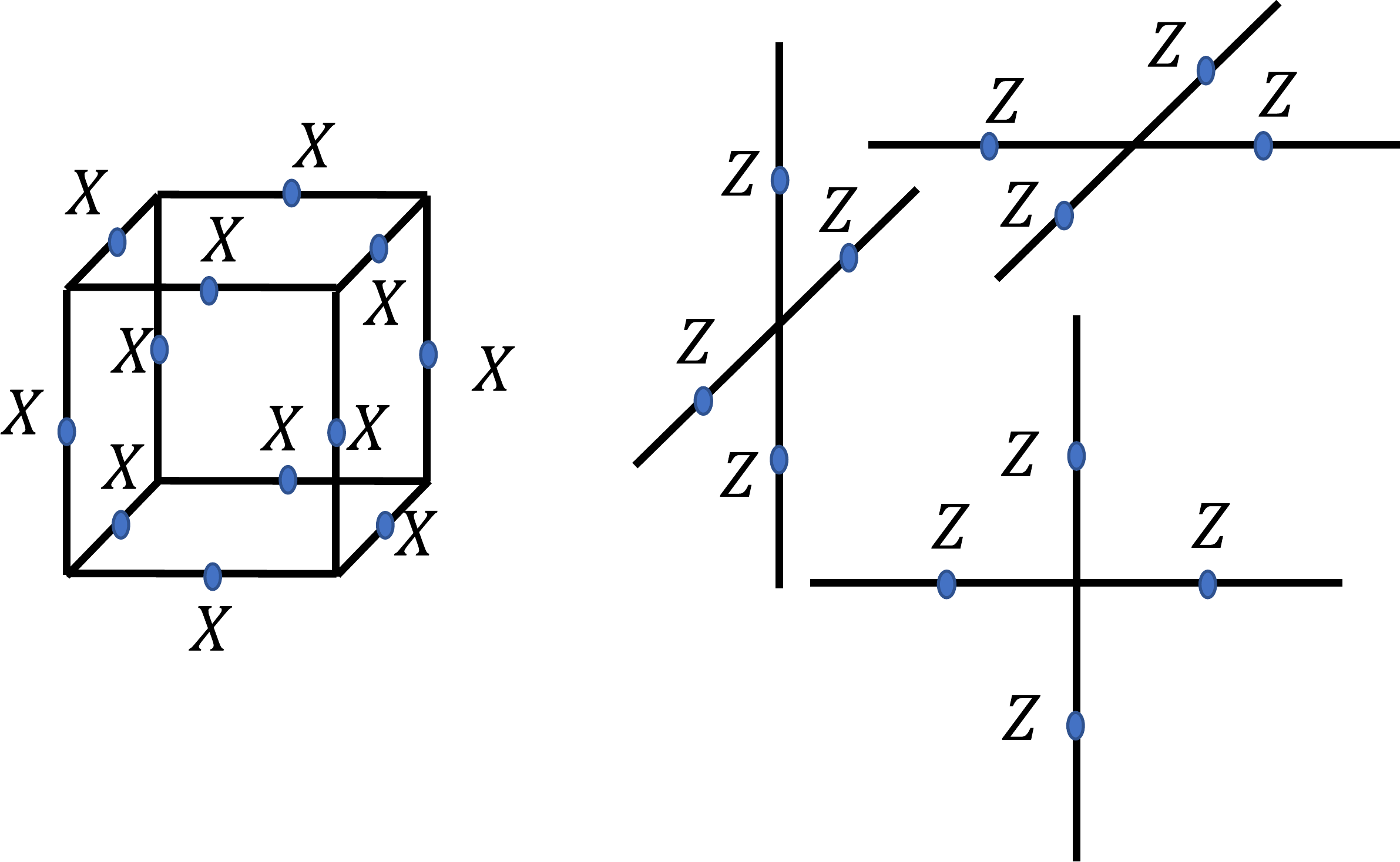}} \hfill
\subfloat[\label{fig:seq_circuit_Xcube_smooth}]{\includegraphics[scale=0.08]{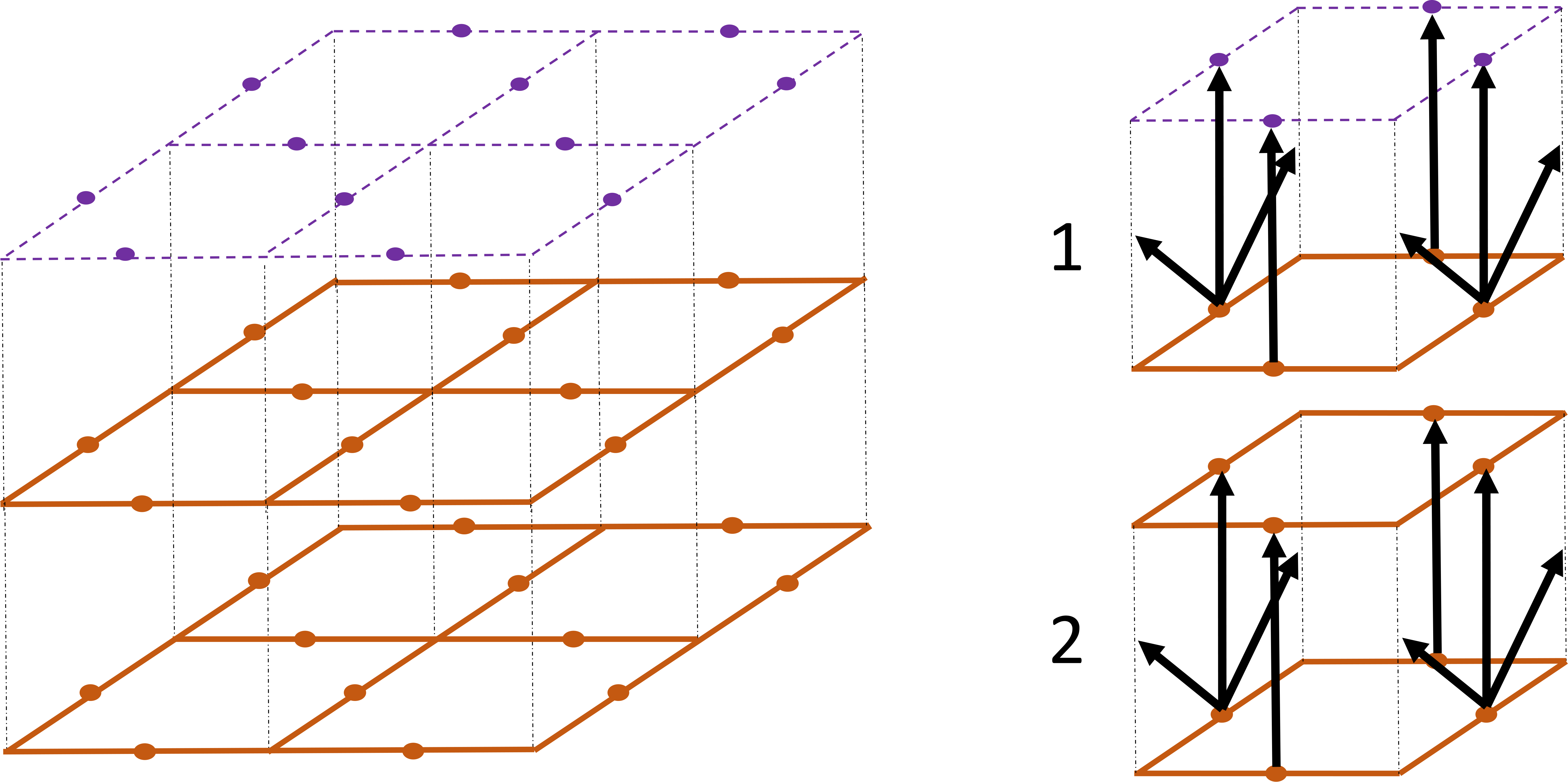}} \hfill
\subfloat[\label{fig:closing_bdry_Xcube}]{\includegraphics[scale=0.08]{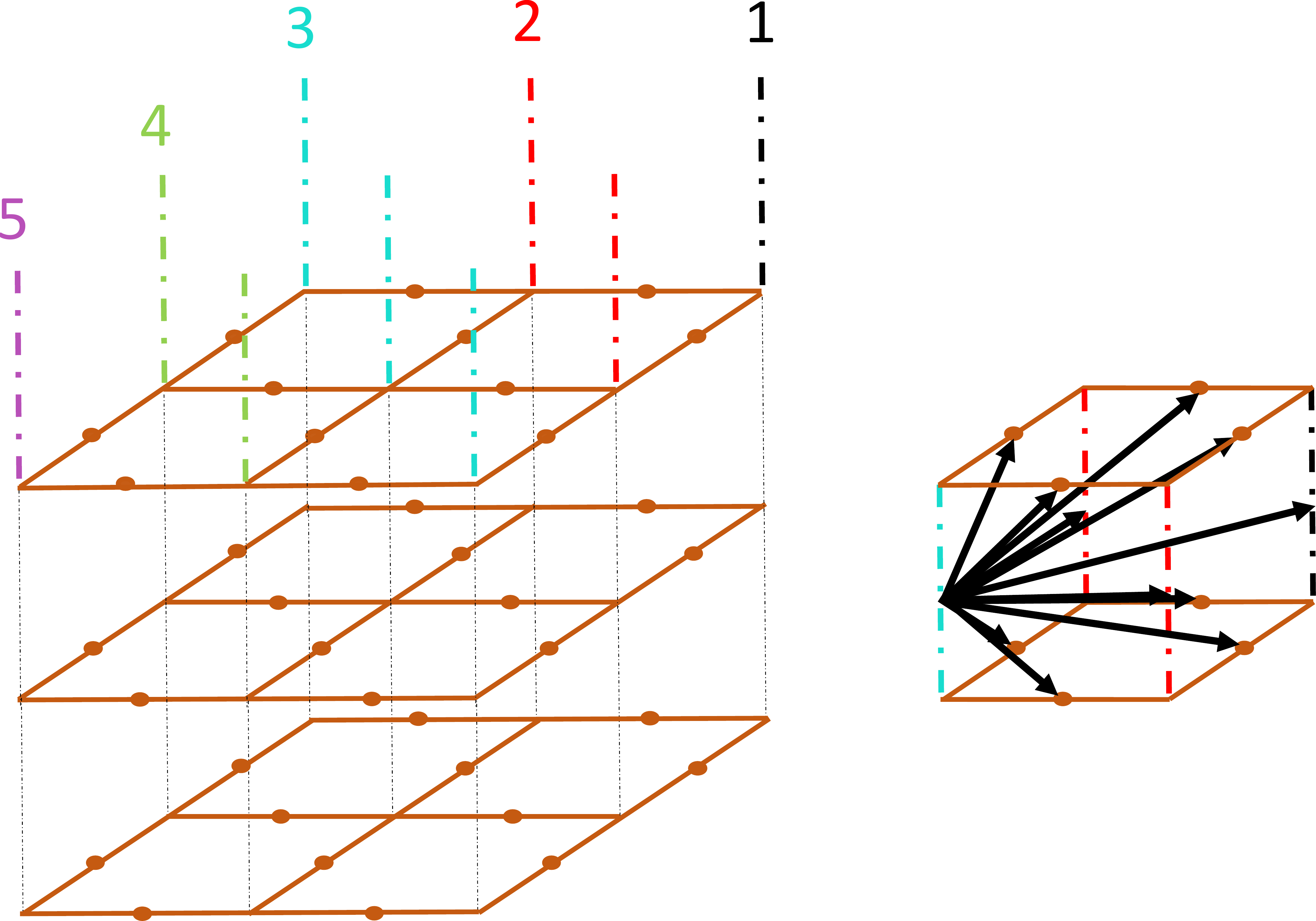}} 
\caption{(a) Pauli stabilizer terms of the X-cube model. The model is defined on a cubic lattice and qubits live on the edges. The X stabilizer term is associated with the cube and the Z stabilizer terms are associated with the vertices. (b) Sequential circuit for X-cube model (of height $L_z=2$ cubes) with smooth boundaries at the top and bottom and periodic boundaries on the sides. Left: We start with $L_z=2$ layers of Toric Code (shown in orange) with periodic boundary conditions in the layer directions and a layer of trivial qubits in state $\ket{0}$ (shown in purple). We add trivial qubits in state $\ket{+}$ on the vertical edges; the dots on the vertical edges are not shown for clarity. Right: On each vertical plaquette, we perform CNOT gates as shown using arrows that go from control to target qubits. We first do these gates as shown in parallel on all the top layer vertical plaquettes, then we do the layer below it, and so on. In the figure, we have $L_z=2$, hence our circuit has depth 2 as shown on the right. (c) Sequential circuit to convert X-cube model with smooth boundaries into an X-cube model on a torus. We add a layer of qubits on vertical edges in state $\ket{+}$ above the top smooth boundary (the top orange layer) such that we can join to the bottom smooth boundary (the bottom orange layer) via these vertical edges to impose periodic boundary conditions; the dots on the vertical edges are not shown for clarity. In one step of the sequential circuit, all qubits on the vertical edges in the same color are used as control qubits. The CNOT gates from each of these control qubits have target qubits on the edges of a single cube. An example is shown on the right. The CNOT gates are shown using arrows that go from control to target qubits.}
\end{figure*}

In this section, we construct sequential circuits for the preparation of the X-cube model, which is a well-known example of fracton topological order \cite{Vijay_2016}. The X-cube model is a commuting projector model where the projectors are Pauli operators, as shown in Fig.~\ref{fig:Xcubestabilizers}. We use two different ways to construct the circuit: the first one is associated with the `foliation' structure of the X-cube model \cite{Shirley2018} while the second one is related to the `p-string condensation' picture \cite{Ma2017,Prem_2019}. 

\subsection{Foliation circuit for X-cube model}
We now state the `foliation' circuit for the X-cube model. We write the circuit to prepare an X-cube model of height $L_z$ cubes with smooth boundaries at the top and bottom and periodic boundary conditions on the sides; however, it can be generalized to other choices of boundary conditions. We start with $L_z$ layers of Toric Code with periodic boundary conditions which were prepared using the sequential circuit shown in Sec.~\ref{sec:2dTC_seq_circuit} and a layer of trivial qubits in state $\ket{0}$. We then combine the Toric Code layers and the layer of trivial qubits into an X-cube model with smooth boundaries using a sequential circuit as shown in  Fig.~\ref{fig:seq_circuit_Xcube_smooth}. The CNOT gates are first applied in the top layer of vertical plaquettes, all in parallel, and then in the next layer, and so on. The overall circuit depth for preparing the X-cube model with smooth boundaries is linear. We apply a circuit of depth $\mathcal{O}(L_x+L_y)$ for the Toric Code layers and then we need a sequential circuit of depth $\mathcal{O}(L_z)$ to combine them into an X-cube model of height $\mathcal{O}(L_z)$ cubes and with smooth boundaries (Fig.~\ref{fig:seq_circuit_Xcube_smooth}). 

The same circuit can be used to prepare the X-cube model with periodic boundary condition in the vertical direction with fewer starting trivial qubits. To be specific, under periodic boundary conditions, the layer of trivial qubits shown in purple is not needed along the vertical direction and the orange layer at the bottom will connect to the top orange layer. The entangling gates as shown in Fig.~\ref{fig:seq_circuit_Xcube_smooth} will then prepare the X-cube model on a 3-torus. 

However, if we first prepare the height $L_z$ X-cube model with smooth boundaries and then convert it to the X-cube model on a 3-torus with height $L_z+1$ by doing local gates at the top layer, we need a sequential circuit as shown in Fig.~\ref{fig:closing_bdry_Xcube}. We add a layer of trivial qubits on vertical edges above the top smooth boundary in state $\ket{+}$; these vertical edges now connect to the bottom smooth boundary due to periodic boundary conditions. Using these vertical qubits as control qubits, we do CNOT gates sequentially to prepare the layer of cube stabilizers. The CNOT gates that are applied in parallel are specified by the choice of the control qubits shown on the vertical edges with the same color and along the diagonal (left of Fig.~\ref{fig:closing_bdry_Xcube}) and by the gates shown (right of Fig.~\ref{fig:closing_bdry_Xcube}). Thus, to convert the X-cube model with smooth boundaries into a 3-torus, we need a sequential circuit of depth $\mathcal{O}(L_x+L_y)$ (Fig.~\ref{fig:closing_bdry_Xcube}).  

A generalization of the X-cube model is the Ising cage-net model which is obtained from a coupled layer construction of double-Ising string-net models instead of Toric Code layers~\cite{Prem_2019}. In Ref.~\onlinecite{wang2022renormalization}, a generalized notion of foliation and a sequential circuit for preparation of the Ising Cage-net model is presented. For the X-cube model, the resource Toric Code layers can be prepared in parallel before being inserted one by one into the bulk to increase the size of the X-cube model. For Ising Cage-net, however, this is not possible. Using the scheme presented in Ref.~\onlinecite{wang2022renormalization}, a thin slab (of small $L_z$) of Ising Cage-net can first be prepared with $\mathcal{O}(L_x+L_y)$ steps. Then the height of the slab $L_z$ can be increased one at a time, each time requiring a circuit of depth $\mathcal{O}(L_x+L_y)$. Therefore, overall a quadratic ($\mathcal{O}(L^2)$) depth circuit is needed to generate the Ising Cage-net model.

Whether one can do better than quadratic circuit depth for the Ising cage-net model is left as an open question. It will also be interesting to construct sequential circuit maps for more general fracton models such as the type-2 fracton models which have fractal-shaped logical operators~\cite{Vijay_2016}.

\begin{figure}
    \centering
    \includegraphics[scale=0.14]{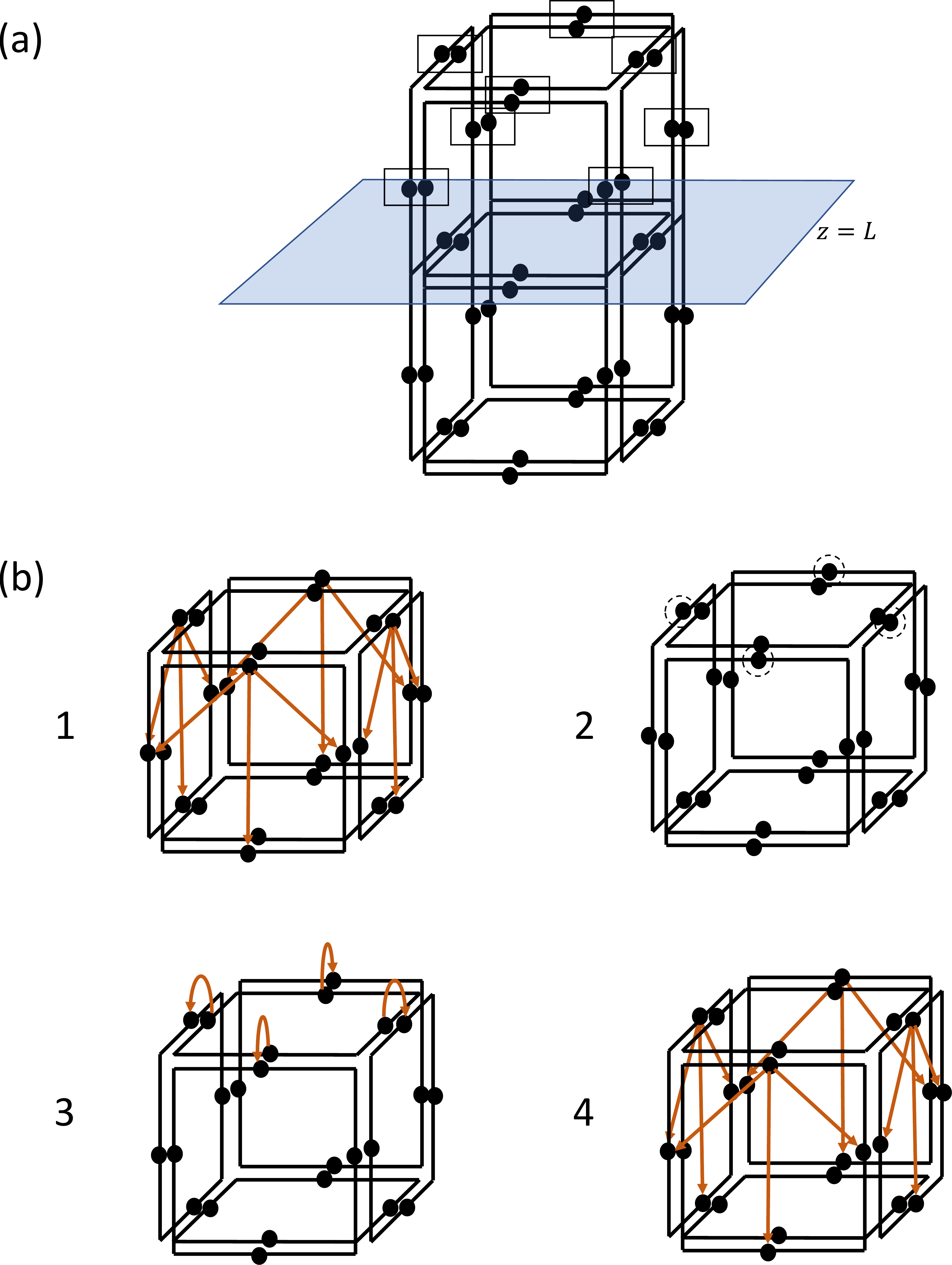}
    \\
    \caption{(a) We start with the 3-foliated stack of 3D Toric Codes such that the resulting lattice has double edges with two qubits each. We add $Z-Z$ stabilizers on the double edges for $z>L$ such that only a product of Toric Code $X$-plaquette stabilizer terms that share the support of $Z-Z$ stabilizers around a cube survives as a stabilizer. In other words, we have performed p-string condensation in the region $z>L$. (b) The sequence of gates that acts on the cubes in the layer below $z=L$ to grow the X-cube-like part from $z>L$ to $z>L-1$. The arrows denote the CNOT gates and the dashed circles denote Hadamard gates. We show the gates on a single cube here but the gates act across the entire layer of cubes. }
    \label{fig:ploop_condensation_circuit}
\end{figure}

\subsection{\MakeLowercase{p}-string condensation circuit for X-cube model}
We now construct a linear-depth sequential circuit for the preparation of the X-cube model using p-string condensation \cite{Ma2017,Prem_2019}.  We first consider three decoupled stacks of Toric Codes in $xy$, $yz$, and $zx$ planes such that a part of this stack $z>L$ has undergone p-string condensation. That is, $Z-Z$ stabilizers have been added to the double edges of the 3-foliated stack lattice for $z>L$, such that only the product of Toric Code $X$ plaquette stabilizers around a cube which share the support of $Z-Z$ double edge stabilizers survive as stabilizers. 
In the part $z\leq L$, we do not add these $Z-Z$ stabilizers and hence, we still have Toric Code vertex and plaquette stabilizers. 
Thus, the part $z>L$ is like the X-cube model, the $z<L$ part is like the decoupled Toric Code stack, and there is a gapped boundary between the two at $z=L$. We assume periodic boundary conditions in the $x$- and $y$-directions, while the boundary conditions in the $z$-direction do not play a role, since we will only consider moving the $z=L$ interface by one lattice constant. 

See Fig.~\ref{fig:ploop_condensation_circuit}(a) for this starting configuration of stabilizers. 
In the X-cube-like part, the stabilizers are the 24-body $X$ stabilizer terms around a cube, $Z-Z$ stabilizers on every composite edge, and the original $Z$ vertex stabilizers of the Toric Code. 
At the $z=L$ plane interface between the X-cube-like part and the decoupled Toric Code part, the $X$-stabilizers supported on the cube are a 20-body $X$ term on the cube above $z=L$ and the Toric Code $X$-plaquette term in the $z=L$ plane.

Our goal is to grow the X-cube portion of the model from $z>L$ to $z>L-1$, \textit{i.e.} to push the gapped boundary between the two models, similar to how our previously defined circuits push a gapped boundary to vacuum.  
In Fig.~\ref{fig:ploop_condensation_circuit}(b), we write the finite-depth circuit that acts on the layer of cubes right below $z=L$ to achieve this. 

The $Z-Z$ stabilizers on the vertical bonds right below $z=L$ follow from the relation among the Toric Code $Z$ vertex stabilizer terms around a ``vertex'' of the 3d stack lattice and the $Z-Z$ double edge stabilizers terms supported on the same qubits as those vertex stabilizers.
In this manner, the  interface which was at $z=L$ has moved to $z=L-1$ after the circuit is applied on the layer of cubes right below $z=L$. 

Starting from Toric Code layers, the circuit depth of the p-string condensation circuit is linear i.e., it scales as $\mathcal{O}(L_z)$ since it takes a finite-depth circuit to grow the X-cube part by one unit length in $L_z$. Since the Toric Code layers in $xy$ planes can be prepared in parallel in circuit depth $\mathcal{O}(L_x+L_y)$, the overall circuit depth scales linearly in the system sizes as $\mathcal{O}(L_x+L_y+L_z)$.

\section{Quantum cellular automata} 
\label{sec:qca}

\begin{figure}
    \centering
    \includegraphics[width=\linewidth]{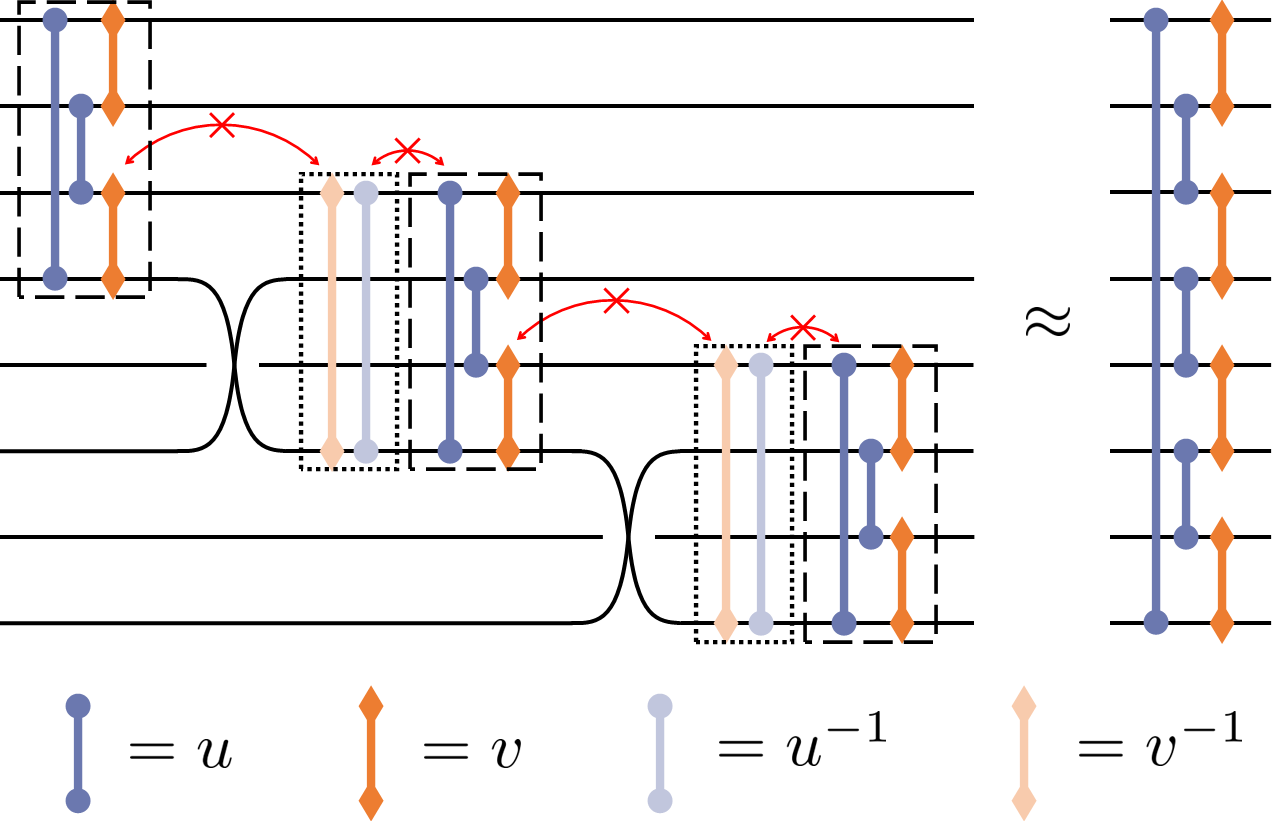}
    \caption{Circuit diagram corresponding to Eq.~\ref{eq:qca_circ} with $N=8$. The sign `$\approx$' means equivalent up to a sequential circuit of SWAP gates. The gates $Q^{(4)}$ and $w$ are shown in dashed and dotted boxes, respectively. Operators in the bulk of the circuit cancel pairwise as indicated by the crosses. Note that this diagram does not depict the differences between the on-site physical dimension $d$ and the internal dimensions $\ell,r$ of the standard form of the QCA.}
    \label{fig:qca}
\end{figure}

In this section, we show that arbitrary quantum cellular automata (QCA) can be realized as sequential circuits. A QCA is a unitary operator $Q$ that preserves locality, meaning that, for any operator $O_i$ supported on lattice site $i$, $QO_iQ^\dagger$ is supported on sites that are at most a distance $c$ away from $i$ for some constant $c$ \cite{Farrelly2020}. We focus on implementing translationally invariant QCA on periodic boundary conditions. Any finite-depth quantum circuit (FDQC) gives an example of a QCA, but there are also QCA that cannot be realized as FDQC, such as the 1D shift operator $S$ which acts on operators on a 1D lattice as $SO_iS^\dagger = O_{i+1}$. However, it is easy to show that $S$ can be realized as a sequential circuit of SWAP gates \cite{Po2016}. Now, we will show that any QCA in any spatial dimension can be realized as a sequential circuit. Additionally, if the QCA $Q$ commutes with some global on-site symmetry $U_g=u_g^{\otimes N}$, the gates in the corresponding sequential circuits will also commute with $U(g)$. Therefore, symmetric QCA are a strict subset of symmetric SQCs. On the other hand, SQC are not always QCAs as they do not always preserve locality.

The construction we use is closely related to the construction used in Ref.~\onlinecite{Stephen2023} to realize QCA as finite-depth circuits with geometrically non-local long-range gates. Let us first consider 1D QCA, and then later generalize to higher dimensions. The starting point is the standard form of a 1D QCA. After blocking sufficiently many sites, any QCA will have unit range, meaning that $QO_iQ^\dagger$ is supported only on sites $i-1,i,i+1$. Then, it was shown in Ref.~\onlinecite{Schumacher2004} (see also Ref.~\onlinecite{Cirac2017}) that the QCA on a ring of $N$ sites can be expressed in the following way,
\begin{equation} \label{eq:standard}
    Q^{(N)} = \left( \prod_{i=1}^{N/2} v_{2i-1,2i} \right)\left( \prod_{i=1} ^{N/2} u_{2i,2i+1} \right).
\end{equation}
where we take periodic boundaries.
Therein, $u$ is a unitary map between a $d\times d$-dimensional Hilbert space to an $\ell\times r$-dimensional Hilbert space where $\ell r=d^2$, and similarly $v$ maps from and $r\times \ell$-dimensional space to a $d\times d$-dimensional space. We emphasize that these internal dimensions $\ell$ and $r$ do not represent any physical degrees of freedom, nor do they appear in the sequential circuits we construct; they are simply a convenient technical tool. Indeed, since the input and output Hilbert spaces differ, $u$ and $v$ do not represent physical operations on their own, but they can be combined to define physical operations such $Q^{(N)}$. As such, Eq.~\ref{eq:standard} is not an FDQC, so it can also represent QCA that are not FDQCs such as the shift QCA.

To realize this QCA as a SQC, we follow closely the idea used for SPT states in Sec.~\ref{sec:SPT}. The fundamental unitary gates in the SQC are the QCA on a ring of length 4,
\begin{equation}
    Q^{(4)}_{i,j,k,l} =v_{i,j}v_{k,l}u_{j,k}u_{l,i} 
\end{equation}
and also the gate,
\begin{equation}
    w_{i,j} = u^{-1}_{j,i}v^{-1}_{i,j}.
\end{equation}
Note that, unlike $u$ and $v$, the operators $Q^{(4)}$ and $w$ are genuine physical unitary operators acting on 4 and 2 sites, respectively. Using these gates, we realize the QCA on a chain of even length $N>4$ as an SQC in the following way,
\begin{align} \label{eq:qca_circ}
    Q^{(N)} &\approx Q^{(4)}_{1,2,3,4} \\
    & \times \prod_{i=1}^{\frac{N-4} {2}}\text{SWAP}_{2i+2,2i+4}w_{2i+1,2i+4}Q^{(4)}_{2i+1,2i+2,2i+3,2i+4} \nonumber
\end{align}
This equality follows from commuting all SWAP gates to one side of the equation, as is shown graphically in Fig.~\ref{fig:qca}. Therein, the `$\approx$' sign means equivalence up to the SWAP gates which can easily be undone in a sequential manner.

\begin{figure}
    \centering
    \includegraphics[width=\linewidth]{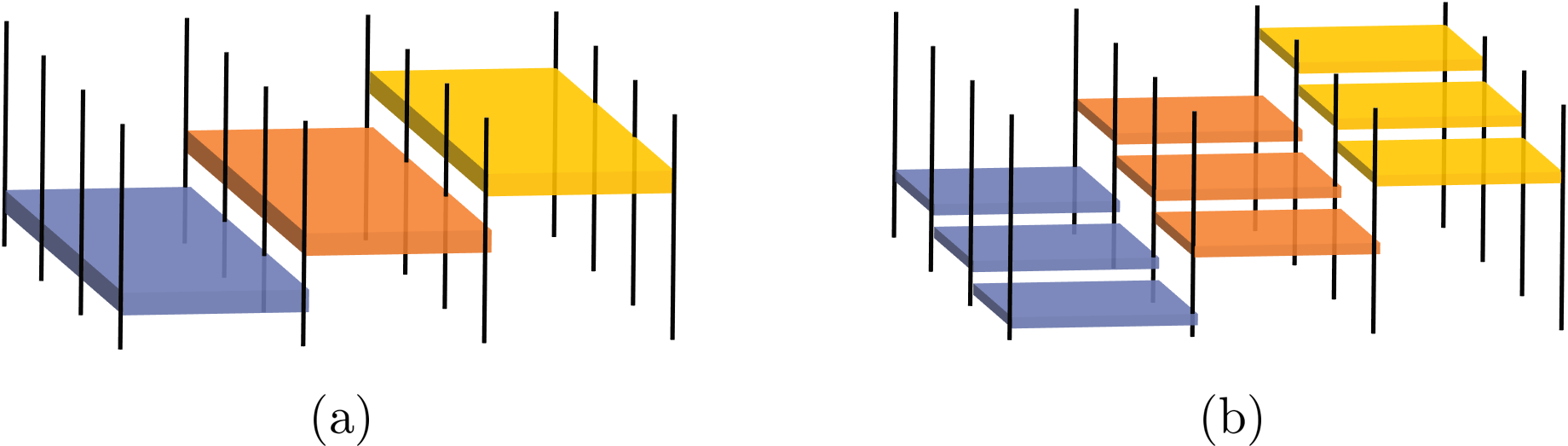}
    \caption{To construct a $D$-dimensional QCA as a sequential circuit, we decompose it in terms of $D-1$-dimensional QCA acting sequentially in one direction (a), then decompose each $D-1$-dimensional QCA into $D-2$-dimensional QCA (b), and so on.}
    \label{fig:2dqca}
\end{figure}

Now we show how to apply the same construction to arbitrary spatial dimension $D$. We assume the topology of a $D$-dimensional hypercubic lattice with periodic boundary conditions. Once again, we can block sites such that $Q$ spreads any operator by at most one site in all directions. Now, suppose we compactify all dimensions except one by viewing the $D$-dimensional system as a ring of supersites, each consisting of ($D-1$)-dimensional tori. Then $Q$ can be viewed as a 1D QCA acting on this compactified system. Therefore, we may use the above result to express it as a SQC in one direction, see Fig.~\ref{fig:2dqca}(a). According to the Eq.~\ref{eq:qca_circ}, the gates in this SQC are $Q^{(4)}$ and $w$ and SWAP. The SWAP gate acts on the compactified systems by swapping corresponding sites between the two tori, so each SWAP of tori is local and can be done in depth 1. Since $Q$ is locality preserving in all dimensions, $Q^{(4)}$ can be viewed as a $D-1$-dimensional QCA. Likewise, $w$, which essentially acts as the QCA on two supersites, is also a ($D-1$)-dimensional QCA, as was proven explicitly in Ref.~\onlinecite{Stephen2023}. Now, we use an inductive argument. We have already shown how to realize any 1D QCA as an SQC. Now, assume we can realize a ($D-1$)-dimensional QCA as an SQC. According to the above discussion, any $D$-dimensional QCA $Q$ can be written as an SQC consisting of ($D-1$)-dimensional QCA. By the inductive assumption, each of these QCA is in turn an SQC, such that $Q$ is itself an SQC. When constructed in this way, the depth of the SQC realizing $Q$ grows like $\mathcal{O}(N)$ where $N$ is the number of sites. This inductive procedure is illustrated in Fig.~\ref{fig:2dqca} for $D=2$.

It turns out that the SQCs constructed above are also composed of symmetric gates, so they are symmetric SQCs. That is, suppose that $Q^{(N)}$ commutes with some global on-site unitary symmetry $u^{\otimes N}$ for all $N$. It is clear that the SWAP gates commute with the global symmetry, as does $Q^{(4)}$ by assumption. It was shown in Ref.~\onlinecite{Stephen2023} that $w$ also commutes with the symmetry. Therefore, each of the local gates in the SQC (Eq.~\ref{eq:qca_circ}) is symmetric, so it is a symmetric SQC. One application of this comes from applying the construction to the FDQCs which create SPT ground states from symmetric product states \cite{Chen2013}. These FDQCs commute with the global symmetry protecting the SPT as a whole, but the individual gates do not commute with the symmetry. Applying our construction gives an SQC consisting of symmetric gates that realizes the same unitary. This gives an alternative, but similar, construction of SQCs for SPT states compared to those given in Sec.~\ref{sec:SPT}. 

It is interesting to consider truncating the SQCs we have constructed. Consider the case of $D=3$, where an important class of QCA is given by those which disentangle ground states of Walker-Wang models \cite{Haah2023,Haah2021,Shirley2022}. 
It has been argued that these QCA are non-trivial, meaning that they cannot be written as a product of FDQCs and translations. The argument is rooted in the fact that, if the QCA acts as an FDQC followed by a translation, then it can be easily truncated to act only in a finite region of space. These truncated circuits could then be used to generate an isolated boundary of the Walker-Wang models that host surface topological order, which is conjectured to be impossible in some cases. Thus, the non-trivial nature of the QCA is tied to the inability to straightforwardly truncate its action to a finite region of space. Since we have constructed SQCs realizing QCA, and these SQCs can be truncated, one may worry that there is a contradiction. However, this is not the case, as truncating the SQC simply implements the QCA exactly on a smaller periodic system, without introducing any boundaries. Therefore, the SQCs cannot be used to isolate boundaries of the Walker-Wang models and the non-triviality of the QCA, as described above, is compatible with the existence of a SQC realizing the QCA.
The sequential circuit shown in Section~\ref{sec:loop_cond3D} for constructing the Walker-Wang wavefunction is not mapped from a QCA and does leave open boundaries if truncated. But as it always generates two surfaces (top and bottom) at the same time, there is no contradiction with the non-triviality of the QCA either.

\section{Summary and outlook}
\label{sec:summary}

In this paper, we discussed the generation of nontrivial gapped quantum states, starting from product states, using Sequential Quantum Circuits. In particular, we discussed how the circuit that generates the (symmetrized) symmetry-breaking state and the SPT states preserves global symmetry but not necessarily locality; how the circuit that generates fractional topological states is associated with gapped boundaries to the vacuum with either charge or loop condensation; and how the circuit that generates fracton states are related to the foliation structure or p-string condensation in the fracton states. One major class of states that are not covered is the invertible chiral topological states like the Integer Quantum Hall. It is not clear what kind of circuit (beyond finite depth circuit) is needed to generate such states and this will be an interesting question to answer.

We can compare four types of many-body unitaries: the finite depth quantum circuit (FDQC), the quantum cellular automata (QCA), the sequential quantum circuit (SQC) and the linear depth quantum circuit (LDQC). Table~\ref{table:comparison} summarizes the similarities and differences between them. 
Based on what we know, we see that they have a strict containment relation
\begin{equation}
\text{FDQC}\ \subset \ \text{QCA} \ \subset \ \text{SQC} \subset \text{LDQC}
\end{equation}


\begin{table}[ht]
\centering
\hspace*{-0.5cm}
\begin{tabular}{ c || c | c | c | c }
preserving    & FDQC & QCA & SQC & LDQC\\ \hline
finite correlation length & Yes & Yes & No & No\\
zero correlation length & Yes & Yes & No & No\\
 short-range entanglement & Yes &  Yes$^*$ & No & No\\
 locality & Yes & Yes & No & No\\ 
 ground state degeneracy & Yes & Yes & No & No\\
 entanglement area law & Yes & Yes\cite{Piroli2020} & Yes & No
\end{tabular}
\caption{Comparison of four types of many-body unitary transformations: Finite Depth Local Unitary Circuit (FDC), Quantum Cellular Automaton (QCA), Sequential Quantum Circuit (SQC), and Linear Depth Quantum Circuit (LDQC). `Yes' means all the unitaries in this class preserve certain properties. `No' means some unitaries in this class do not preserve the property.  $^*$: we consider Walker-Wang models with a modular input category as short-range entangled even though they are generated with a nontrivial QCA.}
 \label{table:comparison}
\end{table}

Table~\ref{table:comparison} summarizes the similarities and differences between the three sets in terms of their effect on global properties of gapped many-body states. Linear depth circuit is the most powerful which can map gapped states to generic entanglement volume law states. The other three types of unitary all preserve entanglement area law, hence mapping gapped states to gapped states. SQC can change short-range correlation into long-range correlation, short-range entanglement into long-range entanglement. It can also change the locality of their parent Hamiltonian and its ground state degeneracy, hence capable of mapping between different gapped phases. They can also generate a non-zero finite correlation length from a state with zero correlation length \footnote{For example, SQCs can generate arbitrary matrix product states on open boundary conditions, which can have finite, non-zero correlation length \cite{Perez-Garcia2006}}. FDQC and QCA, on the other hand, preserve the locality of operator, correlation function and entanglement and cannot change ground state degeneracy of gapped Hamiltonian.

The Multi-scale Entanglement Renormalization procedure (MERA)\cite{Vidal2008,Gu2009,Swingle2016} is another way to map nontrivial entangled states to product states. It takes a many-body entangled wave function, applies one layer of finite depth circuit which disentangles local degrees of freedom (DOF) and maps the wave function back to its original form but with doubled unit cell. Such a step is then repeated at the renormalized length scale until the wave function is completely disentangled after $\log L$ steps. It is different from the sequential circuit in that in later steps of the procedure the degrees of freedoms are very far away from each other and the unitary gates applied are not local any more. This non-locality in unitary gates is also the reason why the MERA procedure does not violate the linear lower bound in circuit depth to generate GHZ or topological states\cite{Bravyi2006}. MERA is more powerful than a sequential circuit as it can map product states to gapless states\cite{Evenbly2016}.


\begin{acknowledgments}
We are indebted to inspiring discussions with Wenjie Ji, Laurens Lootens, Bram Vancraeynest-De Cuiper, Xiao-Gang Wen, and Cenke Xu. X.C. and A.D. are supported by the National Science Foundation under award number DMR-1654340, the Simons Investigator Award (award ID 828078) and the Institute for Quantum Information and Matter at Caltech. X.C. and N.T. are supported by the Walter Burke Institute for Theoretical Physics at Caltech. R.V. is supported by the Belgian American Educational Foundation.  The work of MH on fracton systems was supported by the U.S. Department of Energy, Office of Science, Basic Energy Sciences (BES) under Award number DE-SC0014415. The work of D.T.S., work of M.H. on aspects other than fractons, and (in part) the work of X.C. and A.D. was supported by the Simons Collaboration on Ultra-Quantum
Matter, which is funded by grants from the Simons Foundation (651440, DTS and MH; 651438, XC and AD).
\end{acknowledgments}

\bibliography{references}

\begin{thebibliography}{52}%
\makeatletter
\providecommand \@ifxundefined [1]{%
 \@ifx{#1\undefined}
}%
\providecommand \@ifnum [1]{%
 \ifnum #1\expandafter \@firstoftwo
 \else \expandafter \@secondoftwo
 \fi
}%
\providecommand \@ifx [1]{%
 \ifx #1\expandafter \@firstoftwo
 \else \expandafter \@secondoftwo
 \fi
}%
\providecommand \natexlab [1]{#1}%
\providecommand \enquote  [1]{``#1''}%
\providecommand \bibnamefont  [1]{#1}%
\providecommand \bibfnamefont [1]{#1}%
\providecommand \citenamefont [1]{#1}%
\providecommand \href@noop [0]{\@secondoftwo}%
\providecommand \href [0]{\begingroup \@sanitize@url \@href}%
\providecommand \@href[1]{\@@startlink{#1}\@@href}%
\providecommand \@@href[1]{\endgroup#1\@@endlink}%
\providecommand \@sanitize@url [0]{\catcode `\\12\catcode `\$12\catcode
  `\&12\catcode `\#12\catcode `\^12\catcode `\_12\catcode `\%12\relax}%
\providecommand \@@startlink[1]{}%
\providecommand \@@endlink[0]{}%
\providecommand \url  [0]{\begingroup\@sanitize@url \@url }%
\providecommand \@url [1]{\endgroup\@href {#1}{\urlprefix }}%
\providecommand \urlprefix  [0]{URL }%
\providecommand \Eprint [0]{\href }%
\providecommand \doibase [0]{http://dx.doi.org/}%
\providecommand \selectlanguage [0]{\@gobble}%
\providecommand \bibinfo  [0]{\@secondoftwo}%
\providecommand \bibfield  [0]{\@secondoftwo}%
\providecommand \translation [1]{[#1]}%
\providecommand \BibitemOpen [0]{}%
\providecommand \bibitemStop [0]{}%
\providecommand \bibitemNoStop [0]{.\EOS\space}%
\providecommand \EOS [0]{\spacefactor3000\relax}%
\providecommand \BibitemShut  [1]{\csname bibitem#1\endcsname}%
\let\auto@bib@innerbib\@empty
\bibitem [{\citenamefont {Chen}\ \emph {et~al.}(2010)\citenamefont {Chen},
  \citenamefont {Gu},\ and\ \citenamefont {Wen}}]{Chen2010}%
  \BibitemOpen
  \bibfield  {author} {\bibinfo {author} {\bibfnamefont {X.}~\bibnamefont
  {Chen}}, \bibinfo {author} {\bibfnamefont {Z.-C.}\ \bibnamefont {Gu}}, \ and\
  \bibinfo {author} {\bibfnamefont {X.-G.}\ \bibnamefont {Wen}},\ }\href
  {\doibase 10.1103/PhysRevB.82.155138} {\bibfield  {journal} {\bibinfo
  {journal} {Phys. Rev. B}\ }\textbf {\bibinfo {volume} {82}},\ \bibinfo
  {pages} {155138} (\bibinfo {year} {2010})}\BibitemShut {NoStop}%
\bibitem [{\citenamefont {Bravyi}\ \emph {et~al.}(2006)\citenamefont {Bravyi},
  \citenamefont {Hastings},\ and\ \citenamefont {Verstraete}}]{Bravyi2006}%
  \BibitemOpen
  \bibfield  {author} {\bibinfo {author} {\bibfnamefont {S.}~\bibnamefont
  {Bravyi}}, \bibinfo {author} {\bibfnamefont {M.~B.}\ \bibnamefont
  {Hastings}}, \ and\ \bibinfo {author} {\bibfnamefont {F.}~\bibnamefont
  {Verstraete}},\ }\href {\doibase 10.1103/PhysRevLett.97.050401} {\bibfield
  {journal} {\bibinfo  {journal} {Phys. Rev. Lett.}\ }\textbf {\bibinfo
  {volume} {97}},\ \bibinfo {pages} {050401} (\bibinfo {year}
  {2006})}\BibitemShut {NoStop}%
\bibitem [{\citenamefont {Sch\"on}\ \emph {et~al.}(2005)\citenamefont
  {Sch\"on}, \citenamefont {Solano}, \citenamefont {Verstraete}, \citenamefont
  {Cirac},\ and\ \citenamefont {Wolf}}]{Schon2005}%
  \BibitemOpen
  \bibfield  {author} {\bibinfo {author} {\bibfnamefont {C.}~\bibnamefont
  {Sch\"on}}, \bibinfo {author} {\bibfnamefont {E.}~\bibnamefont {Solano}},
  \bibinfo {author} {\bibfnamefont {F.}~\bibnamefont {Verstraete}}, \bibinfo
  {author} {\bibfnamefont {J.~I.}\ \bibnamefont {Cirac}}, \ and\ \bibinfo
  {author} {\bibfnamefont {M.~M.}\ \bibnamefont {Wolf}},\ }\href {\doibase
  10.1103/PhysRevLett.95.110503} {\bibfield  {journal} {\bibinfo  {journal}
  {Phys. Rev. Lett.}\ }\textbf {\bibinfo {volume} {95}},\ \bibinfo {pages}
  {110503} (\bibinfo {year} {2005})}\BibitemShut {NoStop}%
\bibitem [{\citenamefont {Sch\"on}\ \emph {et~al.}(2007)\citenamefont
  {Sch\"on}, \citenamefont {Hammerer}, \citenamefont {Wolf}, \citenamefont
  {Cirac},\ and\ \citenamefont {Solano}}]{Schon2007}%
  \BibitemOpen
  \bibfield  {author} {\bibinfo {author} {\bibfnamefont {C.}~\bibnamefont
  {Sch\"on}}, \bibinfo {author} {\bibfnamefont {K.}~\bibnamefont {Hammerer}},
  \bibinfo {author} {\bibfnamefont {M.~M.}\ \bibnamefont {Wolf}}, \bibinfo
  {author} {\bibfnamefont {J.~I.}\ \bibnamefont {Cirac}}, \ and\ \bibinfo
  {author} {\bibfnamefont {E.}~\bibnamefont {Solano}},\ }\href {\doibase
  10.1103/PhysRevA.75.032311} {\bibfield  {journal} {\bibinfo  {journal} {Phys.
  Rev. A}\ }\textbf {\bibinfo {volume} {75}},\ \bibinfo {pages} {032311}
  (\bibinfo {year} {2007})}\BibitemShut {NoStop}%
\bibitem [{\citenamefont {Ba\~nuls}\ \emph {et~al.}(2008)\citenamefont
  {Ba\~nuls}, \citenamefont {P\'erez-Garc\'{\i}a}, \citenamefont {Wolf},
  \citenamefont {Verstraete},\ and\ \citenamefont {Cirac}}]{Banuls2008}%
  \BibitemOpen
  \bibfield  {author} {\bibinfo {author} {\bibfnamefont {M.~C.}\ \bibnamefont
  {Ba\~nuls}}, \bibinfo {author} {\bibfnamefont {D.}~\bibnamefont
  {P\'erez-Garc\'{\i}a}}, \bibinfo {author} {\bibfnamefont {M.~M.}\
  \bibnamefont {Wolf}}, \bibinfo {author} {\bibfnamefont {F.}~\bibnamefont
  {Verstraete}}, \ and\ \bibinfo {author} {\bibfnamefont {J.~I.}\ \bibnamefont
  {Cirac}},\ }\href {\doibase 10.1103/PhysRevA.77.052306} {\bibfield  {journal}
  {\bibinfo  {journal} {Phys. Rev. A}\ }\textbf {\bibinfo {volume} {77}},\
  \bibinfo {pages} {052306} (\bibinfo {year} {2008})}\BibitemShut {NoStop}%
\bibitem [{\citenamefont {Wei}\ \emph {et~al.}(2022)\citenamefont {Wei},
  \citenamefont {Malz},\ and\ \citenamefont {Cirac}}]{Wei2022}%
  \BibitemOpen
  \bibfield  {author} {\bibinfo {author} {\bibfnamefont {Z.-Y.}\ \bibnamefont
  {Wei}}, \bibinfo {author} {\bibfnamefont {D.}~\bibnamefont {Malz}}, \ and\
  \bibinfo {author} {\bibfnamefont {J.~I.}\ \bibnamefont {Cirac}},\ }\href
  {\doibase 10.1103/PhysRevLett.128.010607} {\bibfield  {journal} {\bibinfo
  {journal} {Phys. Rev. Lett.}\ }\textbf {\bibinfo {volume} {128}},\ \bibinfo
  {pages} {010607} (\bibinfo {year} {2022})}\BibitemShut {NoStop}%
\bibitem [{\citenamefont {Lin}\ \emph {et~al.}(2021)\citenamefont {Lin},
  \citenamefont {Dilip}, \citenamefont {Green}, \citenamefont {Smith},\ and\
  \citenamefont {Pollmann}}]{Lin2021}%
  \BibitemOpen
  \bibfield  {author} {\bibinfo {author} {\bibfnamefont {S.-H.}\ \bibnamefont
  {Lin}}, \bibinfo {author} {\bibfnamefont {R.}~\bibnamefont {Dilip}}, \bibinfo
  {author} {\bibfnamefont {A.~G.}\ \bibnamefont {Green}}, \bibinfo {author}
  {\bibfnamefont {A.}~\bibnamefont {Smith}}, \ and\ \bibinfo {author}
  {\bibfnamefont {F.}~\bibnamefont {Pollmann}},\ }\href {\doibase
  10.1103/PRXQuantum.2.010342} {\bibfield  {journal} {\bibinfo  {journal} {PRX
  Quantum}\ }\textbf {\bibinfo {volume} {2}},\ \bibinfo {pages} {010342}
  (\bibinfo {year} {2021})}\BibitemShut {NoStop}%
\bibitem [{\citenamefont {Lamata}\ \emph {et~al.}(2008)\citenamefont {Lamata},
  \citenamefont {Le\'on}, \citenamefont {P\'erez-Garc\'{\i}a}, \citenamefont
  {Salgado},\ and\ \citenamefont {Solano}}]{Lamata2008}%
  \BibitemOpen
  \bibfield  {author} {\bibinfo {author} {\bibfnamefont {L.}~\bibnamefont
  {Lamata}}, \bibinfo {author} {\bibfnamefont {J.}~\bibnamefont {Le\'on}},
  \bibinfo {author} {\bibfnamefont {D.}~\bibnamefont {P\'erez-Garc\'{\i}a}},
  \bibinfo {author} {\bibfnamefont {D.}~\bibnamefont {Salgado}}, \ and\
  \bibinfo {author} {\bibfnamefont {E.}~\bibnamefont {Solano}},\ }\href
  {\doibase 10.1103/PhysRevLett.101.180506} {\bibfield  {journal} {\bibinfo
  {journal} {Phys. Rev. Lett.}\ }\textbf {\bibinfo {volume} {101}},\ \bibinfo
  {pages} {180506} (\bibinfo {year} {2008})}\BibitemShut {NoStop}%
\bibitem [{\citenamefont {Saberi}(2011)}]{Saberi2011}%
  \BibitemOpen
  \bibfield  {author} {\bibinfo {author} {\bibfnamefont {H.}~\bibnamefont
  {Saberi}},\ }\href {\doibase 10.1103/PhysRevA.84.032323} {\bibfield
  {journal} {\bibinfo  {journal} {Phys. Rev. A}\ }\textbf {\bibinfo {volume}
  {84}},\ \bibinfo {pages} {032323} (\bibinfo {year} {2011})}\BibitemShut
  {NoStop}%
\bibitem [{\citenamefont {Liu}\ \emph {et~al.}(2022)\citenamefont {Liu},
  \citenamefont {Shtengel}, \citenamefont {Smith},\ and\ \citenamefont
  {Pollmann}}]{Liu2022}%
  \BibitemOpen
  \bibfield  {author} {\bibinfo {author} {\bibfnamefont {Y.-J.}\ \bibnamefont
  {Liu}}, \bibinfo {author} {\bibfnamefont {K.}~\bibnamefont {Shtengel}},
  \bibinfo {author} {\bibfnamefont {A.}~\bibnamefont {Smith}}, \ and\ \bibinfo
  {author} {\bibfnamefont {F.}~\bibnamefont {Pollmann}},\ }\href {\doibase
  10.1103/PRXQuantum.3.040315} {\bibfield  {journal} {\bibinfo  {journal} {PRX
  Quantum}\ }\textbf {\bibinfo {volume} {3}},\ \bibinfo {pages} {040315}
  (\bibinfo {year} {2022})}\BibitemShut {NoStop}%
\bibitem [{\citenamefont {Huang}\ and\ \citenamefont {Chen}(2015)}]{Huang2015}%
  \BibitemOpen
  \bibfield  {author} {\bibinfo {author} {\bibfnamefont {Y.}~\bibnamefont
  {Huang}}\ and\ \bibinfo {author} {\bibfnamefont {X.}~\bibnamefont {Chen}},\
  }\href {\doibase 10.1103/PhysRevB.91.195143} {\bibfield  {journal} {\bibinfo
  {journal} {Phys. Rev. B}\ }\textbf {\bibinfo {volume} {91}},\ \bibinfo
  {pages} {195143} (\bibinfo {year} {2015})}\BibitemShut {NoStop}%
\bibitem [{\citenamefont {Ho}\ and\ \citenamefont {Hsieh}(2019)}]{Ho2019}%
  \BibitemOpen
  \bibfield  {author} {\bibinfo {author} {\bibfnamefont {W.~W.}\ \bibnamefont
  {Ho}}\ and\ \bibinfo {author} {\bibfnamefont {T.~H.}\ \bibnamefont {Hsieh}},\
  }\href {\doibase 10.21468/SciPostPhys.6.3.029} {\bibfield  {journal}
  {\bibinfo  {journal} {SciPost Phys.}\ }\textbf {\bibinfo {volume} {6}},\
  \bibinfo {pages} {029} (\bibinfo {year} {2019})}\BibitemShut {NoStop}%
\bibitem [{\citenamefont {Chen}\ \emph {et~al.}(2022)\citenamefont {Chen},
  \citenamefont {Yan},\ and\ \citenamefont {Cui}}]{Chen2022arxiv}%
  \BibitemOpen
  \bibfield  {author} {\bibinfo {author} {\bibfnamefont {P.}~\bibnamefont
  {Chen}}, \bibinfo {author} {\bibfnamefont {B.}~\bibnamefont {Yan}}, \ and\
  \bibinfo {author} {\bibfnamefont {S.~X.}\ \bibnamefont {Cui}},\ }\href@noop
  {} {\enquote {\bibinfo {title} {Quantum circuits for toric code and x-cube
  fracton model},}\ } (\bibinfo {year} {2022}),\ \Eprint
  {http://arxiv.org/abs/2210.01682} {arXiv:2210.01682 [cond-mat.str-el]}
  \BibitemShut {NoStop}%
\bibitem [{Note1()}]{Note1}%
  \BibitemOpen
  \bibinfo {note} {If we perform the Jordan Wigner transformation and map the
  transverse field Ising model to the Majorana chain, the mapping between the
  two phases can be realized by translation by a single Majorana
  mode.}\BibitemShut {Stop}%
\bibitem [{\citenamefont {Chen}\ \emph {et~al.}(2013)\citenamefont {Chen},
  \citenamefont {Gu}, \citenamefont {Liu},\ and\ \citenamefont
  {Wen}}]{Chen2013}%
  \BibitemOpen
  \bibfield  {author} {\bibinfo {author} {\bibfnamefont {X.}~\bibnamefont
  {Chen}}, \bibinfo {author} {\bibfnamefont {Z.-C.}\ \bibnamefont {Gu}},
  \bibinfo {author} {\bibfnamefont {Z.-X.}\ \bibnamefont {Liu}}, \ and\
  \bibinfo {author} {\bibfnamefont {X.-G.}\ \bibnamefont {Wen}},\ }\href
  {\doibase 10.1103/PhysRevB.87.155114} {\bibfield  {journal} {\bibinfo
  {journal} {Phys. Rev. B}\ }\textbf {\bibinfo {volume} {87}},\ \bibinfo
  {pages} {155114} (\bibinfo {year} {2013})}\BibitemShut {NoStop}%
\bibitem [{\citenamefont {Chen}\ \emph
  {et~al.}(2011{\natexlab{a}})\citenamefont {Chen}, \citenamefont {Gu},\ and\
  \citenamefont {Wen}}]{Chen2011}%
  \BibitemOpen
  \bibfield  {author} {\bibinfo {author} {\bibfnamefont {X.}~\bibnamefont
  {Chen}}, \bibinfo {author} {\bibfnamefont {Z.-C.}\ \bibnamefont {Gu}}, \ and\
  \bibinfo {author} {\bibfnamefont {X.-G.}\ \bibnamefont {Wen}},\ }\href
  {\doibase 10.1103/PhysRevB.83.035107} {\bibfield  {journal} {\bibinfo
  {journal} {Phys. Rev. B}\ }\textbf {\bibinfo {volume} {83}},\ \bibinfo
  {pages} {035107} (\bibinfo {year} {2011}{\natexlab{a}})}\BibitemShut
  {NoStop}%
\bibitem [{\citenamefont {Schuch}\ \emph {et~al.}(2011)\citenamefont {Schuch},
  \citenamefont {P\'erez-Garc\'{\i}a},\ and\ \citenamefont
  {Cirac}}]{Schuch2011}%
  \BibitemOpen
  \bibfield  {author} {\bibinfo {author} {\bibfnamefont {N.}~\bibnamefont
  {Schuch}}, \bibinfo {author} {\bibfnamefont {D.}~\bibnamefont
  {P\'erez-Garc\'{\i}a}}, \ and\ \bibinfo {author} {\bibfnamefont
  {I.}~\bibnamefont {Cirac}},\ }\href {\doibase 10.1103/PhysRevB.84.165139}
  {\bibfield  {journal} {\bibinfo  {journal} {Phys. Rev. B}\ }\textbf {\bibinfo
  {volume} {84}},\ \bibinfo {pages} {165139} (\bibinfo {year}
  {2011})}\BibitemShut {NoStop}%
\bibitem [{\citenamefont {Pollmann}\ \emph {et~al.}(2010)\citenamefont
  {Pollmann}, \citenamefont {Turner}, \citenamefont {Berg},\ and\ \citenamefont
  {Oshikawa}}]{Pollmann2010}%
  \BibitemOpen
  \bibfield  {author} {\bibinfo {author} {\bibfnamefont {F.}~\bibnamefont
  {Pollmann}}, \bibinfo {author} {\bibfnamefont {A.~M.}\ \bibnamefont
  {Turner}}, \bibinfo {author} {\bibfnamefont {E.}~\bibnamefont {Berg}}, \ and\
  \bibinfo {author} {\bibfnamefont {M.}~\bibnamefont {Oshikawa}},\ }\href
  {\doibase 10.1103/PhysRevB.81.064439} {\bibfield  {journal} {\bibinfo
  {journal} {Phys. Rev. B}\ }\textbf {\bibinfo {volume} {81}},\ \bibinfo
  {pages} {064439} (\bibinfo {year} {2010})}\BibitemShut {NoStop}%
\bibitem [{\citenamefont {Chen}\ \emph
  {et~al.}(2011{\natexlab{b}})\citenamefont {Chen}, \citenamefont {Liu},\ and\
  \citenamefont {Wen}}]{Chen2011a}%
  \BibitemOpen
  \bibfield  {author} {\bibinfo {author} {\bibfnamefont {X.}~\bibnamefont
  {Chen}}, \bibinfo {author} {\bibfnamefont {Z.-X.}\ \bibnamefont {Liu}}, \
  and\ \bibinfo {author} {\bibfnamefont {X.-G.}\ \bibnamefont {Wen}},\ }\href
  {\doibase 10.1103/PhysRevB.84.235141} {\bibfield  {journal} {\bibinfo
  {journal} {Phys. Rev. B}\ }\textbf {\bibinfo {volume} {84}},\ \bibinfo
  {pages} {235141} (\bibinfo {year} {2011}{\natexlab{b}})}\BibitemShut
  {NoStop}%
\bibitem [{\citenamefont {Fidkowski}\ \emph {et~al.}(2020)\citenamefont
  {Fidkowski}, \citenamefont {Haah},\ and\ \citenamefont
  {Hastings}}]{Fidkowski2020}%
  \BibitemOpen
  \bibfield  {author} {\bibinfo {author} {\bibfnamefont {L.}~\bibnamefont
  {Fidkowski}}, \bibinfo {author} {\bibfnamefont {J.}~\bibnamefont {Haah}}, \
  and\ \bibinfo {author} {\bibfnamefont {M.~B.}\ \bibnamefont {Hastings}},\
  }\href {\doibase 10.1103/PhysRevB.101.155124} {\bibfield  {journal} {\bibinfo
   {journal} {Phys. Rev. B}\ }\textbf {\bibinfo {volume} {101}},\ \bibinfo
  {pages} {155124} (\bibinfo {year} {2020})}\BibitemShut {NoStop}%
\bibitem [{\citenamefont {Levin}\ and\ \citenamefont
  {Wen}(2005)}]{levin2005string}%
  \BibitemOpen
  \bibfield  {author} {\bibinfo {author} {\bibfnamefont {M.~A.}\ \bibnamefont
  {Levin}}\ and\ \bibinfo {author} {\bibfnamefont {X.-G.}\ \bibnamefont
  {Wen}},\ }\href@noop {} {\bibfield  {journal} {\bibinfo  {journal} {Physical
  Review B}\ }\textbf {\bibinfo {volume} {71}},\ \bibinfo {pages} {045110}
  (\bibinfo {year} {2005})}\BibitemShut {NoStop}%
\bibitem [{\citenamefont {Kitaev}(2003)}]{Kitaev2003}%
  \BibitemOpen
  \bibfield  {author} {\bibinfo {author} {\bibfnamefont {A.}~\bibnamefont
  {Kitaev}},\ }\href {\doibase https://doi.org/10.1016/S0003-4916(02)00018-0}
  {\bibfield  {journal} {\bibinfo  {journal} {Annals of Physics}\ }\textbf
  {\bibinfo {volume} {303}},\ \bibinfo {pages} {2} (\bibinfo {year}
  {2003})}\BibitemShut {NoStop}%
\bibitem [{\citenamefont {Kitaev}\ and\ \citenamefont
  {Kong}(2012)}]{kitaev2012models}%
  \BibitemOpen
  \bibfield  {author} {\bibinfo {author} {\bibfnamefont {A.}~\bibnamefont
  {Kitaev}}\ and\ \bibinfo {author} {\bibfnamefont {L.}~\bibnamefont {Kong}},\
  }\href@noop {} {\bibfield  {journal} {\bibinfo  {journal} {Communications in
  Mathematical Physics}\ }\textbf {\bibinfo {volume} {313}},\ \bibinfo {pages}
  {351} (\bibinfo {year} {2012})}\BibitemShut {NoStop}%
\bibitem [{\citenamefont {Lootens}\ \emph {et~al.}(2022)\citenamefont
  {Lootens}, \citenamefont {Vancraeynest-De~Cuiper}, \citenamefont {Schuch},\
  and\ \citenamefont {Verstraete}}]{lootens2022mapping}%
  \BibitemOpen
  \bibfield  {author} {\bibinfo {author} {\bibfnamefont {L.}~\bibnamefont
  {Lootens}}, \bibinfo {author} {\bibfnamefont {B.}~\bibnamefont
  {Vancraeynest-De~Cuiper}}, \bibinfo {author} {\bibfnamefont {N.}~\bibnamefont
  {Schuch}}, \ and\ \bibinfo {author} {\bibfnamefont {F.}~\bibnamefont
  {Verstraete}},\ }\href@noop {} {\bibfield  {journal} {\bibinfo  {journal}
  {Physical Review B}\ }\textbf {\bibinfo {volume} {105}},\ \bibinfo {pages}
  {085130} (\bibinfo {year} {2022})}\BibitemShut {NoStop}%
\bibitem [{\citenamefont {Wang}\ \emph {et~al.}(2022)\citenamefont {Wang},
  \citenamefont {Ma}, \citenamefont {Stephen}, \citenamefont {Hermele},\ and\
  \citenamefont {Chen}}]{wang2022renormalization}%
  \BibitemOpen
  \bibfield  {author} {\bibinfo {author} {\bibfnamefont {Z.}~\bibnamefont
  {Wang}}, \bibinfo {author} {\bibfnamefont {X.}~\bibnamefont {Ma}}, \bibinfo
  {author} {\bibfnamefont {D.~T.}\ \bibnamefont {Stephen}}, \bibinfo {author}
  {\bibfnamefont {M.}~\bibnamefont {Hermele}}, \ and\ \bibinfo {author}
  {\bibfnamefont {X.}~\bibnamefont {Chen}},\ }\href@noop {} {\bibfield
  {journal} {\bibinfo  {journal} {arXiv preprint arXiv:2301.00103}\ } (\bibinfo
  {year} {2022})}\BibitemShut {NoStop}%
\bibitem [{\citenamefont {Etingof}\ \emph {et~al.}(2016)\citenamefont
  {Etingof}, \citenamefont {Gelaki}, \citenamefont {Nikshych},\ and\
  \citenamefont {Ostrik}}]{etingof2016tensor}%
  \BibitemOpen
  \bibfield  {author} {\bibinfo {author} {\bibfnamefont {P.}~\bibnamefont
  {Etingof}}, \bibinfo {author} {\bibfnamefont {S.}~\bibnamefont {Gelaki}},
  \bibinfo {author} {\bibfnamefont {D.}~\bibnamefont {Nikshych}}, \ and\
  \bibinfo {author} {\bibfnamefont {V.}~\bibnamefont {Ostrik}},\ }\href@noop {}
  {\emph {\bibinfo {title} {Tensor categories}}},\ Vol.\ \bibinfo {volume}
  {205}\ (\bibinfo  {publisher} {American Mathematical Soc.},\ \bibinfo {year}
  {2016})\BibitemShut {NoStop}%
\bibitem [{\citenamefont {Lootens}\ \emph {et~al.}(2021)\citenamefont
  {Lootens}, \citenamefont {Fuchs}, \citenamefont {Haegeman}, \citenamefont
  {Schweigert},\ and\ \citenamefont {Verstraete}}]{lootens2021matrix}%
  \BibitemOpen
  \bibfield  {author} {\bibinfo {author} {\bibfnamefont {L.}~\bibnamefont
  {Lootens}}, \bibinfo {author} {\bibfnamefont {J.}~\bibnamefont {Fuchs}},
  \bibinfo {author} {\bibfnamefont {J.}~\bibnamefont {Haegeman}}, \bibinfo
  {author} {\bibfnamefont {C.}~\bibnamefont {Schweigert}}, \ and\ \bibinfo
  {author} {\bibfnamefont {F.}~\bibnamefont {Verstraete}},\ }\href@noop {}
  {\bibfield  {journal} {\bibinfo  {journal} {SciPost Physics}\ }\textbf
  {\bibinfo {volume} {10}},\ \bibinfo {pages} {053} (\bibinfo {year}
  {2021})}\BibitemShut {NoStop}%
\bibitem [{\citenamefont {Buerschaper}\ \emph {et~al.}(2009)\citenamefont
  {Buerschaper}, \citenamefont {Aguado},\ and\ \citenamefont
  {Vidal}}]{buerschaper2009explicit}%
  \BibitemOpen
  \bibfield  {author} {\bibinfo {author} {\bibfnamefont {O.}~\bibnamefont
  {Buerschaper}}, \bibinfo {author} {\bibfnamefont {M.}~\bibnamefont {Aguado}},
  \ and\ \bibinfo {author} {\bibfnamefont {G.}~\bibnamefont {Vidal}},\
  }\href@noop {} {\bibfield  {journal} {\bibinfo  {journal} {Physical Review
  B}\ }\textbf {\bibinfo {volume} {79}},\ \bibinfo {pages} {085119} (\bibinfo
  {year} {2009})}\BibitemShut {NoStop}%
\bibitem [{\citenamefont {Gu}\ \emph {et~al.}(2009)\citenamefont {Gu},
  \citenamefont {Levin}, \citenamefont {Swingle},\ and\ \citenamefont
  {Wen}}]{gu2009tensor}%
  \BibitemOpen
  \bibfield  {author} {\bibinfo {author} {\bibfnamefont {Z.-C.}\ \bibnamefont
  {Gu}}, \bibinfo {author} {\bibfnamefont {M.}~\bibnamefont {Levin}}, \bibinfo
  {author} {\bibfnamefont {B.}~\bibnamefont {Swingle}}, \ and\ \bibinfo
  {author} {\bibfnamefont {X.-G.}\ \bibnamefont {Wen}},\ }\href@noop {}
  {\bibfield  {journal} {\bibinfo  {journal} {Physical Review B}\ }\textbf
  {\bibinfo {volume} {79}},\ \bibinfo {pages} {085118} (\bibinfo {year}
  {2009})}\BibitemShut {NoStop}%
\bibitem [{\citenamefont {Hamma}\ \emph {et~al.}(2005)\citenamefont {Hamma},
  \citenamefont {Zanardi},\ and\ \citenamefont {Wen}}]{Hamma2005}%
  \BibitemOpen
  \bibfield  {author} {\bibinfo {author} {\bibfnamefont {A.}~\bibnamefont
  {Hamma}}, \bibinfo {author} {\bibfnamefont {P.}~\bibnamefont {Zanardi}}, \
  and\ \bibinfo {author} {\bibfnamefont {X.-G.}\ \bibnamefont {Wen}},\ }\href
  {\doibase 10.1103/PhysRevB.72.035307} {\bibfield  {journal} {\bibinfo
  {journal} {Phys. Rev. B}\ }\textbf {\bibinfo {volume} {72}},\ \bibinfo
  {pages} {035307} (\bibinfo {year} {2005})}\BibitemShut {NoStop}%
\bibitem [{\citenamefont {Walker}\ and\ \citenamefont
  {Wang}(2012)}]{Walker2012}%
  \BibitemOpen
  \bibfield  {author} {\bibinfo {author} {\bibfnamefont {K.}~\bibnamefont
  {Walker}}\ and\ \bibinfo {author} {\bibfnamefont {Z.}~\bibnamefont {Wang}},\
  }\href {\doibase 10.1007/s11467-011-0194-z} {\bibfield  {journal} {\bibinfo
  {journal} {Frontiers of Physics}\ }\textbf {\bibinfo {volume} {7}},\ \bibinfo
  {pages} {150} (\bibinfo {year} {2012})}\BibitemShut {NoStop}%
\bibitem [{\citenamefont {Vijay}\ \emph {et~al.}(2016)\citenamefont {Vijay},
  \citenamefont {Haah},\ and\ \citenamefont {Fu}}]{Vijay_2016}%
  \BibitemOpen
  \bibfield  {author} {\bibinfo {author} {\bibfnamefont {S.}~\bibnamefont
  {Vijay}}, \bibinfo {author} {\bibfnamefont {J.}~\bibnamefont {Haah}}, \ and\
  \bibinfo {author} {\bibfnamefont {L.}~\bibnamefont {Fu}},\ }\href {\doibase
  10.1103/physrevb.94.235157} {\bibfield  {journal} {\bibinfo  {journal}
  {Physical Review B}\ }\textbf {\bibinfo {volume} {94}} (\bibinfo {year}
  {2016}),\ 10.1103/physrevb.94.235157}\BibitemShut {NoStop}%
\bibitem [{\citenamefont {Shirley}\ \emph {et~al.}(2018)\citenamefont
  {Shirley}, \citenamefont {Slagle}, \citenamefont {Wang},\ and\ \citenamefont
  {Chen}}]{Shirley2018}%
  \BibitemOpen
  \bibfield  {author} {\bibinfo {author} {\bibfnamefont {W.}~\bibnamefont
  {Shirley}}, \bibinfo {author} {\bibfnamefont {K.}~\bibnamefont {Slagle}},
  \bibinfo {author} {\bibfnamefont {Z.}~\bibnamefont {Wang}}, \ and\ \bibinfo
  {author} {\bibfnamefont {X.}~\bibnamefont {Chen}},\ }\href {\doibase
  10.1103/PhysRevX.8.031051} {\bibfield  {journal} {\bibinfo  {journal} {Phys.
  Rev. X}\ }\textbf {\bibinfo {volume} {8}},\ \bibinfo {pages} {031051}
  (\bibinfo {year} {2018})}\BibitemShut {NoStop}%
\bibitem [{\citenamefont {Ma}\ \emph {et~al.}(2017)\citenamefont {Ma},
  \citenamefont {Lake}, \citenamefont {Chen},\ and\ \citenamefont
  {Hermele}}]{Ma2017}%
  \BibitemOpen
  \bibfield  {author} {\bibinfo {author} {\bibfnamefont {H.}~\bibnamefont
  {Ma}}, \bibinfo {author} {\bibfnamefont {E.}~\bibnamefont {Lake}}, \bibinfo
  {author} {\bibfnamefont {X.}~\bibnamefont {Chen}}, \ and\ \bibinfo {author}
  {\bibfnamefont {M.}~\bibnamefont {Hermele}},\ }\href {\doibase
  10.1103/PhysRevB.95.245126} {\bibfield  {journal} {\bibinfo  {journal} {Phys.
  Rev. B}\ }\textbf {\bibinfo {volume} {95}},\ \bibinfo {pages} {245126}
  (\bibinfo {year} {2017})}\BibitemShut {NoStop}%
\bibitem [{\citenamefont {Prem}\ \emph {et~al.}(2019)\citenamefont {Prem},
  \citenamefont {Huang}, \citenamefont {Song},\ and\ \citenamefont
  {Hermele}}]{Prem_2019}%
  \BibitemOpen
  \bibfield  {author} {\bibinfo {author} {\bibfnamefont {A.}~\bibnamefont
  {Prem}}, \bibinfo {author} {\bibfnamefont {S.-J.}\ \bibnamefont {Huang}},
  \bibinfo {author} {\bibfnamefont {H.}~\bibnamefont {Song}}, \ and\ \bibinfo
  {author} {\bibfnamefont {M.}~\bibnamefont {Hermele}},\ }\href {\doibase
  10.1103/PhysRevX.9.021010} {\bibfield  {journal} {\bibinfo  {journal} {Phys.
  Rev. X}\ }\textbf {\bibinfo {volume} {9}},\ \bibinfo {pages} {021010}
  (\bibinfo {year} {2019})}\BibitemShut {NoStop}%
\bibitem [{\citenamefont {Farrelly}(2020)}]{Farrelly2020}%
  \BibitemOpen
  \bibfield  {author} {\bibinfo {author} {\bibfnamefont {T.}~\bibnamefont
  {Farrelly}},\ }\href {\doibase 10.22331/q-2020-11-30-368} {\bibfield
  {journal} {\bibinfo  {journal} {{Quantum}}\ }\textbf {\bibinfo {volume}
  {4}},\ \bibinfo {pages} {368} (\bibinfo {year} {2020})}\BibitemShut {NoStop}%
\bibitem [{\citenamefont {Po}\ \emph {et~al.}(2016)\citenamefont {Po},
  \citenamefont {Fidkowski}, \citenamefont {Morimoto}, \citenamefont {Potter},\
  and\ \citenamefont {Vishwanath}}]{Po2016}%
  \BibitemOpen
  \bibfield  {author} {\bibinfo {author} {\bibfnamefont {H.~C.}\ \bibnamefont
  {Po}}, \bibinfo {author} {\bibfnamefont {L.}~\bibnamefont {Fidkowski}},
  \bibinfo {author} {\bibfnamefont {T.}~\bibnamefont {Morimoto}}, \bibinfo
  {author} {\bibfnamefont {A.~C.}\ \bibnamefont {Potter}}, \ and\ \bibinfo
  {author} {\bibfnamefont {A.}~\bibnamefont {Vishwanath}},\ }\href {\doibase
  10.1103/PhysRevX.6.041070} {\bibfield  {journal} {\bibinfo  {journal} {Phys.
  Rev. X}\ }\textbf {\bibinfo {volume} {6}},\ \bibinfo {pages} {041070}
  (\bibinfo {year} {2016})}\BibitemShut {NoStop}%
\bibitem [{\citenamefont {Stephen}\ \emph {et~al.}(2023)\citenamefont
  {Stephen}, \citenamefont {Dua}, \citenamefont {Lavasani},\ and\ \citenamefont
  {Nandkishore}}]{Stephen2023}%
  \BibitemOpen
  \bibfield  {author} {\bibinfo {author} {\bibfnamefont {D.~T.}\ \bibnamefont
  {Stephen}}, \bibinfo {author} {\bibfnamefont {A.}~\bibnamefont {Dua}},
  \bibinfo {author} {\bibfnamefont {A.}~\bibnamefont {Lavasani}}, \ and\
  \bibinfo {author} {\bibfnamefont {R.}~\bibnamefont {Nandkishore}},\
  }\href@noop {} {\enquote {\bibinfo {title} {Non-local finite-depth circuits
  for constructing spt states and quantum cellular automata},}\ } (\bibinfo
  {year} {2023}),\ \Eprint {http://arxiv.org/abs/2212.06844} {arXiv:2212.06844
  [quant-ph]} \BibitemShut {NoStop}%
\bibitem [{\citenamefont {Schumacher}\ and\ \citenamefont
  {Werner}(2004)}]{Schumacher2004}%
  \BibitemOpen
  \bibfield  {author} {\bibinfo {author} {\bibfnamefont {B.}~\bibnamefont
  {Schumacher}}\ and\ \bibinfo {author} {\bibfnamefont {R.~F.}\ \bibnamefont
  {Werner}},\ }\href@noop {} {\enquote {\bibinfo {title} {Reversible quantum
  cellular automata},}\ } (\bibinfo {year} {2004}),\ \Eprint
  {http://arxiv.org/abs/quant-ph/0405174} {arXiv:quant-ph/0405174 [quant-ph]}
  \BibitemShut {NoStop}%
\bibitem [{\citenamefont {Cirac}\ \emph {et~al.}(2017)\citenamefont {Cirac},
  \citenamefont {Perez-Garcia}, \citenamefont {Schuch},\ and\ \citenamefont
  {Verstraete}}]{Cirac2017}%
  \BibitemOpen
  \bibfield  {author} {\bibinfo {author} {\bibfnamefont {J.~I.}\ \bibnamefont
  {Cirac}}, \bibinfo {author} {\bibfnamefont {D.}~\bibnamefont {Perez-Garcia}},
  \bibinfo {author} {\bibfnamefont {N.}~\bibnamefont {Schuch}}, \ and\ \bibinfo
  {author} {\bibfnamefont {F.}~\bibnamefont {Verstraete}},\ }\href {\doibase
  10.1088/1742-5468/aa7e55} {\bibfield  {journal} {\bibinfo  {journal} {Journal
  of Statistical Mechanics: Theory and Experiment}\ }\textbf {\bibinfo {volume}
  {2017}},\ \bibinfo {pages} {083105} (\bibinfo {year} {2017})}\BibitemShut
  {NoStop}%
\bibitem [{\citenamefont {Haah}\ \emph {et~al.}(2023)\citenamefont {Haah},
  \citenamefont {Fidkowski},\ and\ \citenamefont {Hastings}}]{Haah2023}%
  \BibitemOpen
  \bibfield  {author} {\bibinfo {author} {\bibfnamefont {J.}~\bibnamefont
  {Haah}}, \bibinfo {author} {\bibfnamefont {L.}~\bibnamefont {Fidkowski}}, \
  and\ \bibinfo {author} {\bibfnamefont {M.~B.}\ \bibnamefont {Hastings}},\
  }\href {\doibase 10.1007/s00220-022-04528-1} {\bibfield  {journal} {\bibinfo
  {journal} {Communications in Mathematical Physics}\ }\textbf {\bibinfo
  {volume} {398}},\ \bibinfo {pages} {469} (\bibinfo {year}
  {2023})}\BibitemShut {NoStop}%
\bibitem [{\citenamefont {Haah}(2021)}]{Haah2021}%
  \BibitemOpen
  \bibfield  {author} {\bibinfo {author} {\bibfnamefont {J.}~\bibnamefont
  {Haah}},\ }\href {\doibase 10.1063/5.0022185} {\bibfield  {journal} {\bibinfo
   {journal} {Journal of Mathematical Physics}\ }\textbf {\bibinfo {volume}
  {62}},\ \bibinfo {pages} {092202} (\bibinfo {year} {2021})},\ \Eprint
  {http://arxiv.org/abs/https://doi.org/10.1063/5.0022185}
  {https://doi.org/10.1063/5.0022185} \BibitemShut {NoStop}%
\bibitem [{\citenamefont {Shirley}\ \emph {et~al.}(2022)\citenamefont
  {Shirley}, \citenamefont {Chen}, \citenamefont {Dua}, \citenamefont
  {Ellison}, \citenamefont {Tantivasadakarn},\ and\ \citenamefont
  {Williamson}}]{Shirley2022}%
  \BibitemOpen
  \bibfield  {author} {\bibinfo {author} {\bibfnamefont {W.}~\bibnamefont
  {Shirley}}, \bibinfo {author} {\bibfnamefont {Y.-A.}\ \bibnamefont {Chen}},
  \bibinfo {author} {\bibfnamefont {A.}~\bibnamefont {Dua}}, \bibinfo {author}
  {\bibfnamefont {T.~D.}\ \bibnamefont {Ellison}}, \bibinfo {author}
  {\bibfnamefont {N.}~\bibnamefont {Tantivasadakarn}}, \ and\ \bibinfo {author}
  {\bibfnamefont {D.~J.}\ \bibnamefont {Williamson}},\ }\href {\doibase
  10.1103/PRXQuantum.3.030326} {\bibfield  {journal} {\bibinfo  {journal} {PRX
  Quantum}\ }\textbf {\bibinfo {volume} {3}},\ \bibinfo {pages} {030326}
  (\bibinfo {year} {2022})}\BibitemShut {NoStop}%
\bibitem [{\citenamefont {Piroli}\ and\ \citenamefont
  {Cirac}(2020)}]{Piroli2020}%
  \BibitemOpen
  \bibfield  {author} {\bibinfo {author} {\bibfnamefont {L.}~\bibnamefont
  {Piroli}}\ and\ \bibinfo {author} {\bibfnamefont {J.~I.}\ \bibnamefont
  {Cirac}},\ }\href {\doibase 10.1103/PhysRevLett.125.190402} {\bibfield
  {journal} {\bibinfo  {journal} {Phys. Rev. Lett.}\ }\textbf {\bibinfo
  {volume} {125}},\ \bibinfo {pages} {190402} (\bibinfo {year}
  {2020})}\BibitemShut {NoStop}%
\bibitem [{Note2()}]{Note2}%
  \BibitemOpen
  \bibinfo {note} {For example, SQCs can generate arbitrary matrix product
  states on open boundary conditions, which can have finite, non-zero
  correlation length \cite {Perez-Garcia2006}}\BibitemShut {NoStop}%
\bibitem [{\citenamefont {Vidal}(2008)}]{Vidal2008}%
  \BibitemOpen
  \bibfield  {author} {\bibinfo {author} {\bibfnamefont {G.}~\bibnamefont
  {Vidal}},\ }\href {\doibase 10.1103/PhysRevLett.101.110501} {\bibfield
  {journal} {\bibinfo  {journal} {Phys. Rev. Lett.}\ }\textbf {\bibinfo
  {volume} {101}},\ \bibinfo {pages} {110501} (\bibinfo {year}
  {2008})}\BibitemShut {NoStop}%
\bibitem [{\citenamefont {Gu}\ and\ \citenamefont {Wen}(2009)}]{Gu2009}%
  \BibitemOpen
  \bibfield  {author} {\bibinfo {author} {\bibfnamefont {Z.-C.}\ \bibnamefont
  {Gu}}\ and\ \bibinfo {author} {\bibfnamefont {X.-G.}\ \bibnamefont {Wen}},\
  }\href {\doibase 10.1103/PhysRevB.80.155131} {\bibfield  {journal} {\bibinfo
  {journal} {Phys. Rev. B}\ }\textbf {\bibinfo {volume} {80}},\ \bibinfo
  {pages} {155131} (\bibinfo {year} {2009})}\BibitemShut {NoStop}%
\bibitem [{\citenamefont {Swingle}\ and\ \citenamefont
  {McGreevy}(2016)}]{Swingle2016}%
  \BibitemOpen
  \bibfield  {author} {\bibinfo {author} {\bibfnamefont {B.}~\bibnamefont
  {Swingle}}\ and\ \bibinfo {author} {\bibfnamefont {J.}~\bibnamefont
  {McGreevy}},\ }\href {\doibase 10.1103/PhysRevB.93.045127} {\bibfield
  {journal} {\bibinfo  {journal} {Phys. Rev. B}\ }\textbf {\bibinfo {volume}
  {93}},\ \bibinfo {pages} {045127} (\bibinfo {year} {2016})}\BibitemShut
  {NoStop}%
\bibitem [{\citenamefont {Evenbly}\ and\ \citenamefont
  {White}(2016)}]{Evenbly2016}%
  \BibitemOpen
  \bibfield  {author} {\bibinfo {author} {\bibfnamefont {G.}~\bibnamefont
  {Evenbly}}\ and\ \bibinfo {author} {\bibfnamefont {S.~R.}\ \bibnamefont
  {White}},\ }\href {\doibase 10.1103/PhysRevLett.116.140403} {\bibfield
  {journal} {\bibinfo  {journal} {Phys. Rev. Lett.}\ }\textbf {\bibinfo
  {volume} {116}},\ \bibinfo {pages} {140403} (\bibinfo {year}
  {2016})}\BibitemShut {NoStop}%
\bibitem [{\citenamefont {Perez-Garcia}\ \emph {et~al.}(2007)\citenamefont
  {Perez-Garcia}, \citenamefont {Verstraete}, \citenamefont {Wolf},\ and\
  \citenamefont {Cirac}}]{Perez-Garcia2006}%
  \BibitemOpen
  \bibfield  {author} {\bibinfo {author} {\bibfnamefont {D.}~\bibnamefont
  {Perez-Garcia}}, \bibinfo {author} {\bibfnamefont {F.}~\bibnamefont
  {Verstraete}}, \bibinfo {author} {\bibfnamefont {M.~M.}\ \bibnamefont
  {Wolf}}, \ and\ \bibinfo {author} {\bibfnamefont {J.~I.}\ \bibnamefont
  {Cirac}},\ }\href {http://dl.acm.org/citation.cfm?id=2011832.2011833}
  {\bibfield  {journal} {\bibinfo  {journal} {Quantum Info. Comput.}\ }\textbf
  {\bibinfo {volume} {7}},\ \bibinfo {pages} {401} (\bibinfo {year}
  {2007})}\BibitemShut {NoStop}%
\bibitem [{\citenamefont {Schaumann}(2013)}]{schaumann2013traces}%
  \BibitemOpen
  \bibfield  {author} {\bibinfo {author} {\bibfnamefont {G.}~\bibnamefont
  {Schaumann}},\ }\href@noop {} {\bibfield  {journal} {\bibinfo  {journal}
  {Journal of Algebra}\ }\textbf {\bibinfo {volume} {379}},\ \bibinfo {pages}
  {382} (\bibinfo {year} {2013})}\BibitemShut {NoStop}%
\bibitem [{\citenamefont {Hahn}\ and\ \citenamefont
  {Wolf}(2020)}]{hahn2020generalized}%
  \BibitemOpen
  \bibfield  {author} {\bibinfo {author} {\bibfnamefont {A.}~\bibnamefont
  {Hahn}}\ and\ \bibinfo {author} {\bibfnamefont {R.}~\bibnamefont {Wolf}},\
  }\href@noop {} {\bibfield  {journal} {\bibinfo  {journal} {Physical Review
  B}\ }\textbf {\bibinfo {volume} {102}},\ \bibinfo {pages} {115154} (\bibinfo
  {year} {2020})}\BibitemShut {NoStop}%
\end{thebibliography}%

\appendix

\section{Map to symmetry breaking phases of general finite groups}
\label{ap:SBnonabelian}

In this section, we show how the construction in section~\ref{sec:SB} can be generalized from the $Z_2$ group to arbitrary finite groups. 
We start by generalizing the Hilbert space from the $Z_2$ case $\{|0\rangle,|1\rangle\}$ to a finite group $G$ $\{|z\rangle:z\in G\}$ and replacing the $X$ and $(1\pm Z)/2$ operators with 
\begin{equation}\label{eq:genG}
\begin{aligned}
X^g|z\rangle &= |gz\rangle,\quad &g,z\in G \\
T^h |z\rangle &=  \delta_{h,z}|z\rangle,&h,z \in G, 
\end{aligned}
\end{equation}
where $(X^g)^\dagger = X^{g^{-1}}$ is unitary and $T^h$ is a projection onto the state $|h\rangle$. These operators satisfy the commutation relation $X^gT^h=T^{gh}X^g$. 
Using these new operators, we can rewrite the $Z_2$ invariant transverse field Ising model \eqref{eq:IsingHam} as 
\begin{equation}\label{eq:genG_Ham}\
H = -J\sum_{i}\sum_{h\in G}T_i^hT_{i+1}^h - B\frac{1}{|G|}\sum_{i}\sum_{g\in G} X_i^g, 
\end{equation}
which is now invariant under a global symmetry $G$, i.e. 
$(\prod_i X_i^g) H(\prod_i X_i^g)^{\dagger}=H$. 

Similar to the transverse field Ising model, the Hamiltonian above has a symmetric phase ($B\gg J>0$) and  a symmetry breaking phase ($J\gg B>0$). The corresponding fixed-point wave functions are given by 
\begin{equation}
\begin{aligned}
    |\psi_{\mathrm{SY}}\rangle &= |++\cdots+\rangle \\
    |\psi_{\mathrm{SB}}\rangle &= \frac{1}{\sqrt{|G|}}\sum_g|gg\cdots g\rangle,
\end{aligned}
\end{equation}
where $|+\rangle = \frac{1}{\sqrt{|G|}}\sum_g|g\rangle$. 
To construct a quantum circuit that maps $|\psi_{\mathrm{SY}}\rangle$  to $|\psi_{\mathrm{SB}}\rangle$, we define 
\begin{equation}
    R_{ij}(\mathcal{W}) \equiv \sum_{h\in G}T_{i}^hX^h_{j}\mathcal{W}_j X^{h^{-1}}_{j}, 
\end{equation}
where $\mathcal{W}$ is a unitary operator so that $R_{ij}(\mathcal{W})$ is unitary in each subspace projected by $T_i^h$ and, consequently, unitary in the full Hilbert space. One can also check that $R_{ij}(\mathcal{W})$ is invariant under the symmetry operation 
\begin{equation}
\begin{aligned}
    (\prod_kX^g_k)R_{ij}(\mathcal{W})(\prod_kX^g_k)^\dagger
    =&\sum_{h\in G} T_i^{gh}X_{j}^{gh}\mathcal{W}_{j}X_{j}^{(gh)^{-1}}\\=&R_{ij}(\mathcal{W}).
\end{aligned}
\end{equation}
Any circuit made up of $R_{ij}(\mathcal{W})$ gates is hence symmetric. In analogy to Eqs.~\eqref{eq:sy2sb} and \eqref{eq:U1D}, consider a circuit of the following form 
\begin{equation}\label{eq:sy2sb_G}
    \mathcal{U} = R_{N1}(S) \prod_{i=N}^1 U_{i,i+1},
\end{equation}
where
\begin{equation}\label{eq:sy2sb_Gij}
    U_{i,i+1}= R_{i,i+1}(S^\dagger\text{Had}S^\dagger ) R_{i,i+1}(S). 
\end{equation} 
The matrices $R_{i,i+1}(S^\dagger\text{Had}S^\dagger)$ and $R_{i,i+1}(S)$ in \eqref{eq:sy2sb_Gij} are chosen such that they will reduce to $R(X_{i+1})$ and $R(Z_i,Z_{i+1})$ in \eqref{eq:U1D} for the $G=Z_2$ case, respectively. 
To map $|\psi_{\mathrm{SY}}\rangle$ to $|\psi_{\mathrm{SB}}\rangle$, we need to find unitaries $\text{Had}$ and $S$ that map 
\begin{equation}\label{eq:sy2sbHS}
    \text{Had}|+\rangle = |e\rangle, \quad
    S^\dagger |z\rangle = e^{-i\theta(z)}|z\rangle, 
\end{equation}
such that 
\begin{equation}
    \begin{aligned}
        U_{i,i+1}|+,+\rangle = &
        \frac{1}{\sqrt{|G|}}\sum_{h}X_{i+1}^h S^\dagger_{i+1}\text{Had}_{i+1}|h,+\rangle\\
        = & \frac{e^{-i\theta(e)}}{\sqrt{|G|}} \sum_h X^h_{i+1}|h,e\rangle \\
        = &\frac{e^{-i\theta(e)}}{\sqrt{|G|}}\sum_h|h,h\rangle. 
    \end{aligned}
\end{equation}

According to the mapping \eqref{eq:sy2sbHS}, the first row of $\text{Had}$ is restricted to be $\frac{1}{\sqrt{|G|}}(1,\cdots,1)$, and $S$ must be diagonal. In addition to these two constraints, the $\text{Had}$ and $S$ matrices can be selected freely as long as $\text{Had}$ is unitary and $S$ is diagonal.
A particular choice of $\text{Had}$ can be taken as 
\begin{equation}
\label{eq:genHad}
    \text{Had} = \sum_{p,\mu,\nu}\sum_z\sqrt{\frac{\rho_{p}}{|G|}}A^{p}_{\mu\nu}(z)|p,\mu,\nu\rangle\langle z|.
\end{equation}
Here $p$ labels inequivalent irreducible representations of the group $G$, $\mu$ and $\nu$ are the row and column indices of the representing matrix $A^p$, and $\rho_p$ is the dimension of the representation space. Had is a mapping from group elements to entries in the irreducible representation matrices and in particular it should map the identity element to the trivial representation of $G$ in order to ensure the mapping \eqref{eq:sy2sbHS}. 

We can also check the mappings of operators under the circuit \eqref{eq:sy2sb_G} and discuss whether these mappings hold for a general group $G$. First, the symmetry operator remains invariant under the circuit:
\begin{equation}\label{eq:genG_opesym}
    \prod _{i=1}^N X_i^g \rightarrow \prod _{i=1}^N X_i^g.  
\end{equation}
as our unitary circuit converses the symmetry. 
And the transverse magnetic field term is always mapped to the ferromagnetic term: 
\begin{equation}\label{eq:genG_opeX}
    \frac{1}{|G|}\sum_{g\in G}X^g_{i} \rightarrow \sum_h T^h_{i-1}T^h_{i}, \quad i=2,\cdots,N, 
\end{equation}
which can be proved by the following steps 
\begin{equation}
\begin{aligned}
    &U_{i-1,i}\left(\frac{1}{|G|}\sum_{g\in G}X^g_{i}\right)U^\dagger_{i-1,i} \\
    =&\sum_hT^h_{i-1} X_{i}^h S^\dagger_{i}\text{Had}_{i} |+\rangle\langle +| \text{Had}^\dagger_{i}S_{i}X_{i}^{h^{-1}} \\
    = &\sum_h T^h_{i-1}T^h_{i},\\
\end{aligned}
\end{equation}
and 
\begin{equation}
    U_{i,i+1}\left(\sum_h T^h_{i-1}T^h_{i}\right)U^\dagger_{i,i+1} = \sum_h T^h_{i-1}T^h_{i}.
\end{equation}

Unfortunately, the other mappings in the $Z_2$ case in \eqref{eq:sy2sb_ope} do not hold for a general finite group $G$. For instance, the ferromagnetic term $\sum_h T^h_{i}T^h_{i+1}$ generally does not map to the transverse field term $\frac{1}{|G|}\sum_{g\in G}X^g_{i+1}$. 

Things become easier in the case where $G$ is abelian, as the Had matrix in \eqref{eq:genHad} reduces to conventional Fourier transformations of the abelian group $G$. Without loss of generality, we consider the example of a cyclic group $Z_M$, where we can choose 
\begin{equation}\label{eq:sy2sb_abe}
\begin{aligned}
    \text{Had} =& \sum_{\alpha,z=0}^{M-1}\frac{1}{\sqrt{M}}\exp\left(\frac{2\pi i}{M}\alpha z\right)|\alpha\rangle\langle z|,\\
    S =& \sum_{z=0}^{M-1}\exp\left(\frac{\pi i}{M}z^2\right)|z\rangle\langle z|.\\
\end{aligned}
\end{equation}
In the simplest case where $G=Z_2$, the Had and $S$ matrices will reduce to the conventional Hadamard gate and phase gate, respectively. It can also be verified that $R_{ij}(S) = \exp(-i\frac{\pi}{4}(Z_iZ_j-1))$ and $R_{ij}(S^\dagger \text{Had}S^\dagger)=\exp(-i\frac{\pi}{4}X_j)$, which implies that the generalized circuit \eqref{eq:sy2sb_G} will reduce to \eqref{eq:sy2sb} in the $Z_2$ case up to a global phase. 
 
For the abelian groups, all the operator mappings for the $Z_2$ case are preserved as long as we choose the gate as in \eqref{eq:sy2sb_abe}. The mappings in \eqref{eq:sy2sb_ope} will become 
\begin{equation}
\begin{aligned}
\frac{1}{|G|}\sum_gX^g_{i} &\rightarrow \sum_{h}T_{i-1}^hT_i^{h}, \quad i = 2,\cdots, N,\\
\frac{1}{|G|}\sum_gX^g_1 &\rightarrow \sum_g \frac{1}{|G|}\sum_{h,h'}e^{-\frac{2\pi i}{|G|}(h'-h)g}T_1^hT_N^{h'} \prod_{i=1}^NX_i^g, \\
\sum_{h}T^h_iT^h_{i+1} & \rightarrow \frac{1}{|G|}\sum_g X^g_i, \quad i = 2,\cdots,N\\
\sum_{h}T^h_1T^h_{2} & \rightarrow \frac{1}{|G|}\sum_g X^g_1\prod_{i=1}^NX_i^g. 
\end{aligned}
\end{equation}
In the bulk of the system, the transverse field term
maps to the Ising term and vice versa. Near the left endpoint, the correspondence is preserved only under the symmetric subspace $\prod_{i=1}^NX_i^g=1$.

Generalization to higher dimensional versions shows no difference with the $Z_2$ case, as shown in Fig.~\ref{fig:SBcircuit2d}.

\section{Map to string-net states with general gapped boundaries and the torus}
\label{ap:SNstates}

This section contains some of the necessary notions from category theory to define the general sequential circuit in section \ref{sec:string-nets}.\\

Following Kitaev and Kong, gapped boundaries of string-net models are classified by $\mathcal{C}$-module categories $\mathcal{M}$, where $\mathcal{C}$ is the (unitary) fusion category of the string-net model. The boundary can be interpreted as a gapped domain wall between the string-net $\mathcal{C}$ and the vacuum, given by the category of vector spaces $\text{Vec}$, in which case $\mathcal{M}$ can be viewed as a $(\mathcal{C}, \text{Vec})$-bimodule category. At the same time, given $\mathcal{M}$ and $\mathcal{C}$, we can always find the (unique) Morita dual category of $\mathcal{C}$, $\mathcal{D} = \mathcal{C}_{\mathcal{M}}^{*}$, in which case $\mathcal{M}$ is an invertible $(\mathcal{C},\mathcal{D})$-bimodule category. This is the structure we will require to define our quantum circuit. We refer the reader to Ref.~\onlinecite{lootens2021matrix, etingof2016tensor} for more details about the underlying mathematics and will present only some key ingredients here in order to define the sequential circuit. Given the resemblance of the operators needed to define our sequential circuit and the (finite-depth) circuit from Ref~\onlinecite{lootens2022mapping}, we follow the same conventions here. \\

The Hilbert space of string-net models consists of configurations on a hexagonal lattice, with the edges labeled by simple objects $a,b,...$ of a unitary fusion category $\mathcal{C}$. The simple objects obey fusion constraints: $a \times b = \sum_c N_{ab}^c \ c$, which are assumed to be multiplicity free ($N_{ab}^c \in {0,1}$) from now on. There exists a unit object $\mathbf{1}$, such that $\forall \ a, \mathbf{1}\times a = a$. Each simple object has a quantum dimension $d_a$ associated to it. We can diagrammatically write down the resolution of the identity:
\begin{align} \label{eq:resId}
\includegraphics[valign=c,page=13]{figuresSN}.
\end{align}
The fusion category has an associator or  $\F{\mathcal{C}}$-symbol -we use the $\mathcal{C}$-superscript to distinguish the symbol from other associators that will be defined below- expressed in terms of the simple objects as: 
\begin{equation} \label{eq:Fsymbol}
\includegraphics[valign=c,page=1]{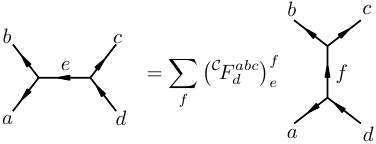}.
\end{equation}
This $\F{\mathcal{C}}$-symbol is simply the usual $F$-symbol for fusion categories and obeys the well-known pentagon equation. We also require the ``bubble pop" identity:
\begin{align}
\includegraphics[valign=c,page=50]{figuresSN}.
\end{align}
A (right) $\mathcal{C}$-module category $\mathcal{M}$, has simple objects that will be denoted by capital Roman letters $(A,B,C,...)$ and a right-action of $\mathcal{C}$ on $\mathcal{M}$, $\triangleleft: \mathcal{M} \times \mathcal{C} \rightarrow \mathcal{M}$. The action is strict: $A \triangleleft \mathbf{1} = A$ and generally we can write $A \triangleleft a = \sum_{B\in\mathcal{M}} N_{Aa}^B\, B$. Note that these new $N_{Aa}^B$ are distinct from the original fusion rules of $\mathcal{C}$ ($N_{ab}^c$) and may carry multiplicities ($N_{Aa}^B > 1$) even if the $N_{ab}^c$ don't. $\mathcal{M}$ has an associator $\F{\triangleleft}$ that implements the following recoupling at the boundary of the string-net:
\begin{equation} \label{eq:Fright}
\includegraphics[valign=c,page=27]{figuresSN},
\end{equation}
where the boundary is now blue, labeled by objects in the module category $\mathcal{M}$. 
A suitable gauge choice can be made such that:
\begin{equation}
\left({}^{\triangleleft}\!F^{A\mathbf{1}b}_B \right)^{B,m}_{b,1k} = \left({}^{\triangleleft}\!F^{Aa\mathbf{1}}_B \right)^{A,m}_{a,k1} = \delta_k^m.
\end{equation}
The inverse symbol is denoted by $\Fi{\triangleleft}$. \\

We now have a second unitary fusion category $\mathcal{D}$ with simple objects labeled by Greek letters ($\alpha, \beta, \delta,...$). In our case, this category will be chosen to be the unique Morita dual of $\mathcal{C}$, given $\mathcal{M}$: $\mathcal{D} = \mathcal{C}_{\mathcal{M}}^{*}$. The $F$-move for $\mathcal{D}$ is defined in the following way:
\begin{equation} \label{eq:Fsymbol2}
\includegraphics[valign=c,page=47]{figuresSN}.
\end{equation}
Just like for $\mathcal{M}$, it is not guaranteed that $\mathcal{D}$ is multiplicity-free, even if $\mathcal{C}$ is. \\

We will now take $\mathcal{M}$ to be a left $\mathcal{D}$-module category, with an associator $\F{\triangleright}$. The bulk string-net $\mathcal{D}$ is now on the left of the boundary:
\begin{equation} \label{eq:Fleft}
\includegraphics[valign=c,page=14]{figuresSN}.
\end{equation}
A similar gauge choice can be made such that:
\begin{equation}
\left({}^{\triangleright}\!F^{\mathbf{1}\beta A}_B \right)^{B,m1}_{\beta ,1k} = \left({}^{\triangleright}\!F^{\alpha\mathbf{1}A}_B \right)^{A,1m}_{a,1k} = \delta_k^m.
\end{equation}
The inverse symbol is denoted by $\Fi{\triangleright}$. For $\mathcal{M}$ to be a $(\mathcal{D},\mathcal{C})$-bimodule category we require one last additional associator $\F{\Join}$:
\begin{equation} \label{eq:Fmiddle}
\includegraphics[valign=c,page=28]{figuresSN}.
\end{equation}
with a similar gauge choice leading to:
\begin{equation}
\left({}^{\Join}\!F^{\mathbf{1}Ab}_B \right)^{b,m1}_{A,1k} = \left({}^{\Join}\!F^{\alpha A\mathbf{1}}_B \right)^{A,1m}_{B,k1} = \delta_k^m.
\end{equation}
The inverse symbol is denoted by $\Fi{\Join}$. There are in total 6 coupled pentagon equations that the set of $F$-symbols should satisfy. These are given in Ref.~\onlinecite{lootens2021matrix}. \\

One final ingredient we need is a resolution of the identity that allows for fusion of two $\mathcal{M}$ lines in terms of a $\mathcal{D}$- and a $\mathcal{C}$ line. Although $\mathcal{M}$ has no intrinsic duality, we can define the `opposite' bimodule category $\overline{\mathcal{M}}$, a $(\mathcal{C},\mathcal{D})$-bimodule category, which contains the dual objects of $\mathcal{M}$. The rigorous mathematical picture behind this is called a Morita context \cite{kitaev2012models, schaumann2013traces}. This allows us to define:
\begin{align} \label{eq:resIdBlue}
\includegraphics[valign=c,page=48]{figuresSN},
\end{align}
and the opposite resolution in terms of a $\mathcal{D}$ line:
\begin{align} \label{eq:resIdBlue2}
\includegraphics[valign=c,page=15]{figuresSN},
\end{align}
where $d_A,d_B$ are quantum dimensions for the objects in $\mathcal{M}$. Note that the invertibility of the bimodule $\mathcal{M}$ guarantees \ref{eq:resIdBlue} and \ref{eq:resIdBlue2}. \\

We are now ready to write down the precise action of the general sequential circuit in figure \ref{generalString-netCircuit} (cylinder case) and \ref{torusCircuit} (torus case) in terms of the individual plaquette actions for the 7 sublattices $L_1,...,L_7$. The original `smooth' boundary case in figure \ref{string-netCircuit} is recovered by choosing $\mathcal{M} = \mathcal{C}$. In this case, the Morita dual is simply $\mathcal{C}$ itself ($\mathcal{D} = \mathcal{C}$) and all $F$-symbols defined above reduce to $\F{\mathcal{C}}$ (up to some suitable permutations of the labels). For simplicity, we will only treat the general case. We will follow the orientation convention as in Ref.~\onlinecite{lootens2022mapping}. The control edges in the sequential circuit are colored red in Eqs. \ref{eq:Sloop}-\ref{eq:L7}. In the full sequential circuit, the control edge has not been acted on by previous operations and is still labeled by the trivial object in $\mathcal{D}$. For the purpose of generality, we assume it to be arbitrary in Eqs. \ref{eq:Sloop}-\ref{eq:L6}. \\

\onecolumngrid

\begin{align} 
\begin{split} \label{eq:Sloop}
\includegraphics[valign=c,page=29,scale=1]{figuresSN} &= \sum_{\{\tilde{E}_i,h_i\}} [{}^{1}\!B_p^{S,\mathcal{M}}]_{\{\epsilon_i,k_i\}}^{\{\tilde{E}_i,h_i\}} \quad \includegraphics[valign=c,page=30,scale=1]{figuresSN},
\\
[{}^{1}\!B_p^{S,\mathcal{M}}]_{\{\epsilon_i,k_i\}}^{\{\tilde{E}_i,h_i\}} = &\sum_{\{n_i\}} \sqrt{\frac{d_{\epsilon_6}d_{\epsilon_5}d_{\epsilon_3}d_{\epsilon_2}d_{\tilde{E}_4}d_{\tilde{E}_1}}{d_{\alpha_5}d_{\alpha_2}d_{\epsilon_4}d_{\epsilon_1}d_{S}^2}}\\
\left( \Fi{\triangleright}^{\alpha_1\epsilon_2S}_{\tilde{E}_1}\right)^{\tilde{E}_2,j_2h_1}_{\epsilon_1,k_1 n_1}
\ &\left( \F{\triangleright}^{\epsilon_2\epsilon_3\tilde{E}_3}_{\tilde{E}_2}\right)^{S,n_3n_2}_{\alpha_2,k_2h_2}
\ \left( \Fi{\triangleright}^{\epsilon_3\alpha_3\tilde{E}_4}_{S}\right)^{\tilde{E}_3,h_3n_3}_{\epsilon_4,k_3n_4} \\
\left( \F{\triangleright}^{\epsilon_5\alpha_4\tilde{E}_4}_{S}\right)^{\tilde{E}_5,h_4n_5}_{\epsilon_4,k_4n_4}
\ &\left( \Fi{\triangleright}^{\epsilon_6\epsilon_5\tilde{E}_5}_{\tilde{E}_6}\right)^{S,n_6n_5}_{\alpha_5,k_5h_5}
\ \left( \F{\triangleright}^{\alpha_6\epsilon_6S}_{\tilde{E}_1}\right)^{\tilde{E}_6,n_6h_6}_{\epsilon_1,k_6n_1}.
\end{split}
\end{align}
The case for the original plaquette operator $B_p^s$ (\ref{eq:Bp}) is recovered by choosing $\mathcal{M} = \mathcal{C} = \mathcal{D}$ in \ref{eq:Sloop}. 
The matrix $[B_p^s]$ we recover in this case is different than the one in the original Levin-Wen string-net paper, because of the different orientation convention and the fact that no tetrahedral symmetry of the $\F{\mathcal{C}}$-symbols is assumed \cite{hahn2020generalized}.
\begin{align} 
\begin{split} \label{eq:L2}
\includegraphics[valign=c,page=33,scale=1]{figuresSN} &= \sum_{\{\tilde{e}_i\tilde{E}_i,h_i\}} [{}^{2}\!B_p^{S,\mathcal{M}}]_{\{\epsilon_i,E_i,k_i\}}^{\{\tilde{e}_i,\tilde{E}_i,h_i\}} \quad \includegraphics[valign=c,page=34,scale=1]{figuresSN},
\\
[{}^{2}\!B_p^{S,\mathcal{M}}]_{\{\epsilon_i,E_i,k_i\}}^{\{\tilde{e}_i,\tilde{E}_i,h_i\}} = &\sum_{\{n_i\}} \sqrt{\frac{d_{\epsilon_2}d_{\epsilon_3}d_{\epsilon_4}d_{\tilde{E}_1}d_{\tilde{E}_3}d_{\tilde{e}_1}d_{\tilde{e}_2}}{d_{\epsilon_1}d_{A_2}d_{A_3}d_{\tilde{E}_2}d_{\tilde{E}_4}d_S^2}}\\
\left( \Fi{\triangleright}^{\alpha_1\epsilon_2S}_{\tilde{E}_1}\right)^{\tilde{E}_2,n_2h_1}_{\epsilon_1,k_1n_1}
\ &\left( \Fi{\Join}^{\epsilon_2S\tilde{e}_1}_{A_1}\right)^{E_1,n_3k_2}_{\tilde{E}_2,n_2h_2}
\ \left( \Fi{\Join}^{\epsilon_3\tilde{E}_3\tilde{e}_1}_{E_1}\right)^{A_2,h_3k_3}_{S,n_4n_3} \\
\left( \F{\Join}^{\epsilon_3\tilde{E}_3\tilde{e}_2}_{E_2}\right)^{A_3,h_4k_4}_{S,n_4n_5}
\ &\left( \F{\Join}^{\epsilon_4S\tilde{e}_2}_{A_4}\right)^{E_2,n_5k_5}_{\tilde{E}_4,n_6h_5}
\ \left( \F{\triangleright}^{\alpha_2\epsilon_4S}_{\tilde{E}_1}\right)^{\tilde{E}_4,n_6h_6}_{e_1,k_6n_1}.
\end{split}
\end{align}
\begin{align} 
\begin{split} \label{eq:L3}
\includegraphics[valign=c,page=35,scale=1]{figuresSN} &= \sum_{\{\tilde{e}_i\tilde{E}_i,h_i\}} [{}^{3}\!B_p^{S,\mathcal{M}}]_{\{\epsilon_i,E_i,k_i\}}^{\{\tilde{e}_i,\tilde{E}_i,h_i\}} \quad \includegraphics[valign=c,page=36,scale=1]{figuresSN},
\\
[{}^{3}\!B_p^{S,\mathcal{M}}]_{\{\epsilon_i,E_i,k_i\}}^{\{\tilde{e}_i,\tilde{E}_i,h_i\}} = &\sum_{\{n_i\}} \sqrt{\frac{d_{\epsilon_2}d_{\epsilon_3}d_{A_2}d_{\tilde{E}_1}d_{\tilde{e}_1}d_{\tilde{e}_3}}{d_{\epsilon_1}d_{E_3}d_{\alpha_2}d_{\tilde{E}_2}d_{\tilde{E}_3}d_{\tilde{e}_2}}}\\
\left( \Fi{\triangleright}^{\alpha_1\epsilon_2S}_{\tilde{E}_1}\right)^{\tilde{E}_2,n_2h_1}_{\epsilon_1,k_1n_1}
\ &\left( \Fi{\Join}^{\epsilon_2S\tilde{e}_1}_{A_1}\right)^{E_1,n_3k_2}_{\tilde{E}_2,n_2h_2}
\ \left( \Fi{\triangleleft}^{S\tilde{e}_1a_2}_{E_2}\right)^{\tilde{e}_2,n_4}_{E_1,n_3k_3} \\
\left( \F{\triangleleft}^{S\tilde{e}_3a_3}_{E_2}\right)^{\tilde{e}_2,n_4}_{E_3,n_5k_4}
\ &\left( \F{\Join}^{\epsilon_3S\tilde{e}_3}_{A_2}\right)^{E_3,n_5k_5}_{\tilde{E}_3,n_6h_3}
\ \left( \F{\triangleright}^{\alpha_4\epsilon_3S}_{E_1}\right)^{\tilde{E}_3,n_6h_4}_{e_1,k_6n_1}.
\end{split}
\end{align}
\begin{align} 
\begin{split} \label{eq:L4}
\includegraphics[valign=c,page=37,scale=1]{figuresSN} &= \sum_{\{\tilde{e}_i\tilde{E}_i,h_i\}} [{}^{4}\!B_p^{S,\mathcal{M}}]_{\{\epsilon_i,E_i,k_i\}}^{\{\tilde{e}_i,\tilde{E}_i,h_i\}} \quad \includegraphics[valign=c,page=38,scale=1]{figuresSN},
\\
[{}^{4}\!B_p^{S,\mathcal{M}}]_{\{\epsilon_i,E_i,k_i\}}^{\{\tilde{e}_i,\tilde{E}_i,h_i\}} = &\sum_{\{n_i\}} \sqrt{\frac{d_{\epsilon_1}d_{E_3}d_{E_1}d_{\tilde{e}_1}d_{\tilde{e}_2}d_{\tilde{e}_4}d_{\tilde{E}_2}}{d_{A_1}d_{E_4}d_{A_4}d_{\tilde{E}_2}d_{\tilde{e}_3}d_S^2}}\\
\left( \Fi{\Join}^{\epsilon_1E_1\tilde{e}_1}_{\tilde{E}_1}\right)^{S,n_1n_2}_{A_1,k_1h_1}
\ &\left( \Fi{\Join}^{\epsilon_1S\tilde{e}_2}_{A_2}\right)^{E_2,n_3k_2}_{\tilde{E}_1,n_2h_2}
\ \left( \Fi{\triangleleft}^{S\tilde{e}_2a_1}_{E_3}\right)^{\tilde{e}_3,n_4}_{E_2,n_3k_3} \\
\left( \F{\triangleleft}^{S\tilde{e}_4a_2}_{E_3}\right)^{\tilde{e}_3,n_4}_{E_4,n_5k_4}
\ &\left( \F{\Join}^{\epsilon_2S\tilde{e}_4}_{A_3}\right)^{E_4,n_5k_5}_{\tilde{E}_2,n_6h_3}
\ \left( \F{\Join}^{\epsilon_2E_1\tilde{e}_1}_{\tilde{E}_2}\right)^{S,n_1n_6}_{A_4,k_6h_4}.
\end{split}
\end{align}
\begin{align} 
\begin{split} \label{eq:L5}
\includegraphics[valign=c,page=39,scale=1]{figuresSN} &= \sum_{\{\tilde{e}_i\tilde{E}_i,h_i\}} [{}^{5}\!B_p^{S,\mathcal{M}}]_{\{\epsilon_i,E_i,k_i\}}^{\{\tilde{e}_i,\tilde{E}_i,h_i\}} \quad \includegraphics[valign=c,page=40,scale=1]{figuresSN},
\\
[{}^{5}\!B_p^{S,\mathcal{M}}]_{\{\epsilon_i,E_i,k_i\}}^{\{\tilde{e}_i,\tilde{E}_i,h_i\}} = &\sum_{\{n_i\}} \sqrt{\frac{d_{\epsilon_1}d_{E_1}d_{E_3}d_{\tilde{e}_2}d_{\tilde{e}_4}d_{\tilde{e}_5}}{d_{A_2}d_{E_2}d_{a_3}d_{\tilde{e}_1}d_S^2}}\\
\left( \Fi{\triangleleft}^{E_1a_1\tilde{e}_2}_{S}\right)^{\tilde{e}_1,n_1}_{E_2,k_1n_2}
\ &\left( \F{\Join}^{\epsilon_1A_1\tilde{e}_2}_{S}\right)^{E_1,h_1n_3}_{E_2,k_2n_2}
\ \left( \F{\Join}^{\epsilon_1\tilde{E}_1\tilde{e}_3}_{E_3}\right)^{A_2,h_2k_3}_{S,n_3n_4} \\
\left( \F{\triangleleft}^{S\tilde{e}_4a_2}_{E_3}\right)^{\tilde{e}_3,n_4}_{E_4,n_5k_4}
\ &\left( \Fi{\triangleleft}^{E_5\tilde{e}_5\tilde{e}_4}_{E_4}\right)^{a_3,k_5}_{S,n_6n_5}
\ \left( \F{\triangleleft}^{E_1a_4\tilde{e}_5}_{S}\right)^{\tilde{e}_1,n_1}_{E_5,k_6n_6}.
\end{split}
\end{align}
\begin{align} 
\begin{split} \label{eq:L6}
\includegraphics[valign=c,page=41,scale=1]{figuresSN} &= \sum_{\{\tilde{e}_i\tilde{E}_i,h_i\}} [{}^{6}\!B_p^{S,\mathcal{M}}]_{\{\epsilon_i,E_i,k_i\}}^{\{\tilde{e}_i,\tilde{E}_i,h_i\}} \quad \includegraphics[valign=c,page=42,scale=1]{figuresSN},
\\
[{}^{6}\!B_p^{S,\mathcal{M}}]_{\{\epsilon_i,E_i,k_i\}}^{\{\tilde{e}_i,\tilde{E}_i,h_i\}} = &\sum_{\{n_i\}} \sqrt{\frac{d_{\epsilon_1}d_{E_3}d_{E_1}d_{\tilde{e}_2}d_{\tilde{e}_4}d_{\tilde{e}_5}}{d_{A_1}d_{E_2}d_{a_3}d_{\tilde{e}_3}d_S^2}}\\
\left( \Fi{\Join}^{\epsilon_1E_1\tilde{e}_1}_{\tilde{E}_1}\right)^{S,n_1n_2}_{A_1,k_1h_2}
\ &\left( \Fi{\Join}^{\epsilon_1S\tilde{e}_2}_{A_2}\right)^{E_2,n_3k_2}_{\tilde{E}_1,n_2h_2}
\ \left( \Fi{\triangleleft}^{S\tilde{e}_2a_1}_{E_3}\right)^{\tilde{e}_3,n_4}_{E_2,n_3k_3} \\
\left( \F{\triangleleft}^{S\tilde{e}_4a_2}_{E_3}\right)^{\tilde{e}_3,n_4}_{E_4,n_5k_4}
\ &\left( \Fi{\triangleleft}^{E_5\tilde{e}_5\tilde{e}_4}_{E_4}\right)^{a_3,k_5}_{S,n_6n_5}
\ \left( \F{\triangleleft}^{E_1a_4\tilde{e}_5}_{S}\right)^{\tilde{e}_1,n_1}_{E_5,k_6n_5}.
\end{split}
\end{align}
\begin{align} 
\begin{split} \label{eq:L7}
\includegraphics[valign=c,page=43,scale=1]{figuresSN} &= \sum_{\{\tilde{e}_i\}} [{}^{7}\!B_p^{S,\mathcal{M}}]_{\{E_i,k_i\}}^{\{\tilde{e}_i\}} \quad \includegraphics[valign=c,page=44,scale=1]{figuresSN},
\\
[{}^{7}\!B_p^{S,\mathcal{M}}]_{\{E_i,k_i\}}^{\{\tilde{e}_i\}} = &\sum_{\{n_i\}} \sqrt{\frac{d_{E_1}d_{E_4}d_{\tilde{e}_2}d_{\tilde{e}_3}d_{\tilde{e}_5}d_{\tilde{e}_6}}{d_{a_2}d_{a_5}d_{\tilde{e}_1}d_{\tilde{e}_4}d_S^2}}\\
\left( \Fi{\triangleleft}^{E_1a_1\tilde{e}_2}_{S}\right)^{\tilde{e}_1,n_1}_{E_2,k_1n_2}
\ &\left( \F{\triangleleft}^{E_2\tilde{e}_2\tilde{e}_3}_{E_3}\right)^{a_2,k_2}_{S,n_2n_3}
\ \left( \Fi{\triangleleft}^{S\tilde{e}_3a_3}_{E_4}\right)^{\tilde{e}_4,n_4}_{E_3,n_3k_3} \\
\left( \F{\triangleleft}^{S\tilde{e}_5a_4}_{E_4}\right)^{\tilde{e}_4,n_4}_{E_5,n_5k_4}
\ &\left( \Fi{\triangleleft}^{E_6\tilde{e}_6\tilde{e}_5}_{E_5}\right)^{a_5,k_5}_{S,n_6n_5}
\ \left( \F{\triangleleft}^{E_1a_6\tilde{e}_6}_{S}\right)^{\tilde{e}_1,n_1}_{E_6,k_6n_6}.
\end{split}
\end{align}
Finally, we show the action of growing an $S$-loop on the minimal torus with trivial edges:
\begin{align} 
\begin{split} \label{eq:minimalTorus}
\includegraphics[valign=c,page=45,scale=0.8]{figuresSN} &= \sum_{\alpha,\beta,\gamma\{h_i\}} \sqrt{\frac{d_{\alpha}d_{\gamma}}{d_{\beta}d_S^2}} \ \left( \F{\triangleright}^{\alpha\gamma S}_{S}\right)^{S,h_3h_1}_{\beta,h_5h_2} \left( \Fi{\triangleright}^{\gamma\alpha S}_{S}\right)^{S,h_1h_3}_{\beta,h_4h_2} \quad \includegraphics[valign=c,page=46,scale=0.8]{figuresSN}.
\end{split}
\end{align}
This action is the first step in the sequential circuit to construct the string-net ground state on the torus (Fig. \ref{torusCircuit}). It guarantees that the state is already in the ground state before the last plaquette operator (\ref{eq:L7}) is performed. Therefore, \ref{eq:L7} is unitary, even though no edge is left to serve as a valid control.

\section{Map to SPTs in two dimensions and higher} \label{app:2dspt}

The sequential circuit in section~\ref{sec:SPT} for generating the CZX state (the 2D $\mathbb{Z}_2$ SPT) from the trivial state can be generalized to group cohomology SPT states with any symmetry and in any dimension \cite{Chen2013}. Following the notation in section~\ref{sec:SPT}, consider fixed point wavefunctions as shown in Fig.~\ref{fig:spt}(c).

There are four spin DOFs in each site with basis states $|g\rangle$, $g\in G$. The symmetry acts on each site as

\begin{equation}
U_g|g_1, g_2, g_3, g_4\rangle \to \frac{\alpha(g_2,g_1) \alpha(g_1,g_4)}{\alpha(g_3,g_4) \alpha(g_2,g_3)}|gg_1, gg_2, gg_3, gg_4\rangle
\end{equation}
where $\alpha(g_i,g_j) = \nu_3(g_i,g_j,g^{-1}g^*,g^*)$. Four spins connected by a black square are in a local entangled state
$\frac{1}{\sqrt{|G|}}\sum_g |gggg\rangle$.
When the entangled states are each contained within a lattice site, the state is a trivial SPT. When the entangled states are between lattice sites, the state is a nontrivial SPT. See Fig.~\ref{fig:spt}(c).

The circuit that maps between the two proceeds with a sequence of SWAP gates, as shown in Fig.~\ref{fig:spt}(d). The sequence of SWAP gates maps between the two wavefunctions. However, after each SWAP gate, the symmetry action changes and needs to be restored.  
Following the notation in Fig.~\ref{fig:spt}(c), after the step of SWAP, Spin $1$ and $5$ are exchanged and spin $4$ and $8$ are exchanged. Therefore, the phase difference that needs to be corrected to restore the symmetry action is
\begin{equation}
\frac{\alpha(g_2,g_1) \cancel{\alpha(g_1,g_4)}}{\alpha(g_3,g_4) \cancel{\alpha(g_2,g_3)}}\frac{\alpha(g_6,g_5) \cancel{\alpha(g_5,g_8)}}{\alpha(g_7,g_8) \cancel{\alpha(g_6,g_7)}} \times \frac{\alpha(g_3,g_8) \cancel{\alpha(g_2,g_3)}}{\alpha(g_2,g_5) \cancel{\alpha(g_5,g_8)}}\frac{\alpha(g_7,g_4) \cancel{\alpha(g_6,g_7)}}{\alpha(g_6,g_1) \cancel{\alpha(g_1,g_4)}}
\end{equation}
which can be separated into two parts for the upper and lower half of the squares
\begin{equation}
\frac{\alpha(g_2,g_1)\alpha(g_6,g_5)}{\alpha(g_2,g_5)\alpha(g_6,g_1)}, \ \ \frac{\alpha(g_3,g_8)\alpha(g_7,g_4)}{\alpha(g_3,g_4)\alpha(g_7,g_8)}
\end{equation}
The two parts can be corrected separately with phase factors acting on $1,2,5,6$ and $4,3,8,7$ respectively. We will focus on the upper part. Using cocycle calculation, we see that
\begin{equation}
\begin{array}{ccc}
\frac{\alpha(g_2,g_1)\alpha(g_6,g_5)}{\alpha(g_2,g_5)\alpha(g_6,g_1)} & = & \frac{\nu_3(g_2,g_1,g^{-1}g^*,g^*)\nu_3(g_6,g_5,g^{-1}g^*,g^*)}{\nu_3(g_2,g_5,g^{-1}g^*,g^*)\nu_3(g_6,g_1,g^{-1}g^*,g^*)} \\
& = &  \frac{\nu_3(g_2,g_1,g^{-1}g^*,g^*)\nu_3(g_6,g_2,g^{-1}g^*,g^*)}{\nu_3(g_6,g_1,g^{-1}g^*,g^*)} \times \frac{\nu_3(g_6,g_5,g^{-1}g^*,g^*)}{\nu_3(g_2,g_5,g^{-1}g^*,g^*)\nu_3(g_6,g_2,g^{-1}g^*,g^*)} \\
& = & \frac{\nu_3(g_6,g_2,g_1,g^*)}{\nu_3(g_6,g_2,g_1,g^{-1}g^*)} \times \frac{\nu_3(g_6,g_2,g_5,g^{-1}g^*)} {\nu_3(g_6,g_2,g_5,g^*)} \\
& = & \frac{\nu_3(g_6,g_2,g_1,g^*)}{\nu_3(gg_6,gg_2,gg_1,g^*)} \times \frac{\nu_3(gg_6,gg_2,gg_5,g^*)} {\nu_3(g_6,g_2,g_5,g^*)}
\end{array}
\end{equation}
which can be achieved by conjugating the symmetry operation by a phase factor of $\frac{\nu_3(g_6,g_2,g_1,g^*)}{\nu_3(g_6,g_2,g_5,g^*)}$.

Similar calculation can be done for the lower part of the squares and for vertical swaps. Therefore, the SWAP gate can be dressed by phase factors to make sure that symmetry operators remain invariant throughout the transformation. Moreover, the phase correction coming from the upper half and the lower half cancel each other when acting on the ground state wavefunction, therefore, the added phases do not change the ground state, so the ground state transforms as we wanted under the SWAP gates. 

A similar construction applies to group cohomology SPT states in higher dimensions as well.

\end{document}